\documentclass[11pt,a4paper]{article} 
\pdfoutput=1
\usepackage{jheppub}
\usepackage{graphicx}
\usepackage{subfigure}
\usepackage{feynmf}

\def\Eq#1{Eq.~(\ref{#1})}
\def\ansatz{{\sl Ansatz}}
\def\ansatze{{\sl Ans\"atze}}
\def\Dim{\text{D}}
\def\pade{{Pad\'e}}
\def\Tproj{{\mathbf T}}
\def\Lproj{{\mathbf L}}
\def\calC{{\mathcal{C}}}
\def\half{\frac{1}{2}}
\def\OO{{\cal O}}

\def\lsim{\mbox{~{\raisebox{0.4ex}{$<$}}\hspace{-1.1em}
        {\raisebox{-0.6ex}{$\sim$}}~}}

\newcommand{\Dphi}{\int \mathcal{D} \phi }
\newcommand{\Gt}{G_3}

\newcommand{\Vgh}{\mathbb{V}}
\newcommand{\Agh}{\mathbb{A}}
\newcommand{\Bgh}{\mathbb{B}}
\newcommand{\tr}{\mathrm{Tr}}
\newcommand{\phibar}{\bar\phi}

\newcommand{\bigO}{\mathcal{O}}

\newcommand{\fmfIR}{\fmfv{decor.shape=circle,d.filled=30,d.size=8mm,l=$\Pi$,l.dist=0mm}}
\newcommand{\fmfcon}{\fmfv{decor.shape=circle,decor.filled=full,decor.size=3mm}}
\newcommand{\fmfHTL}{\fmfv{decor.shape=circle,decor.filled=shaded,decor.size=2.5mm}}

\newcommand{\MSbar}{\overline{\mbox{MS}}}
\def\alphas{\alpha_{\rm s}}
\newcommand{\NPmax}{N^{\text{P}}_{\text{max}}}
\newcommand{\NVmax}{N^{\text{V}}_{\text{max}}}

\title{3-loop 3PI effective action for 3D SU(3) QCD}
\author{Mark C. Abraao York,}
\author{Guy D. Moore,}
\author{Marcus Tassler}
\affiliation{McGill University Department of Physics\\
3600 Rue University\\
Montreal, QC\\
H3A 2T8}

\abstract{The 3PI method is a technique to resum an infinite class of diagrams,
which may be useful in studying nonperturbative thermodynamics and
dynamics in quantum field theory.  But it has never been successfully
applied to gauge theories, where there are serious questions about gauge
invariance breaking.  We show how to perform the 3PI resummation of QCD
in 3 Euclidean spacetime dimensions, a warmup problem to the 4 or 3+1
dimensional case.  We present the complete details of the technical
problems and how they are overcome.  We postpone a comparison of gauge
invariant correlation functions with their lattice-determined
counterparts to a future publication.}

\arxivnumber{1202.4756}

\begin{document}

\maketitle

\section{Introduction}

The early Universe existed in a state of deconfined quark-gluon plasma,
a state which has also recently been produced in the laboratory via
heavy ion collisions.  The thermodynamics of such plasmas are now
relatively well understood.  In the early Universe the gross features
are well described by perturbation theory, and while the strength of the
electroweak phase transition (or crossover) cannot in general be
determined perturbatively, there are powerful lattice methods which can
be brought to bear \cite{KLRS}.  Lattice methods can also describe the
thermodynamics of QCD near
its intrinsic scale $\Lambda_{\rm QCD}$, provided that the chemical
potential is not too large \cite{Karsch, Aoki, Forcrand}.  However, many of the most
interesting questions in each context have to do with dynamics or
unequal time correlations.  For instance, real-time correlations and
nonequilibrium currents are important in understanding whether
electroweak baryogenesis can occur at a first order electroweak phase
transition \cite{Cohen,Cline}.  In the heavy ion context there are many
dynamical quantities we would like to know, such as the viscosity
\cite{Prakash,Muronga,Nakamura,Song}, heavy quark diffusion rate \cite{Rapp1,Rapp2,Svetitsky}, photon
production rate \cite{Steffen,Turbide,David}, and so forth.  Dynamical properties of
QCD are also important at much higher temperatures such as the
electroweak temperature, where they could play a role in baryogenesis
and in various phase transitions.

The problem is that we have no first-principles, intrinsically
nonperturbative technique for theoretically predicting such real-time
properties, even in equilibrium or linear response.  We have models and
phenomenological fitting (for instance, of the viscosity using elliptic
flow in heavy ion collisions \cite{Song}), but the only tool we have
which is close to first principles is perturbation theory.

Perturbation theory is notoriously poorly convergent when applied to hot
gauge theories.  For instance, the expansion for the pressure of QCD as
a series in $\alphas$ is known to order $\alphas^3 \ln \alphas$, but the
series only appears to be useful at temperatures many times the scale
$\Lambda_{\rm QCD}$ \cite{Schroder}.  For real-time quantities, certain
(Hard Thermal Loop or HTL) resummations \cite{HTL}
are necessary even to find leading order results for transport
coefficients \cite{AMY5,AMY6}.  Even so, the perturbative expansion for
real-time quantities appears to be even worse behaved than it is for the
pressure and other thermodynamic quantities, at least if we restrict
attention to transport coefficients, which involve either zero frequency
and momentum or lightlike momentum limits of external 4-momenta.  Only
two such quantities are known beyond leading order; the diffusion
coefficient for a heavy quark \cite{heavy2} and the transverse momentum
diffusion for a fast charge \cite{Simon}.  In each case the first
corrections enter at order $g$, not order $g^2$ as would be normal in a
vacuum perturbative expansion, and the corrections represent of order
$100\%$ shifts in the transport coefficients for $\alphas \sim 0.05$, a
value obtained only at temperatures well above 1 TeV!  Though it is a
little dangerous to extrapolate from two examples, it appears that even
at the electroweak scale, the QCD sector of the Standard Model is
probably not well described by perturbation theory, as far as dynamics
are concerned.

\begin{fmffile}{HTLs}
\begin{figure}
\centering
\begin{tabular}{c@{\hspace{5mm}}c@{\hspace{5mm}}c@{\hspace{5mm}}c}
\parbox{30mm}{\begin{fmfgraph}(30,20)
\fmfforce{0.0w,3mm}{l1}\fmfforce{1.0w,3mm}{r1}
\fmfforce{0.25w,3mm}{v1}\fmfforce{0.75w,3mm}{v2}\fmfforce{0.5w,10.5mm}{vtop}
\fmf{photon,left=0.5}{v1,vtop}\fmf{photon,left=0.5}{vtop,v2}
\fmf{double}{l1,r1}\fmfHTL{vtop}
\end{fmfgraph}}
&&&\\
\parbox{30mm}{\begin{fmfgraph}(30,20)
\fmfforce{0.0w,3mm}{l1}\fmfforce{1.0w,3mm}{r1}
\fmfforce{0.1w,3mm}{v1}\fmfforce{0.35w,10.5mm}{v2}\fmfforce{0.65w,10.5mm}{v3}\fmfforce{0.9w,3mm}{v4}
\fmfforce{0.1732w,8.3033mm}{v5}\fmfforce{0.8268w,8.3033mm}{v6}\fmfforce{0.5w,6mm}{v7}\fmfforce{0.5w,15mm}{v8}
\fmf{photon,left=0.25}{v1,v5,v2}\fmf{photon,left=0.25}{v3,v6,v4}
\fmf{photon,left=0.5}{v2,v8,v3,v7,v2}
\fmfHTL{v2,v3,v5,v6,v7,v8}
\fmf{double}{l1,r1}
\end{fmfgraph}}
&
\parbox{30mm}{\begin{fmfgraph}(30,20)
\fmfforce{0.0w,3mm}{l1}\fmfforce{1.0w,3mm}{r1}
\fmfforce{0.1w,3mm}{v1}\fmfforce{0.9w,3mm}{v2}\fmfforce{0.5w,3mm}{v3}
\fmfforce{0.5w,15mm}{vtop}\fmfforce{0.2172w,11.485mm}{vleft}\fmfforce{0.7828w,11.485mm}{vright}
\fmf{photon,left=0.25}{v1,vleft,vtop,vright,v2}\fmf{photon}{vtop,V1,v3}
\fmfHTL{V1,vtop,vleft,vright}
\fmf{double}{l1,r1}
\end{fmfgraph}}
&
\parbox{30mm}{\begin{fmfgraph}(30,20)
\fmfforce{0.0w,3mm}{l1}\fmfforce{1.0w,3mm}{r1}
\fmfforce{0.25w,3mm}{v1}\fmfforce{0.75w,3mm}{v2}\fmfforce{0.5w,10.5mm}{vtop}
\fmfforce{0.5w,18mm}{vtoptop}
\fmf{photon,left=0.5}{v1,vtop}\fmf{photon,left=0.5}{vtop,v2}
\fmf{photon,left=1}{vtop,vtoptop,vtop}
\fmf{double}{l1,r1}\fmfHTL{vtop,vtoptop}
\end{fmfgraph}}
&
\parbox{30mm}{\begin{fmfgraph}(30,20)
\fmfforce{0.0w,3mm}{l1}\fmfforce{1.0w,3mm}{r1}
\fmfforce{0.25w,3mm}{v1}\fmfforce{0.75w,3mm}{v2}\fmfforce{0.5w,10.5mm}{vtop}
\fmfforce{0.1w,3mm}{V1}\fmfforce{0.9w,3mm}{V2}\fmfforce{0.5w,15mm}{VTOP}
\fmf{photon,left=0.5}{V1,VTOP,V2}\fmf{photon,left=0.5}{v1,vtop,v2}
\fmfHTL{vtop,VTOP}
\fmf{double}{l1,r1}
\end{fmfgraph}}\\
$A$&$B$&$C$&$D$
\end{tabular}
\caption{\label{fig_heavy}NLO heavy quark diffusion.}
\end{figure}
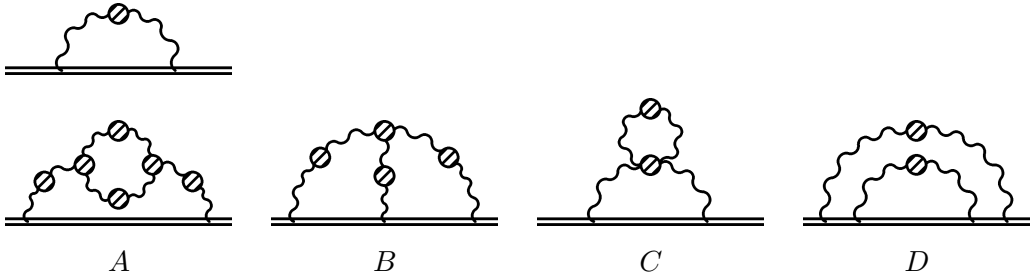

\begin{figure}
\centering
\begin{tabular}{c@{\hspace{5mm}}c@{\hspace{5mm}}c}
\parbox{30mm}{\begin{fmfgraph}(30,20)
\fmfforce{0.0w,3mm}{l1}\fmfforce{1.0w,3mm}{r1}
\fmfforce{0.1w,3mm}{v1}\fmfforce{0.35w,10.5mm}{v2}\fmfforce{0.65w,10.5mm}{v3}\fmfforce{0.9w,3mm}{v4}
\fmfforce{0.1732w,8.3033mm}{v5}\fmfforce{0.8268w,8.3033mm}{v6}\fmfforce{0.5w,6mm}{v7}\fmfforce{0.5w,15mm}{v8}
\fmf{photon,left=0.25}{v1,v5,v2}\fmf{photon,left=0.25}{v3,v6,v4}
\fmf{photon,left=0.5}{v2,v8,v3,v7,v2}
\fmfHTL{v2,v3,v5,v6,v7,v8}
\fmf{double}{l1,r1}
\end{fmfgraph}}
&$\rightarrow$&
\parbox{70mm}{\begin{fmfgraph}(60,20)
\fmfforce{0.0w,3mm}{l1}\fmfforce{1.0w,3mm}{r1}
\fmfforce{0.1w,3mm}{v1}\fmfforce{0.225w,10.5mm}{v2}\fmfforce{0.775w,10.5mm}{v3}\fmfforce{0.9w,3mm}{v4}
\fmfforce{0.1366w,8.3033mm}{v5}\fmfforce{0.8634w,8.3033mm}{v6}
\fmfforce{0.4w,15mm}{vtl}\fmfforce{0.6w,15mm}{vtr}
\fmfforce{0.4w,6mm}{vbl}\fmfforce{0.6w,6mm}{vbr}
\fmf{photon}{vtl,A,vtr,B,vbr,C,vbl,D,vtl}
\fmf{photon}{vbl,E,v2,F,vtl}\fmf{photon}{vbr,G,v3,H,vtr}
\fmf{photon,left=0.25}{v1,v5,v2}\fmf{photon,left=0.25}{v3,v6,v4}
\fmfHTL{v2,v3,v5,v6,vtl,vtr,vbl,vbr,A,B,C,D,E,F,G,H}
\fmf{double}{l1,r1}
\end{fmfgraph}}
\end{tabular}
\caption{\label{fig_heavy2}Inclusion of additional HTLs.}
\end{figure}
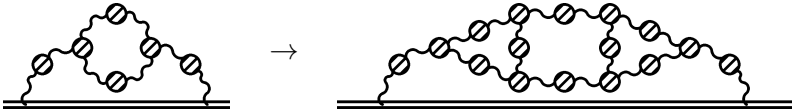
\end{fmffile}

Let us analyze the problem in a little more detail, for the case of
heavy quark diffusion \cite{heavy2}.  The leading-order diagram and the four
next-to-leading order diagrams are shown in Fig.~\ref{fig_heavy}.
Here the double line is the Wilson line heavy quark trajectory, and
hatched blobs represent Hard Thermal Loop (HTL) resummation.  In particular, the
leading-order diagram is already one-loop self-energy resummed.
The contribution from diagram $A$ is 5 to 10 times
as large as that of the other diagrams (in Coulomb gauge, the choice
used in \cite{heavy2}).  Therefore the physics which is problematic for
the perturbative expansion is presumably the physics represented by this
diagram.  This diagram resembles
the leading-order diagram, except that in the leading diagram the
momentum in the self-energy loop is assumed to be large
compared to the propagator momentum, while in diagram $A$ it is allowed
to be of the same order.  When the momentum is soft, it is also necessary
to include HTLs on the propagators and vertices of the
self-energy in diagram $A$.  The large size of diagram $A$ indicates that
the soft loop momentum region in the self-energy is almost as
important as the region where the loop momentum is hard.  This is a
problem because in
this soft region, the HTL corrections must also be
included on the propagators and vertices in diagram $A$.  But why
shouldn't these loops also receive large corrections from their
soft-momentum regions?
Then we also need to include the region where we
replace the HTLs in diagram $A$ with soft momenta, as suggested in
Fig.~\ref{fig_heavy2}.  But when these momenta are soft, there are new
vertices and propagators requiring HTL corrections.  These may also have
large soft corrections, bringing
in more diagrams -- and we are ``off to the races.''

The problem is that the low-momentum ($gT$) region is not really a small
part of phase space, and it is not really weakly coupled.  But the good
news is that the dominance of diagram $A$ above suggests that it is
really only the soft corrections to the HTLs present in the previous
order which are important.  This leads to an infinite number of diagrams
contributing, but only a restricted (infinite) set of diagrams being
``most important'' -- those suggested at in Fig.~\ref{fig_heavy2}.  If
these diagrams could be resummed somehow, then we would likely capture
all the most important corrections, and the range of validity of
perturbation theory might be significantly expanded.

There {\em is} a procedure for performing an iterative resummation of
all one-loop self-energy and 3-point vertex corrections, as suggested in
Fig.~\ref{fig_heavy2}.  It is the 3-loop, 3-particle irreducible (3PI)
resummation scheme \cite{Cornwall,Berges}.  The above discussion
suggests that a 3PI resummation may capture the most important
higher-order physics and greatly improve the convergence of the
perturbative expansion -- even capturing some nonperturbative
information.  But this is by no means guaranteed, since the 3PI approach
performs an incomplete resummation of higher-order effects, which
furthermore is not gauge invariant.  We would like, eventually, to
perform a 3-loop 3PI resummation of real-time, 3+1 dimensional QCD (or
the full Standard Model) at small or intermediate coupling.  However, as
yet the 3PI technique has never been applied to nonabelian gauge
theories. And we would also like to {\sl test} whether the resummation
technique is effective and reliable in a somewhat more controlled
setting.  Therefore we
feel it is necessary to consider a slightly simpler problem as a warm-up
exercise, and as a testing ground for whether the 3PI technique is
effective in nonabelian gauge theory.  Therefore, in this work we will
investigate the 3-loop, 3PI resummation of {\em 3-dimensional} (3D),
SU(3) Yang-Mills theory, an endeavor we will call the
``all threes'' problem (3D, 3-loop 3PI for SU(3)).

Working in 3 Euclidean dimensions simplifies our problem in two ways.
First, the UV behavior of 3D Yang-Mills theory is much milder than in
4D, since the theory is super-renormalizable.  The second simplification
is that the vacuum Euclidean theory has fewer Lorentz invariants than the
finite-temperature, Minkowski theory (or the Euclidean theory with
periodic time direction).  For instance, in (vacuum)
Euclidean space the propagator $G^{\mu\nu}$ is built from two tensorial
structures and is a function of one invariant;
$G^{\mu\nu}(p) = \Tproj^{\mu\nu} G_T(p^2) + \Lproj^{\mu\nu} G_L(p^2)$.
In real time at finite temperature, or in imaginary time with periodic
boundary conditions, it has more tensorial structures and is a function
of {\sl two} variables, $p^2$ and the energy $p^0$.

Note however that 3D Yang-Mills theory is far from trivial.
The flip side of super-renormalizability is that,
since the gauge coupling is dimensionful, it establishes a scale
(momentum scale $p \sim g^2$ or length scale $\lambda \sim 1/g^2$) where
we expect strongly coupled, nonperturbative behavior.  Therefore 3D
Yang-Mills theory displays both weak or strong coupling, depending on
the energy scale.  3D Yang-Mills theory is also physically interesting.
At the thermodynamic level, the infrared behavior of thermal 4D
Yang-Mills theory (with any fermionic matter content) is 3D Yang-Mills
with an adjoint scalar \cite{Appelquist,Braaten,Kajantie}, a slight
extension of the problem we consider.  The nonperturbative scale
$g^2 \equiv g^2_{3\Dim}$ corresponds to the
scale $g^2_{4\Dim} T$ of the full theory (at leading perturbative order).
It is believed that the poor convergence of perturbation
theory at intermediate couplings in QCD is due to the nonperturbative
physics of the 3D theory.  Therefore in a sense studying the 3D gauge
theory by nonperturbative means is treating most of the physics which
makes thermal QCD poorly behaved (at intermediate couplings).

The goal of this paper is to produce a complete solution to the
3-loop truncation of the 3PI effective action for 3D QCD.
As outlined above, this should be viewed as a warm-up problem to what we
would really like to do, which is to apply the resummation in 3+1
dimensions in a thermal (or even nonequilibrium) context.  We feel that
this first step is well motivated.  As already emphasized, the 3D theory
is a subset of the 3+1D theory (the theory we really want to solve).
And the 3D theory can be studied nonperturbatively on a lattice, which
means we will be able to test the 3PI resummation procedure in a
nonperturbative context by seeing whether its predictions are
successful.  Such a test is necessary because the 3PI resummation
captures only an incomplete and gauge-noninvariant subset of diagrams,
so there is no hard guarantee that it will successfully reproduce the
nonperturbative IR physics.  This paper will concentrate on the
resummation; we will return to the comparison with nonperturbative
lattice studies in a follow-up publication.

\section{The 3PI Effective Action}
The 3PI effective action is obtained by a Legendre transform of the generating
functional of connected diagrams
\begin{equation}
 W[J,K,L] = - \log \Dphi ~ e^{-\big(S[\phi] + J\phi + \frac{1}{2}K\phi^2 +
\frac{1}{6}L \phi^3\big)}
\end{equation}
with one, two and three-particle sources, $J$, $K$ and $L$ (our notation
is schematic, with fields generically denoted by $\phi$
and matrix indices and space integrations suppressed).
Conjugate to $J$, $K$ and $L$ are
the variables $\phibar$, $G$ and $\Gt$, which are labelled as such since it can
be proven that they are equal to the connected one, two and three-point functions.  The functional derivatives of $W$ are given by
\begin{eqnarray}
 \label{eq:deltaWdeltaJ}\frac{\delta W}{\delta J} &=& \phibar\\
 \label{eq:deltaWdeltaK}\frac{\delta W}{\delta K} &=& \frac{1}{2}(G +
\phibar^2)\\
 \label{eq:deltaWdeltaL}\frac{\delta W}{\delta L} &=& \frac{1}{6}(\Gt + 3 G\phibar + \phibar^3),
\end{eqnarray}
and the 3PI effective action follows,
\begin{equation}
 \nonumber \Gamma[\phibar,G,\Gt] = J \frac{\delta W}{\delta J} + K
\frac{\delta W}{\delta K} + L \frac{\delta W}{\delta L} - W[J,K,L] .
\end{equation}
{}From Eqs.~(\ref{eq:deltaWdeltaK}) and (\ref{eq:deltaWdeltaL}), we see that $W$
generates connected as well as disconnected diagrams%
\footnote{For instance, $\delta \Gamma / \delta K$ generates $\phibar^2
  \equiv \langle \phi \rangle^2$.},
so it does not, strictly
speaking, generate cumulants. However, $W$ is still equal to the
logarithm of a moment generating functional, hence once can expect that it as
well as its Legendre transform $\Gamma$ have well defined extrema. In terms of $\Gamma$ the equations of motion for $\phibar$, $G$ and $\Gt$ read
\begin{equation}
\label{eq:deltaGammaGeneric}
 \frac{\delta \Gamma}{\delta \phibar} = \frac{\delta \Gamma}{\delta G} =
\frac{\delta \Gamma}{\delta \Gt} = 0 .
\end{equation}

We will now specialize this procedure to the case of QCD.
Since we consider pure-glue QCD there is no Higgs mechanism and we
expect the one-point functions to vanish at the extremum%
\footnote{When Grassman fields such as ghosts are present, the extremum
  is generally a saddle-point rather than a maximum.};
hence we can set the VEVs of
$A^\mu$, $\bar{c}$ and $c$ to zero and work only with the two and three-point functions.  To write down the exact form of $\Gamma$ relevant
to the QCD$_3$ field content, it is useful to define the propagators and
vertex functions%
\footnote{Throughout this work, we will generally suppress color indices
  and will frequently also suppress Lorentz indices, when the indices
  can be inferred from context.  For instance,
  $G = G_{\mu\nu} = G^{ab}_{\mu\nu}$,
  $V_{\mu\nu\rho} = V^{abc}_{\mu\nu\rho}$ {\it etc.} }
listed in Table \ref{tab:QCDFeynRules}.  The one-particle-irreducible
3-vertex $V$ is related to the three-point function via $\Gt = G^3 V$;
extremizing with respect to $G$ and $\Gt$ is equivalent to extremizing
with respect to $G$ and $V$, which we find to be more convenient
variables.  The 3PI effective action is given in terms of  $G$,
$\Delta$, $V$ and $\Vgh$ by
\begin{fmffile}{gamma}
\begin{table}
\centering
\begin{tabular}{l@{\hspace{5mm}}|@{\hspace{5mm}}l@{\hspace{5mm}}|@{\hspace{5mm}}
l}
& Resummed & Bare \\
\hline
Gluon Propagator &
$G^{ab}_{\mu\nu}$ = \parbox{15mm}{\begin{fmfgraph}(15,10)
\fmfleft{l1}\fmfright{r1}
\fmf{dbl_wiggly}{l1,r1}
\end{fmfgraph}} &
$G^{(0)ab}_{\mu\nu}$ = \parbox{15mm}{\begin{fmfgraph}(15,10)
\fmfleft{l1}\fmfright{r1}
\fmf{photon}{l1,r1}
\end{fmfgraph}} \\[4mm]
Ghost Propagator &
$\Delta^{ab}$ = \parbox{15mm}{\begin{fmfgraph}(15,10)
\fmfleft{l1}\fmfright{r1}
\fmf{plain_arrow}{l1,r1}
\end{fmfgraph}} &
$\Delta^{(0)ab}$ = \parbox{15mm}{\begin{fmfgraph}(15,10)
\fmfleft{l1}\fmfright{r1}
\fmf{scalar}{l1,r1}
\end{fmfgraph}}\\[4mm]
Gluon 3-Vertex &
$V^{abc}_{\mu\nu\rho}$ = \parbox{10mm}{\begin{fmfgraph}(10,10)
\fmfleft{lt,lb}\fmfright{r1}
\fmf{dbl_wiggly}{lt,v1}\fmf{dbl_wiggly}{lb,v1}\fmf{dbl_wiggly}{r1,v1}
\fmfcon{v1}
\end{fmfgraph}} ~or \parbox{10mm}{\begin{fmfgraph}(10,10)
\fmfleft{lt,lb}\fmfright{r1}
\fmf{photon}{lt,v1}\fmf{photon}{lb,v1}\fmf{photon}{r1,v1}
\fmfcon{v1}
\end{fmfgraph}} &
$gV^{(0)abc}_{\mu\nu\rho}$ = \parbox{10mm}{\begin{fmfgraph}(10,10)
\fmfleft{lt,lb}\fmfright{r1}
\fmf{dbl_wiggly}{lt,v1}\fmf{dbl_wiggly}{lb,v1}\fmf{dbl_wiggly}{r1,v1}
\end{fmfgraph}} ~or \parbox{10mm}{\begin{fmfgraph}(10,10)
\fmfleft{lt,lb}\fmfright{r1}
\fmf{photon}{lt,v1}\fmf{photon}{lb,v1}\fmf{photon}{r1,v1}
\end{fmfgraph}}\\[4mm]
Ghost-Gluon Vertex &
$\Vgh^{abc}_{\mu}$ = \parbox{10mm}{\begin{fmfgraph}(10,10)
\fmfleft{lt,lb}\fmfright{r1}
\fmf{dbl_wiggly}{lb,v1}\fmf{plain}{lt,v1}\fmf{plain_arrow}{v1,r1}
\fmfcon{v1}
\end{fmfgraph}} ~or \parbox{10mm}{\begin{fmfgraph}(10,10)
\fmfleft{lt,lb}\fmfright{r1}
\fmf{photon}{lb,v1}\fmf{dashes}{lt,v1}\fmf{scalar}{v1,r1}
\fmfcon{v1}
\end{fmfgraph}} &
$g\Vgh^{(0)abc}_{\mu}$ = \parbox{10mm}{\begin{fmfgraph}(10,10)
\fmfleft{lt,lb}\fmfright{r1}
\fmf{dbl_wiggly}{lb,v1}\fmf{plain}{lt,v1}\fmf{plain_arrow}{v1,r1}
\end{fmfgraph}} ~or \parbox{10mm}{\begin{fmfgraph}(10,10)
\fmfleft{lt,lb}\fmfright{r1}
\fmf{photon}{lb,v1}\fmf{dashes}{lt,v1}\fmf{scalar}{v1,r1}
\end{fmfgraph}}\\[4mm]
Gluon 4-Vertex & &
$g^2V^{(0)abcd}_{\mu\nu\rho\tau}$ = \parbox{10mm}{\begin{fmfgraph}(10,10)
\fmfleft{l1,l2}\fmfright{r1,r2}
\fmf{dbl_wiggly}{l1,v1}\fmf{dbl_wiggly}{l2,v1}
\fmf{dbl_wiggly}{v1,r1}\fmf{dbl_wiggly}{v1,r2}
\end{fmfgraph}} ~or \parbox{10mm}{\begin{fmfgraph}(10,10)
\fmfleft{l1,l2}\fmfright{r1,r2}
\fmf{photon}{l1,v1}\fmf{photon}{l2,v1}
\fmf{photon}{v1,r1}\fmf{photon}{v1,r2}
\end{fmfgraph}}
\end{tabular}
\caption{\label{tab:QCDFeynRules}Feynman Rules for QCD$_3$}
\end{table}
\begin{eqnarray}
\label{eq:Gamma}
\nonumber \Gamma &=& S[A^\mu,\bar c,c] + \frac{1}{2}\tr \log G -
\frac{1}{2} \tr [G^{(0)}]^{-1} G - \tr\log\Delta
+ \tr [\Delta^{(0)}]^{-1}\Delta \\
\nonumber &+& \frac{1}{6}~
\parbox{20mm}{\begin{fmfgraph}(20,20)
\fmfforce{1mm,0.5h}{v1}\fmfforce{19mm,0.5h}{v2}
\fmf{dbl_wiggly,left,tension=0.4}{v1,v2,v1}
\fmf{dbl_wiggly}{v1,v2}
\fmfcon{v1}
\end{fmfgraph}} ~- \frac{1}{12}~
\parbox{20mm}{\begin{fmfgraph}(20,20)
\fmfforce{1mm,0.5h}{v1}\fmfforce{19mm,0.5h}{v2}
\fmfcon{v1,v2}
\fmf{dbl_wiggly,left,tension=0.4}{v1,v2,v1}
\fmf{dbl_wiggly}{v1,v2}
\end{fmfgraph}} ~+ \frac{1}{8}~
\parbox{20mm}{\begin{fmfgraph}(20,20)
\fmfforce{1mm,0.5h}{v1}\fmfforce{19mm,0.5h}{v3}
\fmfforce{0.5w,0.5h}{v2}
\fmf{dbl_wiggly,right,tension=0.4}{v1,v2,v1}
\fmf{dbl_wiggly,left,tension=0.4}{v3,v2,v3}
\end{fmfgraph}} \\
\nonumber &-&~
\parbox{20mm}{\begin{fmfgraph}(20,20)
\fmfforce{1mm,0.5h}{v1}\fmfforce{19mm,0.5h}{v2}
\fmf{plain_arrow,left,tension=0.4}{v1,v2,v1}
\fmf{dbl_wiggly}{v1,v2}
\fmfcon{v1}
\end{fmfgraph}} ~+ \frac{1}{2}~
\parbox{20mm}{\begin{fmfgraph}(20,20)
\fmfforce{1mm,0.5h}{v1}\fmfforce{19mm,0.5h}{v2}
\fmfcon{v1,v2}
\fmf{plain_arrow,left,tension=0.4}{v1,v2,v1}
\fmf{dbl_wiggly}{v1,v2}
\end{fmfgraph}}\\
\nonumber &+& \frac{1}{48}~
\parbox{20mm}{\begin{fmfgraph}(20,20)
\fmfforce{1mm,0.5h}{v1}\fmfforce{19mm,0.5h}{v2}
\fmf{dbl_wiggly,left,tension=0.4}{v1,v2,v1}
\fmf{dbl_wiggly,left=0.5}{v1,v2}
\fmf{dbl_wiggly,right=0.5}{v1,v2}
\end{fmfgraph}} ~+ \frac{1}{24}~
\parbox{20mm}{\begin{fmfgraph}(20,20)
\fmfforce{10mm,10mm}{v4}\fmfforce{2.206mm,5.5mm}{v1}
\fmfforce{17.794mm,5.5mm}{v2}
\fmfforce{10mm,19mm}{v3}
\fmfcon{v1,v2,v3,v4}
\fmf{dbl_wiggly,left=0.55}{v1,v3,v2,v1}
\fmf{dbl_wiggly}{v1,v4,v3}
\fmf{dbl_wiggly}{v2,v4,v3}
\fmf{dbl_wiggly}{v1,v4,v2}
\end{fmfgraph}} ~+ \frac{1}{8}~
\parbox{20mm}{\begin{fmfgraph}(20,20)
\fmfforce{1mm,10mm}{v1}\fmfforce{19mm,10mm}{v2}
\fmfforce{10mm,19mm}{v3}\fmfforce{10mm,1mm}{v4}
\fmfcon{v1,v2}
\fmf{dbl_wiggly,right=0.5,tension=1}{v4,v2,v3,v1,v4}
\fmf{dbl_wiggly,right=0.5,tension=0.4}{v1,v3}
\fmf{dbl_wiggly,left=0.5,tension=0.4}{v2,v3}
\end{fmfgraph}} \\
&-& \frac{1}{3}~
\parbox{20mm}{\begin{fmfgraph}(20,20)
\fmfforce{10mm,10mm}{v4}\fmfforce{2.206mm,5.5mm}{v1}
\fmfforce{17.794mm,5.5mm}{v2}
\fmfforce{10mm,19mm}{v3}
\fmfcon{v1,v2,v3,v4}
\fmf{plain_arrow,left=0.6,tension=0.4}{v1,v3,v2,v1}
\fmf{dbl_wiggly}{v1,v4,v3}
\fmf{dbl_wiggly}{v2,v4,v3}
\fmf{dbl_wiggly}{v1,v4,v2}
\end{fmfgraph}} ~- \frac{1}{4}~
\parbox{20mm}{\begin{fmfgraph}(20,20)
\fmfforce{10mm,10mm}{v4}\fmfforce{2.206mm,5.5mm}{v1}
\fmfforce{17.794mm,5.5mm}{v2}
\fmfforce{10mm,19mm}{v3}
\fmfcon{v1,v2,v3,v4}
\fmf{plain_arrow,left=0.575,tension=0.4}{v1,v3,v2}
\fmf{dbl_wiggly,left=0.575,tension=0.4}{v2,v1}
\fmf{plain_arrow}{v4,v1}
\fmf{plain_arrow}{v2,v4}
\fmf{dbl_wiggly}{v3,v4}
\end{fmfgraph}},
\end{eqnarray}
\end{fmffile}%
(where we have explicitly written symmetry factors associated with
diagrams and signs associated with ghost loops for clarity, as we will
throughout),
from which we have the following four equations of motion
\begin{eqnarray}
\label{eq:deltaGamma}
\frac{\delta \Gamma}{\delta G} = \frac{\delta \Gamma}{\delta \Delta}
= \frac{\delta
\Gamma}{\delta V} =
\frac{\delta \Gamma}{\delta \Vgh} = 0 .
\end{eqnarray}

To illustrate the physics of \Eq{eq:deltaGamma}, consider
$\delta \Gamma / \delta G(p)$.  Performing the variation using the
expression in \Eq{eq:Gamma}, we find
\begin{fmffile}{pi1}
\begin{eqnarray}
\label{prop_resum}
G^{-1}(p) & = & [G^0]^{-1}(p) - \Pi(p)\,, \\
\Pi^{(1)}(p) & = &
\parbox{20mm}{\begin{fmfgraph}(20,20)
\fmfleft{l1}\fmfright{r1}
\fmfforce{0.25w,0.5h}{v1}\fmfforce{0.75w,0.5h}{v3}
\fmf{dbl_wiggly}{l1,v1}\fmf{dbl_wiggly,left=1,tension=0.4}{v1,v3,v1}
\fmf{dbl_wiggly}{r1,v3}
\fmfcon{v1}
\end{fmfgraph}}
 - \frac{1}{2}~
\parbox{20mm}{\begin{fmfgraph}(20,20)
\fmfleft{l1}\fmfright{r1}
\fmfforce{0.25w,0.5h}{v1}\fmfforce{0.75w,0.5h}{v3}
\fmf{dbl_wiggly}{l1,v1}\fmf{dbl_wiggly,left=1,tension=0.4}{v1,v3,v1}
\fmf{dbl_wiggly}{r1,v3}
\fmfcon{v1,v3}
\end{fmfgraph}}
 + \frac{1}{2}~
\parbox{20mm}{\begin{fmfgraph}(20,20)
\fmfleft{l1}\fmfright{r1}\fmftop{t1}\fmfforce{0.5w,0.5h}{v1}
\fmf{dbl_wiggly}{l1,v1}\fmf{phantom,tension=5.0}{t1,v2}
\fmf{dbl_wiggly,left=1,tension=0.4}{v1,v2,v1}
\fmf{dbl_wiggly}{r1,v1}
\end{fmfgraph}}
 - 2~
\parbox{20mm}{\begin{fmfgraph}(20,20)
\fmfleft{l1}\fmfright{r1}
\fmfforce{0.25w,0.5h}{v1}\fmfforce{0.75w,0.5h}{v3}
\fmf{dbl_wiggly}{l1,v1}\fmf{plain_arrow,left=1,tension=0.4}{v1,v3,v1}
\fmf{dbl_wiggly}{r1,v3}
\fmfcon{v1}
\end{fmfgraph}}
 + ~
\parbox{20mm}{\begin{fmfgraph}(20,20)
\fmfleft{l1}\fmfright{r1}
\fmfforce{0.25w,0.5h}{v1}\fmfforce{0.75w,0.5h}{v3}
\fmf{dbl_wiggly}{l1,v1}\fmf{plain_arrow,left=1,tension=0.4}{v1,v3,v1}
\fmf{dbl_wiggly}{r1,v3}
\fmfcon{v1,v3}
\end{fmfgraph}}  \nonumber
\end{eqnarray}\end{fmffile}%
which we recognize as the resummed one-loop self-energy
($\Pi=\Pi^{(1)} + \Pi^{(2)}$, we have not shown the similar graphical representation of $\Pi^{(2)}$).
Similarly, variation with respect to $V$ gives
\begin{fmffile}{vert1}
\begin{equation}
\parbox{20mm}{\begin{fmfgraph}(20,20)
\fmfleft{l1}\fmfright{r1,r2,r3}
\fmf{dbl_wiggly}{l1,v1}\fmf{dbl_wiggly}{r1,v1}\fmf{dbl_wiggly}{r3,v1}
\fmfcon{v1}
\end{fmfgraph}}
{} = {}~
\parbox{20mm}{\begin{fmfgraph}(20,20)
\fmfleft{l1}\fmfright{r1,r2,r3}
\fmf{dbl_wiggly}{l1,v1}\fmf{dbl_wiggly}{r1,v1}\fmf{dbl_wiggly}{r3,v1}
\end{fmfgraph}}
+ {}~
\parbox{20mm}{\begin{fmfgraph}(20,20)
\fmfleft{l1}\fmfright{r1,r2,r3}
\fmf{dbl_wiggly}{l1,v1}\fmf{dbl_wiggly}{r1,v2}\fmf{dbl_wiggly}{r3,v3}
\fmf{dbl_wiggly,tension=0.3}{v1,v2,v3,v1}
\fmfcon{v1,v2,v3}
\end{fmfgraph}}
~+ {} \frac{3}{2}~
\parbox{20mm}{\begin{fmfgraph}(20,20)
\fmfleft{l1}\fmfright{r1,r2,r3}
\fmf{dbl_wiggly}{l1,v1}\fmf{dbl_wiggly}{r1,v2}\fmf{dbl_wiggly}{r3,v2}
\fmf{dbl_wiggly,left=1,tension=0.4}{v1,v2,v1}
\fmfcon{v1}
\end{fmfgraph}}
~- {}2~
\parbox{20mm}{\begin{fmfgraph}(20,20)
\fmfleft{l1}\fmfright{r1,r2,r3}
\fmf{dbl_wiggly}{l1,v1}\fmf{dbl_wiggly}{r1,v2}\fmf{dbl_wiggly}{r3,v3}
\fmf{plain_arrow,tension=0.3}{v1,v2,v3,v1}
\fmfcon{v1,v2,v3}
\end{fmfgraph}}
\label{vert_resum}
\end{equation}
\end{fmffile}%
which we recognize as a Schwinger-Dyson equation for vertex
resummation.

Note that the propagators appearing in all diagrams in
\Eq{prop_resum} and \Eq{vert_resum}, as well as all vertices (except
certain vertices in the one-loop self-energies), are the full objects.
Therefore these equations must be solved {\sl self-consistently}.
The self-consistent solution of these equations represents our main
challenge.  We face two chief difficulties:
\begin{itemize}
\item
Decomposing the propagator into its transverse and longitudinal parts,
\begin{eqnarray}
G^{\mu\nu}(p) &=& G^{\mu\nu}_T(p)\Tproj^{\mu\nu} + G^{\mu\nu}_L(p)\Lproj^{\mu\nu} \,, \nonumber \\
\Lproj^{\mu\nu} & \equiv & \frac{p^\mu p^\nu}{p^2} \,, \qquad
\Tproj^{\mu\nu} \equiv g^{\mu\nu} - \Lproj^{\mu\nu} \,,
\end{eqnarray}
the propagators are determined in terms of three {\sl arbitrary}
functions of one continuous variable,
$G_T(p), G_L(p)$ and $\Delta(p)$ with $p=\sqrt{p^2}$.  Similarly, the vertex
$V_{\mu_1\mu_2\mu_3}(p_1,p_2,p_3)$ (with $p_{3\mu}=-p_{1\mu}-p_{2\mu}$) can be expressed
in terms of {\sl six} independent tensorial structures (see below), each
multiplying an undetermined function of the {\sl three} invariants
$p_1^2$, $p_2^2$, and $p_1 \cdot p_2$ (or equivalently $p_1^2$, $p_2^2$,
and $p_3^2$).

The challenge is that we are not merely solving for a few numbers, but
self-consistently solving for unknown functions of one to three
continuous variables.
\item
The one-loop gluon self-energy diagrams are linearly divergent, and the
two-loop gluon self-energy diagrams are individually logarithmically
divergent.  These divergences must be regulated in a manner which
respects gauge invariance,%
\footnote{One might argue that, since the 3PI technique truncated to 3 loops is not gauge invariant, the use of a gauge invariant regulator
  is unnecessary.  But we believe that it is necessary; first, our
  approach at least retains gauge invariance to low loop order, which
  would be lost without a gauge invariant regularization.  And second, a
  gauge non-invariant regularization at 1-loop order would allow
  divergent masses, which fundamentally damage the physics.}
such as dimensional regularization.  However
since the propagators and vertices appearing in the diagrams are general
functions of momentum which are presumably only known numerically and
only in $\Dim=3$ dimensions, we will have to perform these integrations
numerically.
\end{itemize}

The issue of divergences in self-energies is a technical issue which can
be handled rather easily in 3 dimensions.  The key fact is that at large
momenta $G,\Delta,V$ and $\Vgh$ approach their free values up to power
suppressed corrections.  We therefore know the exact form of the UV
divergences.  If we can find an expression with the same UV divergent
behavior which is simple enough to integrate using dimensional
regularization, we can add and subtract it.  The subtraction renders the
numerical integration of the full self-energy expressions finite, while
the added version is integrated using dimensional regularization.  We
will explain this procedure in more detail in the next section.

In order to fit arbitrary functions of one or a few real variables, we
will write down a sufficiently flexible \ansatz\ for each function, with
some set of variational parameters.  That is, we take
$G_T(p) = G_T(c_i,p)$ where $c_i$ are coefficients -- in practice, we
take $G_T(p)$ to be a rational function of $p$, and the
$c_i$ are coefficients of this rational function.
Extremization of $\Gamma$ with respect to $G(p)$ is then replaced by its
extremization with respect to the coefficients $c_i$ of each propagator
and vertex function.  Under this procedure, some set of integral moments
of the Schwinger-Dyson equations will be satisfied, rather than the
equations being satisfied at every momentum value.  We can determine the
quality and limitations of this approach by seeing how the determined
correlation functions change as the sizes of the variational \ansatze\
are changed; and we can directly test how well the Schwinger-Dyson
equations are obeyed by computing directly the self-energies and vertex
corrections at various momenta and comparing to the \ansatz, or
measuring an integrated mean squared failure of the Schwinger-Dyson
equations.

\section{Divergences and Regularization}

\label{sec:divergence}

As discussed above, variation of $\Gamma$ with respect to a propagator
gives rise to a Schwinger-Dyson equation involving self-energies
written in terms of $G$ and $V$. These will be, in general, complicated functions.
Yet the self-energies may be UV divergent and so they must be
regulated.  Even after taming these self-energy divergences, the
variation with respect to a propagator or vertex \ansatz\ coefficient
$\delta\Gamma/\delta c_i$ may lead to a divergent integral over the propagator
momentum $p$.  We must also ensure that such divergences do not occur.
We will handle these two problems in turn.
Throughout we denote the momentum entering a self-energy as $p$, and use
$k$ and $q$ for internal loop momenta.

\subsection{Divergences in self-energies}

\label{sec:divergeself}

In 3 dimensions the only divergent subdiagrams are gluon self-energies.
To handle these divergences we must work in a regularization scheme
which renders the self-energy diagrams finite {\em and} preserves gauge
invariance.  Therefore we will perform all integrals in dimensional
regularization (DR), so $\int d^3 q \rightarrow \int d^\Dim q$ with
$\Dim=3+2\epsilon$.  Unfortunately the self-energies contain the functions
$G$ and $V$, which are complicated and are only known in 3 dimensions.
However, for any integral which is finite and well behaved in 3
dimensions, the $\epsilon\rightarrow 0$ limit of the DR value is the
same as the value directly computed in 3 dimensions.  Therefore we will start with
identifying the UV divergent behavior of the full integrals containing $G$
and $V$ so that we can subtract and add simple integrals with the same divergences. We can then perform the (finite) subtracted versions numerically in 3
dimensions, and finish off by adding back the simple integrals using DR.

Dressed vertices and propagators are well behaved in the IR, hence the
only divergences that we expect to see arise from the region of momentum
space where $q$ is large. Therefore we need to determine the asymptotic
behaviors of $G$ and $V$.  Our theory is super-renormalizable, meaning
that the coupling $g^2$ carries dimension, $[g^2]=1=[q]$.  For large
$q$, $g^2$ is small compared to the relevant scale $q$, so the large $q$
region is weakly coupled and has a perturbative expansion.  Further,
powers of $g^2$ in the expansion must be balanced against powers of $q$
on dimensional grounds.  Therefore the leading and first subleading
behavior of the propagator in 3 dimensions is
\begin{equation}
\label{eq:Gtaylorexp}
      G_{\mu\nu}(q) = \frac{1}{q^2}\big(
      \Tproj_{\mu\nu}(q) +\xi \Lproj_{\mu\nu}(q)\big)
      + \frac{g^2 \Pi^{\text{B}(1)}}{q^3}\Tproj_{\mu\nu}(q) + \bigO (q^{-4})\,.
\end{equation}
Similarly, the vertex goes as
\begin{equation}
 V \sim q + g^2 q^0 + \bigO (q^{-1}).
\end{equation}
The specific form of the $g^2$ correction to $V$ is known, see
Appendix \ref{sec:vertex}; but as we see in a moment we do not need it
here.  The one-loop correction to the gluon self-energy is also
known.  It is purely transverse and equals \cite{Appelquist}
\begin{equation}
\Pi^{\text{B}(1)}_{\mu\nu}(q) = q \frac{g^2 N}{64} (\xi^2 + 2\xi + 11)
                        \Tproj_{\mu\nu}(q),
\end{equation}
hence $\Pi^{\text{B}(1)}$ introduced in \Eq{eq:Gtaylorexp} is
\begin{equation}
\Pi^{\text{B}(1)} = \frac{N}{64}(\xi^2 + 2\xi + 11) \,.
\end{equation}

Now, consider the diagram\begin{fmffile}{dimregprocedure}
\begin{equation}
\label{eq:oneloopdressed}
\parbox{20mm}{\begin{fmfgraph}(20,20)
\fmfleft{l1}\fmfright{r1}\fmfforce{0.25w,0.5h}{v1}\fmfforce{0.75w,0.5h}{v3}
\fmf{dbl_wiggly}{l1,v1}\fmf{dbl_wiggly,left=1,tension=0.4}{v1,v3,v1}
\fmf{dbl_wiggly}{r1,v3}\fmfcon{v1,v3}\end{fmfgraph}} ~ =
\int \frac{d^\Dim q}{(2\pi)^\Dim} ~ V_{\mu\alpha\delta}V_{\nu\beta\kappa} G^{\alpha\beta}(p+q) G^{\delta\kappa}(q) ,
\end{equation}
where traces over internal color indices are implied (and hence the
overall diagram is proportional to the color identity). Expanding the
integrand in powers of $q$, for $\Dim=3$ the large $q$ region of the
integral behaves as
\begin{equation}
\label{eq:linandlog}
\parbox{20mm}{\begin{fmfgraph}(20,20)
\fmfleft{l1}\fmfright{r1}\fmfforce{0.25w,0.5h}{v1}\fmfforce{0.75w,0.5h}{v3}
\fmf{dbl_wiggly}{l1,v1}\fmf{dbl_wiggly,left=1,tension=0.4}{v1,v3,v1}
\fmf{dbl_wiggly}{r1,v3}\fmfcon{v1,v3}\end{fmfgraph}} ~ \sim
g^2
\int d^3 q\left[ (q)^2\frac{1}{(q^2)^2}  \\
+ \underbrace{2g^2(q)^2 \frac{\Pi^{\text{B}(1)}}{q^3}\frac{1}{q^2}}_{\text{NLO propagator}}
+ \underbrace{2g^2(q)\frac{1}{(q^2)^2}}_{\text{NLO~vertex}}
+ \underbrace{\bigO (g^4q^{-4})}_{\text{NLO}^2 + \text{NNLO}} + ~ ...\right] .
\end{equation}
The first term arises from the leading order (bare) terms in the
vertices and propagators, and the next two terms originate from the
one-loop corrections (as marked). These first three integrals diverge,
so we will have to add and subtract something to cancel their divergent
behavior.

Actually, the NLO vertex corrections above will cancel when we sum over
the one-loop self-energy corrections.  To see this, consider the two
diagrams, with one and with two full vertices:
\begin{equation}
\label{eq:NumSelfEnergy2}\Bigg [
~\parbox{20mm}{\begin{fmfgraph}(20,20)
\fmfleft{l1}\fmfright{r1}\fmfforce{0.25w,0.5h}{v1}\fmfforce{0.75w,0.5h}{v3}
\fmf{dbl_wiggly}{l1,v1}\fmf{dbl_wiggly,left=1,tension=0.4}{v1,v3,v1}
\fmf{dbl_wiggly}{r1,v3}\fmfcon{v1}\end{fmfgraph}}~
- \frac{1}{2} ~
\parbox{20mm}{\begin{fmfgraph}(20,20)
\fmfleft{l1}\fmfright{r1}\fmfforce{0.25w,0.5h}{v1}\fmfforce{0.75w,0.5h}{v3}
\fmf{dbl_wiggly}{l1,v1}\fmf{dbl_wiggly,left=1,tension=0.4}{v1,v3,v1}
\fmf{dbl_wiggly}{r1,v3}\fmfcon{v1,v3}\end{fmfgraph}}~\Bigg ]_{q\gg g^2}
\sim \frac{1}{2} ~
\parbox{20mm}{\begin{fmfgraph}(20,20)
\fmfleft{l1}\fmfright{r1}\fmfforce{0.25w,0.5h}{v1}\fmfforce{0.75w,0.5h}{v3}
\fmf{dbl_wiggly}{l1,v1}\fmf{dbl_wiggly,left=1,tension=0.4}{v1,v3,v1}
\fmf{dbl_wiggly}{r1,v3}\end{fmfgraph}} ~ \Bigg |_{q\gg g^2}
\end{equation}
The diagram with one full vertex enters with $-2$ times the weight of
the diagram with two full vertices.  Therefore the NLO vertex
contributions from these two diagrams cancel, and the UV behavior is the
same at NLO as the behavior of a loop with no vertex corrections.
Provided that we perform the two diagrams by adding their integrands
inside the integration, this cancellation takes place at the level of
the integrand and does not lead to a log divergence in the integral in 3
dimensions.
(This cancellation does not mean that the NLO vertex correction
disappears; instead this correction will be accounted for explicitly
when we include two-loop self-energy corrections.)

Next consider the bare part of \Eq{eq:linandlog}, which is linearly
divergent in 3 dimensions.  We will add and subtract a diagram made
out of the bare vertex and propagator functions,
\begin{equation}
\label{eq:oneloopbare}
\parbox{20mm}{\begin{fmfgraph}(20,20)
\fmfleft{l1}\fmfright{r1}\fmfforce{0.25w,0.5h}{v1}\fmfforce{0.75w,0.5h}{v3}
\fmf{dbl_wiggly}{l1,v1}\fmf{photon,left=1,tension=0.4}{v1,v3,v1}
\fmf{dbl_wiggly}{r1,v3}\end{fmfgraph}} ~ =
g^2\int_q ~ V^{(0)}_{\mu\alpha\delta}V^{(0)}_{\nu\beta\kappa} G^{(0)\alpha\beta}(p+q) G^{(0)\delta\kappa}(q)
\end{equation}
with $G^{(0)}$'s are $V^{(0)}$'s denoting bare propagators and
vertices.  The difference
\begin{equation}
\frac{1}{2} ~ \parbox{20mm}{\begin{fmfgraph}(20,20)
\fmfleft{l1}\fmfright{r1}\fmfforce{0.25w,0.5h}{v1}\fmfforce{0.75w,0.5h}{v3}
\fmf{dbl_wiggly}{l1,v1}\fmf{dbl_wiggly,left=1,tension=0.4}{v1,v3,v1}
\fmf{dbl_wiggly}{r1,v3}\end{fmfgraph}} ~ - \frac{1}{2} ~
\parbox{20mm}{\begin{fmfgraph}(20,20)
\fmfleft{l1}\fmfright{r1}\fmfforce{0.25w,0.5h}{v1}\fmfforce{0.75w,0.5h}{v3}
\fmf{dbl_wiggly}{l1,v1}\fmf{photon,left=1,tension=0.4}{v1,v3,v1}
\fmf{dbl_wiggly}{r1,v3}\end{fmfgraph}}
\end{equation}
is only logarithmically divergent in 3 dimensions.
Moreover, \Eq{eq:oneloopbare} is finite when computed in DR and
its $\Dim\rightarrow 3$ limit is
\begin{equation}
\label{eq:oneloopbareDR}
\half \:
\parbox{20mm}{\begin{fmfgraph}(20,20)
\fmfleft{l1}\fmfright{r1}\fmfforce{0.25w,0.5h}{v1}\fmfforce{0.75w,0.5h}{v3}
\fmf{dbl_wiggly}{l1,v1}\fmf{photon,left=1,tension=0.4}{v1,v3,v1}
\fmf{dbl_wiggly}{r1,v3}\end{fmfgraph}} ~\Bigg\vert_{\text{DR}} =
p\frac{g^2 N}{64}\left((\xi^2 + 2\xi + 11)\Tproj_{\mu\nu}(p)
-g_{\mu\nu} - \frac{p_\mu p_\nu}{p^2}\right)  \,.
\end{equation}
This diagram, plus the bare ghost diagram which cancels the
non-transverse piece above, gives rise to $\Pi^{\text{B}(1)}$ stated earlier.

Lastly, we must subtract something with the same NLO ``propagator''
behavior remaining in \Eq{eq:linandlog}.  Naively, we could do this by
defining
\begin{equation}
G^{(1,\epsilon)}_{\mu\nu}(p) =
  G^{(0)}_{\mu\alpha}(p)\Pi^{\text{B}(1,\epsilon)\alpha\beta}(p)
            G^{(0)}_{\beta\nu}(p)
=
\frac{\Pi^{\text{B}(1,\epsilon)}}{\mu^{2\epsilon}}\frac{1}{p^{3-2\epsilon}}
           \Tproj_{\mu\nu}(p)
\end{equation}
with $\Pi^{\text{B}(1,\epsilon)}_{\alpha\beta}(p)$ and $\Pi^{\text{B}(1,\epsilon)}$
defined by Eqs.~(\ref{eq:Pi1emunu}) and (\ref{eq:Pi1e}), and then
``adding and subtracting'' the following diagram:
\begin{eqnarray}
\label{eq:Pi2IRinfinite}
 \parbox{20mm}{\begin{fmfgraph*}(20,20)
\fmfleft{l1}\fmfright{r1}
\fmfforce{0.25w,0.5h}{v1}\fmfforce{0.75w,0.5h}{v3}\fmfforce{0.5w,0.25h}{v2}
\fmfforce{0.5w,0.75h}{v4}\fmf{dbl_wiggly}{l1,v1}
\fmf{photon,left=0.5,tension=0.4}{v1,v4,v3,v2,v1}\fmf{dbl_wiggly}{r1,v3}
\fmfv{decor.shape=circle,d.filled=empty,d.size=8mm,l=$\Pi$,l.dist=0mm}{v4}
\end{fmfgraph*}} ~
&=&  g^2\int_q ~ V^{(0)}_{\mu\alpha\delta}V^{(0)}_{\nu\beta\kappa}
      G^{(0)\alpha\beta}(p+q) G^{(1,\epsilon)\delta\kappa}(q)
\nonumber \\
& = & \frac{\mathcal{A}}{\epsilon}g_{\mu\nu}
    - \frac{\mathcal{B}}{\epsilon}\left(g_{\mu\nu}
    - \frac{p_\mu p_\nu}{p^2}\right) +
\text{finite} \,.
\end{eqnarray}
The problem is that, as the above equation shows, the diagram is not
only UV divergent (as expected, with coefficient ${\cal A}$ which we give
below), but also IR divergent (with coefficient ${\cal B}$, whose exact
value will not be relevant).  If we add and subtract this diagram, we
will cause an IR divergence where none should appear.  Instead, we will
add and subtract an appropriately regulated self-energy corrected
propagator,
\begin{equation}
\label{eq:GIRReg}G^{(1,\epsilon)\text{IR Reg}}_{\mu\nu} (p) = \frac{\Pi^{\text{B}(1,\epsilon)}}{\mu^{2\epsilon}}\frac{1}{(p^2 + m^2)^{\frac{3}{2}-\epsilon}} \left(  g_{\mu\nu} - \frac{p_\mu p_\nu}{p^2 + m^2}\right)
\end{equation}
so that the integration in
\begin{equation}
\label{eq:Pi2IRfinite}
\parbox{20mm}{\begin{fmfgraph*}(20,20)
\fmfleft{l1}\fmfright{r1}
\fmfforce{0.25w,0.5h}{v1}\fmfforce{0.75w,0.5h}{v3}\fmfforce{0.5w,0.25h}{v2}
\fmfforce{0.5w,0.75h}{v4}\fmf{dbl_wiggly}{l1,v1}
\fmf{photon,left=0.5,tension=0.4}{v1,v4,v3,v2,v1}\fmf{dbl_wiggly}{r1,v3}
\fmfIR{v4}\end{fmfgraph*}} ~ = g^2\int_q ~ V^{(0)}_{\mu\alpha\delta}V^{(0)}_{\nu\beta\kappa} G^{(0)\alpha\beta}(p+q) G^{(1,\epsilon)\text{IR Reg}~\delta\kappa}(q)
\end{equation}
can still be performed analytically in DR. Upon integration, this diagram has
the following form:
\begin{equation}
\label{eq:Pi2IRfinite2}
\parbox{20mm}{\begin{fmfgraph*}(20,20)
\fmfleft{l1}\fmfright{r1}
\fmfforce{0.25w,0.5h}{v1}\fmfforce{0.75w,0.5h}{v3}\fmfforce{0.5w,0.25h}{v2}
\fmfforce{0.5w,0.75h}{v4}\fmf{dbl_wiggly}{l1,v1}
\fmf{photon,left=0.5,tension=0.4}{v1,v4,v3,v2,v1}\fmf{dbl_wiggly}{r1,v3}
\fmfIR{v4}\end{fmfgraph*}} ~ \Bigg |_{\text{DR}}
= \frac{\mathcal{A}}{\epsilon}g_{\mu\nu} + \text{finite} .
\end{equation}
The coefficient
that multiplies the UV $1/\epsilon$,
\begin{equation}
\mathcal{A} = - \frac{g^4 N^2}{768 \pi^2}\frac{p^{4\epsilon}}{\mu^{4\epsilon}}(\xi + 4)(\xi^2 + 2\xi + 11) ,
\end{equation}
is identical to that of \Eq{eq:Pi2IRinfinite} due to the simple fact that
\begin{equation}
\lim_{q\rightarrow \infty} G^{(1,\epsilon)\text{IR Reg}}_{\mu\nu}(q) = \lim_{q\rightarrow\infty}G^{(1,\epsilon)}_{\mu\nu}(q) .
\end{equation}
Naturally, the finite parts of \Eq{eq:Pi2IRinfinite}
and \Eq{eq:Pi2IRfinite} will differ.

It may worry some readers that we have introduced an IR mass regulator.
But we emphasize that we are {\sl not} adding such a regulator to the
full propagator $G^{\mu\nu}(p)$.  We are only adding an IR mass regulator
to a term which we add and subtract, for reasons of computational
convenience.  Hence, the value of the regulator -- in fact, the effect
of the whole term which we are adding and subtracting -- exactly cancels
when we combine the (analytic) result of \Eq{eq:Pi2IRfinite} and the
(numerical) result of the full but subtracted diagram in \Eq{eq:PiSD}.
We have naturally checked that the value of the regulator in
\Eq{eq:GIRReg} has no effect on our results for the full self-energy and
therefore for the determined value of the full propagator.

Returning to \Eq{eq:linandlog}, we now have
\begin{eqnarray}
\label{eq:Pi2LogSubtracted}
&& \Bigg[
~\parbox{20mm}{\begin{fmfgraph}(20,20)
\fmfleft{l1}\fmfright{r1}\fmfforce{0.25w,0.5h}{v1}\fmfforce{0.75w,0.5h}{v3}
\fmf{dbl_wiggly}{l1,v1}\fmf{dbl_wiggly,left=1,tension=0.4}{v1,v3,v1}
\fmf{dbl_wiggly}{r1,v3}\fmfcon{v1}\end{fmfgraph}}~
- \frac{1}{2} ~
\parbox{20mm}{\begin{fmfgraph}(20,20)
\fmfleft{l1}\fmfright{r1}\fmfforce{0.25w,0.5h}{v1}\fmfforce{0.75w,0.5h}{v3}
\fmf{dbl_wiggly}{l1,v1}\fmf{dbl_wiggly,left=1,tension=0.4}{v1,v3,v1}
\fmf{dbl_wiggly}{r1,v3}\fmfcon{v1,v3}\end{fmfgraph}}
 ~ - \frac{1}{2} ~
\parbox{20mm}{\begin{fmfgraph}(20,20)
\fmfleft{l1}\fmfright{r1}\fmfforce{0.25w,0.5h}{v1}\fmfforce{0.75w,0.5h}{v3}
\fmf{dbl_wiggly}{l1,v1}\fmf{photon,left=1,tension=0.4}{v1,v3,v1}
\fmf{dbl_wiggly}{r1,v3}\end{fmfgraph}} ~ - ~
\parbox{20mm}{\begin{fmfgraph*}(20,20)
\fmfleft{l1}\fmfright{r1}
\fmfforce{0.25w,0.5h}{v1}\fmfforce{0.75w,0.5h}{v3}\fmfforce{0.5w,0.25h}{v2}
\fmfforce{0.5w,0.75h}{v4}\fmf{dbl_wiggly}{l1,v1}
\fmf{photon,left=0.5,tension=0.4}{v1,v4,v3,v2,v1}\fmf{dbl_wiggly}{r1,v3}
\fmfIR{v4}\end{fmfgraph*}} ~ \Bigg] \nonumber\\
&+& \Bigg[ \frac{1}{2} ~ \parbox{20mm}{\begin{fmfgraph}(20,20)
\fmfleft{l1}\fmfright{r1}\fmfforce{0.25w,0.5h}{v1}\fmfforce{0.75w,0.5h}{v3}
\fmf{dbl_wiggly}{l1,v1}\fmf{photon,left=1,tension=0.4}{v1,v3,v1}
\fmf{dbl_wiggly}{r1,v3}\end{fmfgraph}} ~ + ~
\parbox{20mm}{\begin{fmfgraph*}(20,20)
\fmfleft{l1}\fmfright{r1}
\fmfforce{0.25w,0.5h}{v1}\fmfforce{0.75w,0.5h}{v3}\fmfforce{0.5w,0.25h}{v2}
\fmfforce{0.5w,0.75h}{v4}\fmf{dbl_wiggly}{l1,v1}
\fmf{photon,left=0.5,tension=0.4}{v1,v4,v3,v2,v1}\fmf{dbl_wiggly}{r1,v3}
\fmfIR{v4}\end{fmfgraph*}} ~ \Bigg]
\sim \frac{\mathcal{A}}{\epsilon}g_{\mu\nu} + \text{finite}\,.
\end{eqnarray}
The first line is the only part which contains full propagators; but it
is finite at $\Dim=3$.  Therefore its value in DR in the $\Dim\rightarrow 3$
limit simply equals its finite value in 3 dimensions, which we can find
by numerical integration.  The second line is divergent in 3D but can be
carried out relatively easily in DR.  It gives rise to the
${\cal A} g_{\mu\nu}/\epsilon$ contribution, and to some of the finite terms presented
in \Eq{eq:Pi2IRReg} (\Eq{eq:Pi2IRReg} is the sum of \Eq{eq:Pi2IRfinite} as well as a similarly IR regulated Snowcone diagram).
\end{fmffile}

The procedure for handling the 2-loop graphs is similar.  While the
graphs are more complicated, the procedure is simpler, since in every
case the only UV divergences arise when all components of the graph take
their bare values.  Further, no 2-loop graph we need, when built out of bare
quantities, is IR divergent.  Therefore we may simply subtract from each
2-loop graph, built using $G$ and $V$, the same graph built using $G^{(0)}$
and $V^{(0)}$.  The bare 2-loop graphs can be performed in DR and will also give rise
to a $g_{\mu\nu}/\epsilon$ divergence plus a finite part; the sum of these diagrams is given by \Eq{eq:2loopbare}. In the end, we find that all of the $1/\epsilon$'s cancel between two-loop diagrams: the $\bigO (g^4)$ IR regulated gluon self-energy is UV finite in DR (the complete expression for which is given by Eqs.~(\ref{eq:2loopbare}) and (\ref{eq:Pi2IRReg})).

With this procedure in place, the Schwinger-Dyson equation,
\Eq{prop_resum}, becomes
\begin{fmffile}{gammaglderivative}
\begin{eqnarray}
\label{eq:PiSD}G^{-1}(p) & = & G_{0}^{-1}(p) - \Pi^{(1)}(p) - \Pi^{(2)}(p) \,, \\
\nonumber\Pi^{(1)}_{\mu\nu} + \Pi^{(2)}_{\mu\nu} & = &
\parbox{20mm}{\begin{fmfgraph}(20,20)
\fmfleft{l1}\fmfright{r1}
\fmfforce{0.25w,0.5h}{v1}\fmfforce{0.75w,0.5h}{v3}
\fmf{dbl_wiggly}{l1,v1}\fmf{dbl_wiggly,left=1,tension=0.4}{v1,v3,v1}
\fmf{dbl_wiggly}{r1,v3}
\fmfcon{v1}
\end{fmfgraph}}
~ - \frac{1}{2}~
\parbox{20mm}{\begin{fmfgraph}(20,20)
\fmfleft{l1}\fmfright{r1}
\fmfforce{0.25w,0.5h}{v1}\fmfforce{0.75w,0.5h}{v3}
\fmf{dbl_wiggly}{l1,v1}\fmf{dbl_wiggly,left=1,tension=0.4}{v1,v3,v1}
\fmf{dbl_wiggly}{r1,v3}
\fmfcon{v1,v3}
\end{fmfgraph}}
~ + \frac{1}{2}~
\parbox{20mm}{\begin{fmfgraph}(20,20)
\fmfleft{l1}\fmfright{r1}\fmftop{t1}\fmfforce{0.5w,0.5h}{v1}
\fmf{dbl_wiggly}{l1,v1}\fmf{phantom,tension=5.0}{t1,v2}
\fmf{dbl_wiggly,left=1,tension=0.4}{v1,v2,v1}
\fmf{dbl_wiggly}{r1,v1}
\end{fmfgraph}}
~ - 2~
\parbox{20mm}{\begin{fmfgraph}(20,20)
\fmfleft{l1}\fmfright{r1}
\fmfforce{0.25w,0.5h}{v1}\fmfforce{0.75w,0.5h}{v3}
\fmf{dbl_wiggly}{l1,v1}\fmf{plain_arrow,left=1,tension=0.4}{v1,v3,v1}
\fmf{dbl_wiggly}{r1,v3}
\fmfcon{v1}
\end{fmfgraph}}
\\ \nonumber &&
~ + ~
\parbox{20mm}{\begin{fmfgraph}(20,20)
\fmfleft{l1}\fmfright{r1}
\fmfforce{0.25w,0.5h}{v1}\fmfforce{0.75w,0.5h}{v3}
\fmf{dbl_wiggly}{l1,v1}\fmf{plain_arrow,left=1,tension=0.4}{v1,v3,v1}
\fmf{dbl_wiggly}{r1,v3}
\fmfcon{v1,v3}
\end{fmfgraph}} \:
- \frac{1}{2}~
\parbox{20mm}{\begin{fmfgraph}(20,20)
\fmfleft{l1}\fmfright{r1}
\fmfforce{0.25w,0.5h}{v1}\fmfforce{0.75w,0.5h}{v3}
\fmf{dbl_wiggly}{l1,v1}
\fmf{photon,left=1,tension=0.4}{v1,v3,v1}
\fmf{dbl_wiggly}{r1,v3}
\end{fmfgraph}}
~ - ~
\parbox{20mm}{\begin{fmfgraph*}(20,20)
\fmfleft{l1}\fmfright{r1}
\fmfforce{0.25w,0.5h}{v1}\fmfforce{0.75w,0.5h}{v3}
\fmfforce{0.5w,0.25h}{v2}\fmfforce{0.5w,0.75h}{v4}
\fmf{dbl_wiggly}{l1,v1}\fmf{photon,left=0.5,tension=0.4}{v1,v4,v3,v2,v1}
\fmf{dbl_wiggly}{r1,v3}
\fmfIR{v4}
\end{fmfgraph*}}
~ - \frac{1}{2}~
\parbox{20mm}{\begin{fmfgraph}(20,20)
\fmfleft{l1}\fmfright{r1}\fmftop{t1}\fmfforce{0.5w,0.5h}{v1}
\fmf{dbl_wiggly}{l1,v1}\fmf{phantom,tension=5.0}{t1,v2}
\fmf{photon,left=1,tension=0.4}{v1,v2,v1}
\fmf{dbl_wiggly}{r1,v1}
\end{fmfgraph}}
\\ \nonumber  &&
~ - \frac{1}{2}~
\parbox{20mm}{\begin{fmfgraph*}(20,20)
\fmfleft{l1}\fmfright{r1}\fmftop{t1}\fmfforce{0.5w,0.5h}{v1}
\fmf{dbl_wiggly}{l1,v1}\fmf{phantom,tension=5.0}{t1,v2}
\fmf{photon,left=1,tension=0.4}{v1,v2,v1}
\fmf{dbl_wiggly}{r1,v1}\fmffreeze
\fmfIR{v2}
\end{fmfgraph*}}
+ \parbox{20mm}{\begin{fmfgraph}(20,20)
\fmfleft{l1}\fmfright{r1}
\fmfforce{0.25w,0.5h}{v1}\fmfforce{0.75w,0.5h}{v3}
\fmf{dbl_wiggly}{l1,v1}\fmf{scalar,left=1,tension=0.4}{v1,v3,v1}
\fmf{dbl_wiggly}{r1,v3}
\end{fmfgraph}}
~ + 2~
\parbox{20mm}{\begin{fmfgraph}(20,20)
\fmfleft{l1}\fmfright{r1}\fmfforce{0.25w,0.5h}{v1}\fmfforce{0.75w,0.5h}{v2}
\fmftop{vt}\fmfforce{0.5w,2.5mm}{vb}
\fmf{dbl_wiggly}{l1,v1}
\fmf{dbl_wiggly}{r1,v2}
\fmf{dashes,left=0.5,tension=3}{v1,v3}\fmf{dashes,left=0.5,tension=3}{v4,v2}
\fmf{photon,left=1,tension=0.4}{v3,v4}\fmf{dashes,left=1,tension=0.4}{v4,v3}
\fmf{scalar,left=1}{v2,v1}
\fmf{phantom,tension=3}{v3,vt,v4}
\end{fmfgraph}}\\
\nonumber &&+
\frac{1}{6}~
\parbox{20mm}{\begin{fmfgraph}(20,20)
\fmfleft{l1}\fmfright{r1}\fmfforce{0.25w,0.5h}{v1}\fmfforce{0.75w,0.5h}{v2}
\fmf{dbl_wiggly}{l1,v1}\fmf{dbl_wiggly,left=1}{v1,v2,v1}
\fmf{dbl_wiggly}{v1,v2}
\fmf{dbl_wiggly}{r1,v2}
\end{fmfgraph}}
~ + \frac{1}{2}~
\parbox{20mm}{\begin{fmfgraph}(20,20)
\fmfleft{l1}\fmfright{r1}\fmfforce{0.25w,0.5h}{v1}\fmfforce{0.75w,0.5h}{v2}
\fmfforce{0.5w,0.75h}{vt}\fmfforce{0.5w,0.25h}{vb}
\fmf{dbl_wiggly}{l1,v1}
\fmf{dbl_wiggly,left=0.5}{v1,vt,v2,vb,v1}
\fmf{dbl_wiggly}{r1,v2}\fmf{dbl_wiggly}{vt,vb}
\fmfcon{v1,v2,vt,vb}
\end{fmfgraph}}
~ + ~
\parbox{20mm}{\begin{fmfgraph}(20,20)
\fmfleft{l1}\fmfright{r1}\fmfforce{0.25w,0.5h}{v1}\fmfforce{0.75w,0.5h}{v2}
\fmfforce{0.5w,0.75h}{vt}\fmfforce{0.5w,0.25h}{vb}
\fmf{dbl_wiggly}{l1,v1}
\fmf{dbl_wiggly,left=0.5}{v1,vt,v2,vb,v1}
\fmf{dbl_wiggly}{r1,v2}
\fmf{dbl_wiggly,right=0.5}{vt,v2}
\fmfcon{v1,vt}
\end{fmfgraph}}
~ + \frac{1}{4}~
\parbox{20mm}{\begin{fmfgraph}(20,20)
\fmfleft{l1}\fmfright{r1}\fmfforce{0.25w,0.5h}{v1}\fmfforce{0.75w,0.5h}{v3}
\fmfforce{0.5w,0.5h}{v2}
\fmf{dbl_wiggly}{l1,v1}
\fmf{dbl_wiggly,left=1}{v1,v2,v1}
\fmf{dbl_wiggly,left=1}{v3,v2,v3}
\fmf{dbl_wiggly}{r1,v3}
\fmfcon{v1,v3}
\end{fmfgraph}}\\
\nonumber &&-
\parbox{20mm}{\begin{fmfgraph}(20,20)
\fmfleft{l1}\fmfright{r1}\fmfforce{0.25w,0.5h}{v1}\fmfforce{0.75w,0.5h}{v2}
\fmfforce{0.5w,0.75h}{vt}\fmfforce{0.5w,0.25h}{vb}
\fmf{dbl_wiggly}{l1,v1}
\fmf{plain_arrow,left=0.5}{v1,vt}
\fmf{plain,left=0.5}{vt,v2}
\fmf{plain_arrow,left=0.5}{v2,vb}
\fmf{plain,left=0.5}{vb,v1}
\fmf{dbl_wiggly}{r1,v2}\fmf{dbl_wiggly}{vt,vb}
\fmfcon{v1,v2,vb,vt}
\end{fmfgraph}}
~ - 2~
\parbox{20mm}{\begin{fmfgraph}(20,20)
\fmfleft{l1}\fmfright{r1}\fmfforce{0.25w,0.5h}{v1}\fmfforce{0.75w,0.5h}{v2}
\fmfforce{0.5w,0.75h}{vt}\fmfforce{0.5w,0.25h}{vb}
\fmf{dbl_wiggly}{l1,v1}
\fmf{plain,left=0.5}{v1,vt}
\fmf{plain,left=0.5}{vb,v1}
\fmf{dbl_wiggly,left=0.5}{vt,v2,vb}
\fmf{dbl_wiggly}{r1,v2}\fmf{plain_arrow}{vt,vb}
\fmfcon{v1,v2,vt,vb}
\end{fmfgraph}}\\
\nonumber &&- \frac{1}{6}~
\parbox{20mm}{\begin{fmfgraph}(20,20)
\fmfleft{l1}\fmfright{r1}\fmfforce{0.25w,0.5h}{v1}\fmfforce{0.75w,0.5h}{v2}
\fmf{dbl_wiggly}{l1,v1}\fmf{photon,left=1}{v1,v2,v1}
\fmf{photon}{v1,v2}
\fmf{dbl_wiggly}{r1,v2}
\end{fmfgraph}}
~ - \frac{1}{2}~
\parbox{20mm}{\begin{fmfgraph}(20,20)
\fmfleft{l1}\fmfright{r1}\fmfforce{0.25w,0.5h}{v1}\fmfforce{0.75w,0.5h}{v2}
\fmfforce{0.5w,0.75h}{vt}\fmfforce{0.5w,0.25h}{vb}
\fmf{dbl_wiggly}{l1,v1}
\fmf{photon,left=0.5}{v1,vt,v2,vb,v1}
\fmf{dbl_wiggly}{r1,v2}\fmf{photon}{vt,vb}
\end{fmfgraph}}
~ - ~
\parbox{20mm}{\begin{fmfgraph}(20,20)
\fmfleft{l1}\fmfright{r1}\fmfforce{0.25w,0.5h}{v1}\fmfforce{0.75w,0.5h}{v2}
\fmfforce{0.5w,0.75h}{vt}\fmfforce{0.5w,0.25h}{vb}
\fmf{dbl_wiggly}{l1,v1}
\fmf{photon,left=0.5}{v1,vt,v2,vb,v1}
\fmf{dbl_wiggly}{r1,v2}
\fmf{photon,right=0.5}{vt,v2}
\end{fmfgraph}}
~ - \frac{1}{4}~
\parbox{20mm}{\begin{fmfgraph}(20,20)
\fmfleft{l1}\fmfright{r1}\fmfforce{0.25w,0.5h}{v1}\fmfforce{0.75w,0.5h}{v3}
\fmfforce{0.5w,0.5h}{v2}
\fmf{dbl_wiggly}{l1,v1}
\fmf{photon,left=1}{v1,v2,v1}
\fmf{photon,left=1}{v3,v2,v3}
\fmf{dbl_wiggly}{r1,v3}
\end{fmfgraph}}\\
\nonumber &&+~
\parbox{20mm}{\begin{fmfgraph}(20,20)
\fmfleft{l1}\fmfright{r1}\fmfforce{0.25w,0.5h}{v1}\fmfforce{0.75w,0.5h}{v2}
\fmfforce{0.5w,0.75h}{vt}\fmfforce{0.5w,0.25h}{vb}
\fmf{dbl_wiggly}{l1,v1}
\fmf{scalar,left=0.5}{v1,vt}
\fmf{dashes,left=0.5}{vt,v2}
\fmf{scalar,left=0.5}{v2,vb}
\fmf{dashes,left=0.5}{vb,v1}
\fmf{dbl_wiggly}{r1,v2}\fmf{photon}{vt,vb}
\end{fmfgraph}}
~ + 2~
\parbox{20mm}{\begin{fmfgraph}(20,20)
\fmfleft{l1}\fmfright{r1}\fmfforce{0.25w,0.5h}{v1}\fmfforce{0.75w,0.5h}{v2}
\fmfforce{0.5w,0.75h}{vt}\fmfforce{0.5w,0.25h}{vb}
\fmf{dbl_wiggly}{l1,v1}
\fmf{dashes,left=0.5}{v1,vt}
\fmf{dashes,left=0.5}{vb,v1}
\fmf{photon,left=0.5}{vt,v2,vb}
\fmf{dbl_wiggly}{r1,v2}\fmf{scalar}{vt,vb}
\end{fmfgraph}}\\
&&+~ {\prod}^{\text{B}(1,0)}_{\mu\nu} \big(p\big)+\lim_{\epsilon \rightarrow 0} \bigg ( {\prod}^{\text{B}(2,\epsilon) \text{UV}}_{\mu\nu} \big ( p \big) +
 {\prod}^{\text{B}(2,\epsilon) \text{IR Reg}}_{\mu\nu}\big(p\big)
\bigg)\nonumber . \label{eq:GammaGPDeriv}
\end{eqnarray}\end{fmffile}%
where the $\Pi^{\text{B}(a,b)}$'s represent the sum
all bare diagrams computed analytically in dimensional regularization (their
values can be read from Appendix. \ref{sec:gluonself}, noting that
$\Pi^{\text{B}(2,\epsilon)\text{IR Reg}}_{\mu\nu}(p)$ is exactly as stated in
\Eq{eq:Pi2IRReg}, with no large $p$ limit taken).  In writing this
equation we have suppressed the Lorentz indices, as written one should
take $\half \Tproj_{\mu\nu} \mathbf{\Pi}^{\mu\nu}$ to get the transverse part of the
self-energy needed to resum $G_T$ and $\Lproj_{\mu\nu} \mathbf{\Pi}^{\mu\nu}$ to get
  the self-energy needed to resum $G_L$.

All ghost self-energies are finite after angular averaging, and the
vertex loops are power-counting finite, so no similar subtractions are
needed in these cases.  Nevertheless, we will still encounter
divergences when it comes time to integrate these Schwinger-Dyson
equations over propagator or vertex momenta.

\subsection{Divergences upon $p$-integration}

\label{sec:divergep}

Our plan is to find an approximate extremum of $\Gamma$ by writing
variational \ansatze\ for the propagators and vertices and to vary with
respect to the \ansatz\ parameters.  For instance, one could assume that
the transverse propagator $G_T(p)$ is the sum of a set of test functions
with unknown coefficients, $G_T(p) = \sum_i c_i \phi_i(p)$.  More
generally, we choose $G_T(p)$ to have some functional form with a set of
variational parameters $c_i$; we will give our specific choice in
Section \ref{sec:ansatz}.  Then
variation of $\Gamma$ with respect to $c_j$ would yield
\begin{equation}
\label{eq:intSD}
\frac{\delta \Gamma}{\delta c_j} = \int d^3 p
     \frac{\delta G_T(p)}{\delta c_j} \frac{\delta \Gamma}{\delta G_T(p)}
 = \int d^3 p \frac{\delta G_T(p)}{\delta c_j}
 \left( G_T^{-1}(p) - G^{(0)-1}_{T}(p) + \Pi_T(p) \right) \,,
\end{equation}
and similarly for $G_L$ and $\Delta$.  In the
last section we ensured that the integrals involved in all self-energies
$\Pi$ are finite.  But this does not guarantee that the $p$ integral
above will be finite.  For instance, if we chose the \ansatz\
\begin{equation}
G_T[p,c_i,\mbox{example}] = \frac{c_1}{p^2+m^2} +
      \frac{c_2}{(p^2+m^2)^{\frac 32}}
     + \frac{c_3}{(p^2+m^2)^2}
\end{equation}
then $\delta G_T(p)/\delta c_1 = \frac{1}{p^2 + m^2}$.  And if
$c_1 \neq 1$, then for large $p$,
$G_T^{-1}(p) - G_{T}^{(0)-1}(p) + \Pi_T(p) \sim p^2$.  In this case
\Eq{eq:intSD} would be cubically divergent.  Physically this means
that if we allow $G(p)$ to vary from its correct value in a way which
does not die away fast in the UV, then $\Gamma$ will be divergently far
from its extremum.

Continuing with the same example, if we fix $c_1=1$, forcing the
propagator to have the correct free-theory limit in the UV, then the
most severe UV divergence arises from $c_2$.  For large $p$ we have
$\delta G_T(p) / \delta c_2 \sim p^{-3}$ and
$G_T^{-1}(p) - G_{T}^{(0)-1}(p) + \Pi_T(p) \sim p$ (by virtue of the
cancellation of the $p^2$ terms in $G_T^{-1}$ and $G_{T}^{(0)-1}$).
In this case the integral is $\sim \int d^3p (1/p^3) (p)$ which is
linearly divergent.  This is better but still unacceptable.
To ensure a finite answer we must choose an \ansatz\ which automatically
enforces the right $\OO(p^2)$ {\sl and} $\OO(p)$ behavior in
$G_T^{-1}(p)$, namely,
\begin{equation}
\label{eq:GTtaylorexp}
G_T(p) = \frac{1}{p^2} + \frac{g^2 \Pi^{\text{B}(1)}}{p^3}
        +\mbox{(\ansatz\ starting at $\OO(p^{-4})$)} \,.
\end{equation}
In this case, $\delta G(p) / \delta c_i \lsim p^{-4}$
automatically, and
$G_{T}^{-1}(p) - G_{T}^{(0)-1}(p) + \Pi_T(p) \sim p^0$.
This is sufficient to ensure that the integral in \Eq{eq:intSD}
will be UV finite.
The same argument applies to $G_L$ and $\Delta$; in each case we must
build in the correct $1/p^2$ {\sl and} $1/p^3$ behavior of the
propagator (or $\OO(p^2)$ and $\OO(p)$ behavior in the inverse
propagator) into our \ansatz; but having done so, the variations
$\delta \Gamma / \delta c_i$ will all automatically be finite (unless
there are IR problems).

Applying the same reasoning to the vertices, the variation of $\Gamma$
with respect to a generic coefficient $d_i$ determining $V$ gives rise
to a correction of order (the phase space is explained in
Appendix \ref{sec:setsun})
\begin{equation}
\frac{\delta \Gamma}{\delta d_i}
\sim \int p dp \: q dq \: k dk
     \: \frac{\delta V}{\delta d_i} \:  G(p) G(q) G(k)
     \: \Big( V - V_{\text{bare}} - \delta V \Big) \,.
\end{equation}
If we allow our \ansatz\ to change $V$ on the scale of its leading
behavior $\sim p,k,q$ then this expression is quadratically divergent.
Therefore our \ansatz\ must be restricted such that $V$ takes on its
correct (free) asymptotic limiting behavior.  Even so, in this case
$\delta V / \delta d_i$ and $V-V_{\rm bare}$ will be $\OO(p^0)$, giving
rise to a log divergence.  Therefore we must compute and implement the
first subleading behavior of $V$ and only allow our \ansatz\ to change
$V$ at NNLO, $\OO(p^{-1},k^{-1},q^{-1})$.  This will ensure finite
variations of $\Gamma$ with respect to the parameters $d_i$ (again
assuming there are no infrared issues).

It is not necessary to determine the NNLO behavior of either self-energy
or vertex corrections in order to avoid potential divergences.  This is
a good thing, because the $\OO(p^{-4})$ propagator correction
(or $\OO(p^0)$ self-energy correction) is where nonperturbative physics
first arises.   To see this, consider the one-loop self-energy diagram
in \Eq{eq:oneloopdressed}.  Let us estimate the contribution when the
external momentum $p$ is large but one of the internal propagators is at
a small momentum $q \sim g^2$.  There is a factor of $g^2 p^2$ from the
vertices, $1/p^2$ from the hard propagator, and
$\int d^3 q / q^2 \sim q \sim g^2$ from the momentum integration and
soft propagator in the soft region.  So the contribution when one
propagator is soft is of order $\Pi_{\rm soft} \sim g^4$.  This
contribution is nonperturbative because the behavior of the propagator
at small momentum is.  Therefore we actually {\sl cannot} determine the
NNLO behavior of the propagator at large momenta; the perturbative
expansion we alluded to earlier actually fails at this order.
Fortunately, determining this order turns out to be unnecessary to
eliminate divergences and render our extremization problem well posed.

Note that the elimination of divergences, both in subdiagrams and in the
final variation of $\Gamma$ with respect to variational \ansatz\
parameters, is much easier in $\Dim=3$ spacetime dimensions than it would
be in $\Dim=4$.  In that case, self-energies would be quadratically
divergent at all loop orders and $\delta \Gamma / \delta c_i$ would
generically be quartically divergent.  It is therefore not completely
clear to us how our procedure could be extended to four dimensions without
some changes or restrictions.  We will not address this
issue further at this time.

\section{Variational \ansatze}

\label{sec:ansatz}

We will now start to actually solve, rather than discuss, the
problem by writing out the variational \ansatze\ used for all
propagators and vertices. The only dimensionful constant in 3D
Yang-Mills theory is $g$. Moreover, a three-loop truncation of $\Gamma$
only contains planar diagrams, and the only subleading in $N$ correction
which enters when evaluating them is an overall factor of
$(N^2{-}1)$; that is, the coupling and group theory factor for an $m$-loop
bubble diagram is $(N^2{-}1) (g^2 N)^{m-1}$.  Therefore, to the loop order we
work, the coupling expansion is strictly an expansion in the 't Hooft
coupling $g^2 N$.  In $\Dim=3$ dimensions the 't Hooft coupling has dimensions
of energy, and it therefore sets the natural energy scale in
the problem.  Therefore we will factor out the overall $(N^2{-}1)$ and
will scale all dimensionful quantities by the appropriate power of
$g^2 N$, {\it i.e.}, quantities with dimension $[\text{mass}]^\alpha$, are
expressed in units of $(g^2 N)^\alpha$.  For the most part, this eliminates any explicit
reference to $g^2$ or $N$ in what follows.

Variational coefficients will be generically denoted by $c_i$. This is a
slight abuse of notation, and one should recall throughout that the
$c_i$ are independent for each function. For instance, it should not be
interpreted from expressions like $G_T(c_i;p)$ and $G_L(c_i;p)$ that
$G_T$ and $G_L$ are defined by the same set of parameters.
In practice we will use rational functions (\pade\ approximants) for our
variational \ansatze; we distinguish coefficients in numerators from
those in denominators by labelling the former by $a_i$ and the latter by
$b_i$, so that $\{c_i\}=\{a_i,b_i\}$ (or $\{c_i\}=\{a_{ijk},b_{ijk}\}$ for
vertex function coefficients). Finally, it is now always implied that
when we refer to a correlation function, we are specifically referring
to its \ansatz. We will drop the $c_i$ from the arguments of these
functions, so $G(p)$ implies $G(c_i;p)$.

Continuing on, $G$ is first decomposed into transverse and longitudinal
components
\begin{equation}
 G^{\mu\nu}(p) = G_T(p)\Tproj^{\mu\nu}(p) + G_L(p) \Lproj^{\mu\nu}(p) .
\end{equation}
{}From the arguments of the previous section, we know that whatever we
write down for $G_T(p)$ has to converge to \Eq{eq:GTtaylorexp} at
large $p$ (and likewise for $G_L(p)$ and $\Delta(p)$).  Hence,
\begin{eqnarray}\label{eq:GTsimpleansatz}
G_T(p) = \frac{1}{p^2-\Pi_T(p)},\qquad \Pi_T(p) &=& \calC_T^{(1)} p
        +\mbox{(\ansatz\ starting at $\OO(p^{0})$)}\\
G_L(p) = \frac{\xi}{p^2-\xi\Pi_L(p)},\qquad \Pi_L(p) &=& \calC_L^{(1)} p
        +\mbox{(\ansatz\ starting at $\OO(p^{0})$)}\\
\label{eq:Deltasimpleansatz}\Delta(p) = \frac{1}{p^2(1-\Sigma(p)/p^2)},\qquad \Sigma(p)/p^2 &=& \frac{\calC_\Delta^{(1)}}{p + \omega} +\mbox{(\ansatz\ starting at $\OO(p^{-2})$)}
\end{eqnarray}
with
\begin{equation}
\calC_T^{(1)}= \frac{\xi^2+\xi+11}{64} , \qquad \calC_L^{(1)} = 0 , \qquad \calC_\Delta^{(1)} = \frac{1}{16} .
\end{equation}
We see that $\calC_{T/L/\Delta}^{(1)}$ must be measured in units of $g^2N$ since $[\calC_{T/L/\Delta}^{(1)}] = [\text{mass}]$, while $[G_T] = [\text{mass}]^{-2}$.

The \ansatz\ for $\Delta$ differs from $G_T$ and $G_L$; namely, we
assume that $\Sigma(p) \propto p^2$, and therefore
$\Delta \propto p^{-2}$,  at small $p$.  This is a condition which
arises due to the structure of ghost vertices, which we will discuss in
a little more detail when we present the ghost vertices.
The parameter $\omega$ is not treated as
a variational parameter; instead its value is fixed to $\omega=1$
(really $\omega=g^2 N$).  This choice should not be important, provided
the variational \ansatz\ is flexible enough.

We will use \pade\ approximants for the
propagator \ansatze. The ``order'' of these \ansatze\ will be denoted by
$\NPmax$, which refers to the highest power of momentum appearing in the
numerator and denominators. With this choice, Eqs.~(\ref{eq:GTsimpleansatz}) -
(\ref{eq:Deltasimpleansatz}) read
\begin{eqnarray}
 \label{eq:GT} \Pi_T(p) &=& \calC^{(1)}_T p + \Pi_T^{\text{NP}}(p) = \calC^{(1)}_T p
+ \frac{\sum_{i=0}^{\NPmax} a^{\{G_T\}}_ip^i}{\sum_{i=1}^{\NPmax} b^{\{G_T\}}_i
p^i + 1}\\
 \label{eq:GL} \Pi_L(p) &=& \calC^{(1)}_L p + \Pi_L^{\text{NP}}(p) = \frac{\sum_{i=0}^{\NPmax} a^{\{G_L\}}_ip^i}{\sum_{i=1}^{\NPmax} b^{\{G_L\}}_i
p^i + 1}\\
\label{eq:DP} \Sigma(p) &=& \calC^{(1)}_\Delta p
 + \Sigma^{\text{NP}}(p) =
\frac{\calC^{(1)}_\Delta p^2}{p+\omega} +
 \frac{\sum_{i=2}^{\NPmax} a^{\{\Delta\}}_ip^i}
  {\sum_{i=1}^{\NPmax} b^{\{\Delta\}}_i p^i + 1}.
\end{eqnarray}
In each case we define $\Pi^{\text{NP}}$ as the self-energy minus its
one-loop perturbative (linear in momentum) part.  In the case of
$\Sigma$ this is {\sl not} the same as the part determined by the
variational \ansatz.

In constructing the most general gluon 3-vertex $V$ (where it is assumed
that momentum flows out of a vertex), six independent tensor structures
require consideration.
We will adopt the basis used in Refs.~\cite{Ball,Davydychev}, which is
\begin{eqnarray}
\label{BallChiu}
\mathbf{A}_{\mu_1\mu_2\mu_3} &=& g_{\mu_1\mu_2}(p_1 - p_2)_{\mu_3}
\nonumber \\
\mathbf{B}_{\mu_1\mu_2\mu_3} &=& g_{\mu_1\mu_2}(p_1 + p_2)_{\mu_3} \nonumber \\
\mathbf{C}_{\mu_1\mu_2\mu_3} &=& \big [p_1\cdot p_2 g_{\mu_1\mu_2} -
p_{1\mu_2}p_{2\mu_1}\big ](p_1 - p_2)_{\mu_3} \nonumber \\
\mathbf{F}_{\mu_1\mu_2\mu_3} &=& \big [ p_1\cdot p_2 g_{\mu_1\mu_2} -
p_{1\mu_2}p_{2\mu_1}\big] \big[ (p_2\cdot p_3)p_{1\mu_3}
  - (p_1\cdot p_3) p_{2\mu_3} \big]
\nonumber \\
\mathbf{H}_{\mu_1\mu_2\mu_3} &=& g_{\mu_1\mu_2}\big [
    ( p_1 \cdot p_3)p_{2\mu_3} -
    (p_2 \cdot p_3)p_{1\mu_3} \big]
  + \frac{1}{3}(p_{1\mu_3} p_{2\mu_1} p_{3\mu_2}
  - p_{1\mu_2} p_{2\mu_3} p_{3\mu_1})  \nonumber \\
\mathbf{S}_{\mu_1\mu_2\mu_3} &=& p_{1\mu_3} p_{2\mu_1} p_{3\mu_2} + p_{1\mu_2}
p_{2\mu_3} p_{3\mu_1} .
\end{eqnarray}
Color dependence can be factored out of the vertex function,
\begin{equation}
V_{\mu_1 \mu_2 \mu_3}^{a_1a_2a_3} (p_1,p_2,p_3) = F^{a_1a_2a_3} V_{\mu_1 \mu_2 \mu_3} (p_1,p_2,p_3)
\end{equation}
where $F^{abc} = -if_{abc}$ is the adjoint representation matrix,
satisfying $F^{iab} F^{jba} = C_A \delta^{ij}$, $C_A = N$ for
SU($N$). In a similar fashion to the propagators, our \ansatz\ for $V$
is designed so that it can be easily made to converge to its
perturbative form at large momenta. We will separate the various
contributions to $V$ much like we did with the propagators,
\begin{equation}
V_{\mu_1 \mu_2 \mu_3} = g \big (V^{(0)}_{\mu_1 \mu_2 \mu_3} + V^{(1)}_{\mu_1 \mu_2 \mu_3} + V^{\text{NP}}_{\mu_1 \mu_2 \mu_3} \big ) .
\end{equation}
The bare and one-loop corrections are denoted by $V^{(0)}$ and $V^{(1)}$, while $V^{\text{NP}}$ denotes the nonperturbative correction to the vertex that has to solved for self-consistently by finding the stationary point of $\Gamma$. The bare term, $V^{(0)}$ is simply
\begin{equation}
V^{(0)}_{\mu_1 \mu_2 \mu_3} = (p_2-p_3)_{\mu_1}g_{\mu_2\mu_3} + (p_3-p_1)_{\mu_2} g_{\mu_1\mu_3} +
(p_1-p_2)_{\mu_3} g_{\mu_1\mu_2}
\end{equation}
which is expressed entirely in terms of cyclic permutations of the
$\mathbf{A}$ tensor, {\it i.e.}
\begin{equation}
V^{(0)}_{\mu_1 \mu_2 \mu_3} = A^{(0)} \mathbf{A}_{\mu_1 \mu_2 \mu_3} + \text{cyclic perms.}
\end{equation}
with $A^{(0)} = 1$. The one-loop correction to the vertex $V^{(1)}$ has a much more intricate tensor structure
\begin{eqnarray}
\nonumber V^{(1)}_{\mu_1\mu_2\mu_3} (p_1,p_2,p_3) &=& A^{(1)}(p_1,p_2;p_3)
\mathbf{A}_{\mu_1\mu_2\mu_3} + B^{(1)}(p_1,p_2;p_3)
\mathbf{B}_{\mu_1\mu_2\mu_3} \\
\nonumber &+& C^{(1)}(p_1,p_2;p_3) \mathbf{C}_{\mu_1\mu_2\mu_3}+
F^{(1)}(p_1,p_2;p_3)\mathbf{F}_{\mu_1\mu_2\mu_3}\\
\nonumber &+& H^{(1)}(p_1,p_2,p_3)
\mathbf{H}_{\mu_1\mu_2\mu_3} + S^{(1)}(p_1,p_2,p_3)
\mathbf{S}_{\mu_1\mu_2\mu_3}\\
&+& \text{cyclic perms.} \label{eq:VStrucFuncs}
\end{eqnarray}
and likewise for $V^{\text{NP}}$.  As we discussed above, we need the
explicit forms of the 1-loop vertex corrections in order to eliminate
logarithmic UV divergences in the variational problem.  We present
explicit results for the 1-loop vertices in Appendix \ref{sec:vertex}
and we use those results in the following\footnote{Setting $\omega^{(1)} = 1/4$; this parameter serves an analogous purpose to $\omega$ as it appears in the ghost propagator, see \Eq{eq:OneLoopVertexIrReg}.}.  Note in particular that
$S^{(1)} = 0$.  However it remains to write variational \ansatze\ for
$V^{\text{NP}}$.  Here we make no assumptions about the vanishing of the
coefficients for any of the tensorial structures.
The functions $A$, $C$ and $F$ are symmetric in their first two
arguments, $B$ is antisymmetric in its first two arguments, $H$ is fully
symmetric and $S$ is fully antisymmetric. We respect these symmetry properties by choosing the following
\ansatze:
\begin{eqnarray}
 A^{\text{NP}}(p_1, p_2; p_3) &=&
\frac{1}{p_1^{2} + p_2^{2} + p_3^{2} + \omega^{2}}
    \frac{ \sum_{i\geq j} a^{\{A\}}_{ijk}
         (p_1^i p_2^j + p_1^j p_2^i)p_3^k}
{\sum_{i\geq j} b^{\{A\}}_{ijk}(p_1^i p_2^j + p_1^j p_2^i)p_3^k} 
\\
 B^{\text{NP}}(p_1, p_2; p_3) &=&
\frac{1}{p_1^{2} + p_2^{2} + p_3^{2} + \omega^{2}}
     \frac{\sum_{i>j} a^{\{B\}}_{ijk}
         (p_1^i p_2^j - p_1^j p_2^i)p_3^k}
    {\sum_{i\geq j} b^{\{B\}}_{ijk}(p_1^i p_2^j + p_1^j p_2^i)p_3^k}
\\
 C^{\text{NP}}(p_1, p_2; p_3) &=&
\frac{1}{p_1^{4} + p_2^{4} + p_3^{4} + \omega^{4}}
    \frac{ \sum_{i\geq j} a^{\{C\}}_{ijk}
     (p_1^i p_2^j + p_1^j p_2^i)p_3^k}
    {\sum_{i\geq j} b^{\{C\}}_{ijk}(p_1^i p_2^j + p_1^j p_2^i)p_3^k}
 \\
 F^{\text{NP}}(p_1, p_2; p_3) &=&
\frac{1}{p_1^{6} + p_2^{6}+ p_3^{6} + \omega^{6}}
    \frac{ \sum_{i\geq j} a^{\{F\}}_{ijk}
   (p_1^i p_2^j + p_1^j p_2^i)p_3^k}
  {\sum_{i\geq j} b^{\{F\}}_{ijk}(p_1^i p_2^j + p_1^j p_2^i)p_3^k}
 \\
 H^{\text{NP}}(p_1, p_2, p_3) &=&
     \frac{1}{p_1^{4} + p_2^{4} + p_3^{4} + \omega^{4}}
    \frac{\sum_{i\geq j\geq k} a^{\{H\}}_{ijk}
  (p_1^i p_2^j p_3^k + \text{perms.})}
  {\sum_{i\geq j\geq k} b^{\{H\}}_{ijk}(p_1^i p_2^j p_3^k + \text{perms.})}
    \qquad \\
 S^{\text{NP}}(p_1, p_2, p_3) &=&
\frac{1}{p_1^{4} + p_2^{4} + p_3^{4} + \omega^{4}}
    \frac{\sum_{i>j>k} a^{\{S\}}_{ijk}
   (\epsilon_{xyz} p_x^i p_y^j p_z^k)}
  {\sum_{i\geq j\geq k} b^{\{S\}}_{ijk}(p_1^i p_2^j p_3^k + \text{perms.})}
\end{eqnarray}
which automatically have these symmetries built into them
($\epsilon_{xyz}$ is the permutation symbol).  Each sum is
truncated so that $(i{+}j{+}k) \leq \NVmax$.

Notice that not all six tensors are of the same dimension; $\mathbf{A}$
and $\mathbf{B}$ have dimensions of $[\text{mass}]$,
$\mathbf{C}$, $\mathbf{H}$ and $\mathbf{S}$ are $[\text{mass}]^3$ and
$\mathbf{F}$ is $[\text{mass}]^5$. In every case the UV behavior of
the vertex function must satisfy $V^{(1)} \propto 1$ and
$V^{\text{NP}} \propto p^{-1}$.
We enforce the correct momentum scaling for the vertex functions by hand, so that the \pade\ Approximants are all $\mathcal{O}(1)$; this way, despite the individual dimensionalities of the Vertex functions, all of the \ansatze\ are the same ``size.''

The ghost-gluon vertex $\Vgh$ is somewhat simpler.  Factoring out
explicit color dependence,
\begin{equation}
\Vgh^{a_1a_2a_3}_{\mu_3}(p_1,p_2,p_3) = gF^{a_1a_2a_3}
                                        \Vgh_{\mu_3}(p_1,p_2,p_3)
\end{equation}
with the outgoing ghost and gluon indexed by $(a_1,p_1)$ and $(a_3,\mu_3,p_3)$
respectively ($p_1$ flows outwards), we have
\begin{equation}
 \Vgh_{\mu_3}(p_1,p_2,p_3) =  \Agh(p_1,p_2,p_3) p_{1\mu_3} +
                              \Bgh(p_1,p_2,p_3) p_{2\mu_3} \,,
\end{equation}
where
\begin{eqnarray}
\Agh(p_1, p_2, p_3) &=& \Agh^{(0)} + \Agh^{(1)} + \Agh^{\text{NP}}\\
\Bgh(p_1, p_2, p_3) &=& \Bgh^{(1)} +  \Bgh^{\text{NP}},
\end{eqnarray}
and
\begin{eqnarray}
\Agh^{\text{NP}} &=& \frac{1}{p_1^{2} + p_2^{2}
         + p_3^{2} + \omega^2} \frac{\sum a^{\{\Agh\}}_{ijk}p_1^i p_2^j p_3^k}{\sum b^{\{\Agh\}}_{ijk}p_1^i p_2^j p_3^k} \\
\Bgh^{\text{NP}} &=&
    \frac{1}{p_1^{2} + p_2^{2}
         + p_3^{2} + \omega^2}\frac{p_1 \sum a^{\{\Bgh\}}_{ijk} p_1^i p_2^j p_3^k
    + (p_2 - p_3) \sum a^{\{\Bgh\}}_{ij} p_2^j p_3^k}{\sum b^{\{\Bgh\}}_{ijk}p_1^i p_2^j p_3^k}
\end{eqnarray}
In this case $\Agh^{(0)} = 1$, and as above, $\Agh^{(1)}$ and
$\Bgh^{(1)}$ are one-loop
corrections (read directly from Eqs.~(\ref{eq:GhAFunction}) and
(\ref{eq:GhBFunction}), modulo $g^2 N$).

The \ansatz\ for $\Bgh$ is chosen
so as to guarantee that
$\lim_{p_1\rightarrow 0} \Bgh = 0$, linearly in $p_1$.  Let us briefly
discuss this assumption, and our similar assumption that the ghost
self-energy $\Sigma(p) \propto p^2$ at small $p$.  Both properties arise
because the tree-level ghost vertex
$\Vgh^{(0)}_{\mu}(p_1,p_2,p_3) \propto p_{1\mu}$ the outgoing ghost
momentum.  In any loop diagram modifying $\Vgh$, with arbitrarily many
loops, the outgoing ghost line, with momentum $p_1$,
always encounters a bare $\Vgh$, leading to a proportionality of the
full diagram to $p_1$.  This proportionality is automatic in the
$\Agh$ term; we are also enforcing it in the $\Bgh$ term.  One could
argue that this argument assumes a strict diagrammatic expansion and
might be violated somehow when we fully resum.  However, it is at least
self-consistent that $\Bgh \propto p_1$ (note that
$\lim_{p_1 \rightarrow 0}(p_2{-}p_3) = 0$ linearly in $p_1$).  And if it
is, this is enough to ensure that any vertex correction, with resummed
as well as bare vertices, is still always proportional to $p_1$.
Since the vertices are always proportional to $p_1$, the self-energy
must also vanish at least linearly in $p$; but assuming that $\Sigma(p)$
is smooth at small $p$ (which is true provided that there are no IR
divergences in ghost self-energy corrections), $\Sigma(p)$ must in fact
vanish quadratically in $p$.  Building these properties into our
\ansatze\ improves the stability of the numerical extremization; however
we have also tried to solve the variational problem without these
assumptions (using \ansatze\ which allow $\Bgh \propto p_1^0$ and
$\Sigma(p) \propto p^0$), with results which are consistent with the
assumed $\propto p_1$ and $\propto p^2$ behaviors.

\subsection{Numerical Implementation}

Obtaining the solution in terms of the variational coefficients involves
performing three non-trivial tasks, which together can be referred to as
the numerical implementation. These tasks are
\begin{itemize}
\item Tensor contraction and diagram generation
\item Numerical integration over a 9D phase-space (three loops)
\item Using an extremization algorithm to locate the extremum of $\Gamma$.
\end{itemize}

Concerning diagram generation, the purely gluonic \textit{Mercedes-Benz}
is by far the most complicated diagram.  Each propagator has 2 tensorial
structures, each vertex has 14 (the permutations of the 6 structures
described in the last subsection).  An inefficient tensor contraction
would therefore contain $2^{6} 14^4 = 2458624$ terms.
Therefore it is important to perform
the tensorial contractions carefully, building intermediate structures
with the minimum number of terms.  For instance, the
\textit{Mercedes-Benz} can be regarded as
\begin{fmffile}{MBdec}
\begin{equation}
\parbox{20mm}{\begin{fmfgraph}(20,20)
\fmfforce{10mm,10mm}{v4}\fmfforce{2.206mm,5.5mm}{v1}
\fmfforce{17.794mm,5.5mm}{v2}
\fmfforce{10mm,19mm}{v3}
\fmfcon{v1,v2,v3,v4}
\fmf{dbl_wiggly,left=0.55}{v1,v3,v2,v1}
\fmf{dbl_wiggly}{v1,v4,v3}
\fmf{dbl_wiggly}{v2,v4,v3}
\fmf{dbl_wiggly}{v1,v4,v2}
\end{fmfgraph}}
 ~ = ~
\parbox{20mm}{\begin{fmfgraph}(20,20)
\fmfleft{l1}\fmfright{r1,r2,r3}
\fmf{dbl_wiggly}{l1,v1}\fmf{dbl_wiggly}{r1,v1}\fmf{dbl_wiggly}{r3,v1}
\fmfcon{v1}
\end{fmfgraph}}
~ \times ~
\parbox{20mm}{\begin{fmfgraph}(20,20)
\fmfleft{l1}\fmfright{r1,r2,r3}
\fmf{dbl_wiggly}{l1,v1}\fmf{dbl_wiggly}{r1,v2}\fmf{dbl_wiggly}{r3,v3}
\fmf{dbl_wiggly,tension=0.3}{v1,v2,v3,v1}
\fmfcon{v1,v2,v3}
\end{fmfgraph}} ~ ,
\end{equation}
which we will write as \textit{vertex} contracted with \textit{triangle}. The
\textit{triangle} can be represented in terms of a basis of 36 tensors
(all three-index objects that can be constructed out of three momenta
[two external and one loop] and the metric), and likewise, the resummed
vertex
contains 14 distinct tensors (all three-index objects that can be
constructed out two momenta and the metric, due to momentum
conservation). Specifically, the vertex {\sl after} contracting the
tensors associated with the propagators is of form
\begin{eqnarray}
\nonumber V_{\mu_1 \mu_2 \mu_3} &=& Z_{001} g_{\mu_1 \mu_2} p_{1\mu_3} {+} Z_{010} g_{\mu_1 \mu_3} p_{1\mu_2} {+} Z_{100}g_{\mu_2 \mu_3} p_{1\mu_1} \\
\nonumber & {+}& Z_{002} g_{\mu_1 \mu_2} p_{2\mu_3} {+} Z_{020} g_{\mu_1 \mu_3} p_{2\mu_2} {+} Z_{200} g_{\mu_2 \mu_3} p_{2\mu_1}\\
\nonumber& {+}&   Z_{112}  p_{1\mu_1} p_{1\mu_2} p_{2\mu_3} {+}  Z_{121}  p_{1\mu_1} p_{2\mu_2} p_{1\mu_3} {+}  Z_{211}  p_{2\mu_1} p_{1\mu_2} p_{1\mu_3} {+} Z_{111}  p_{1\mu_1} p_{1\mu_2} p_{1\mu_3}\\
& {+}& Z_{221} p_{2\mu_1} p_{2\mu_2} p_{1\mu_3} {+} Z_{212} p_{2\mu_1}
p_{1\mu_2} p_{2\mu_3} {+} Z_{122} p_{1\mu_1} p_{2\mu_2} p_{2\mu_3} {+}
Z_{222} p_{2\mu_1} p_{2\mu_2} p_{2\mu_3} . \qquad \;
\end{eqnarray}
With the
\textit{vertex} and \textit{triangle} factored as such, standard
algebraic packages can perform all of the remaining tensor contractions and
simplifications.\end{fmffile}

In addition to drastically simplifying diagram construction, the use of
these bases allows for an economical use of floating point
operations. The \textit{triangle} is contained in all gluonic
derivatives of $\Gamma$; furthermore; the 36 \textit{triangle}
$Z$-coefficients are by far the largest polynomials contained within
the problem. With the triangle expressed as such, each of these 36
functions need only to be computed a single time at each point in a 6D
space of integration variables.  This is important because
high-dimensional numerical integration will require of order $10^5$
evaluations of each diagram {\sl per step} in the extremization
procedure for $\Gamma$.

With the Lorentz algebra in hand, we turn to the problem of
multidimensional numerical integration.
The first step is to choose a convenient basis for integration.  Our
choice is described in Appendix \ref{phasespace}.
Performing the global (Eulerian) angular integrations, two-loop diagrams
require a 3D integration and three-loop integrals require a 6D
integration.  These happen to be the same as the number of propagators
in the bubble diagrams built with 3-point vertices.  And it is possible,
and convenient, to choose integration variables which are precisely the
magnitudes of the momenta on each propagator.  (This is a special
feature of phase space integration in 3 dimensions, discussed in the
appendix.)  The most numerically challenging integration is again the
{\sl Mercedes-Benz} topology.  In the notation of Appendix
\ref{phasespace}, the three finite-range (angular) integrations,
over $k'$, $q'$, and $l$, as well as the $p$ integration, were performed
using Gaussian quadratures, while
the (infinite) $k$ and $q$ momentum axes were rescaled into the
unit interval and sampled using an array of points
constructed with a quasi-random hopping Halton series.%
\footnote{%
    More specifically, we first write $k = p x/(1-x)$ with
$x\in [0,1)$.  Then we write $x = 3y^2-2y^3$ with $y \in [0,1)$
chosen with uniform weight.  The former transformation ensures that
$k,q,p$ are of comparable magnitude; the latter transformation increases
the sampling at the top and bottom relative to the middle of the range.}
Our quadratures procedure is symmetric over intervals, which ensures
that certain cancellations on angular integration are preserved, and it
avoids edgepoint evaluations.  Also, since neither algorithm uses random
or pseudorandom numbers or dynamic mesh refinement, each integration
evaluation is over exactly the
same distribution of phase-space points.  This ensures that the
effective action $\Gamma$ is not ``noisy'' in the sense that it does not
fluctuate between evaluations with the same or almost the same choices
of propagator and vertex functions, a feature which is essential for
conjugate-gradient and other differential extremum
seeking algorithms.

\begin{figure}
\centering
\includegraphics[scale=0.8]{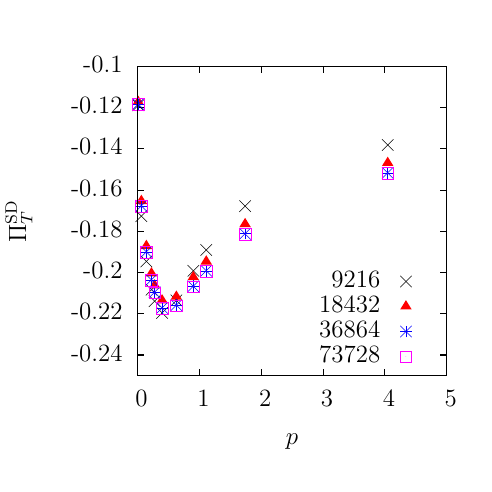}
\caption{\label{fig:IntPointsDep}Non-perturbative correction to
  $\Pi_T$. viz. \Eq{eq:PiTSD}, as a function of the total number of integration
  points along the $k$, $q$, $k'$, $q'$ and $l$ integration axes
  (labelled in the bottom right corner).}
\end{figure}

To test the stability of our algorithm against changes in the number of
integration points, we computed the 2-loop self-energy for particular
values of full propagators and vertices with varying numbers of
integration points.  As illustrated in
Fig.~\ref{fig:IntPointsDep}, the results converge when sufficiently
many integration points are used. Finally, as an additional check, our numerical procedure for performing 3-loop integrals was tested against the known result for the bare massive 3-loop \textit{Mercedes-Benz} in 3D \cite{Rajantie}, with which we find agreement.

Now that we have explained how numerical integrations can be performed,
we turn to the problem of extremizing the effective action $\Gamma$.
One challenge is that, because of gauge fixing and ghosts, the extremum
is actually a saddle, rather than a maximum or minimum.  This can be
easily seen from the first line of \Eq{eq:Gamma}, where the gluon
propagator $G$ and the ghost propagator $\Delta$ enter with opposite
sign.  With the sign conventions chosen there, the extremum of the
1-loop action with respect to $G$ is a maximum; with respect to $\Delta$
it is a minimum.  This rules out any straightforward application of the
conjugate gradient algorithm.  The Newton-Raphson algorithm can find
general extrema, but it is inefficient and tends to converge well only
in rather small basins of attraction.  So some hybrid approach is
needed.

Fortunately, despite it being not at all \textit{a priori} obvious, one
observes for the most part that each individual function that makes up the extremization
problem has relatively little effect on the others.  This opens up the
possibility of iteratively extremizing each constituent function
($G$, $\Delta$, $V$ and $\Vgh$).  Our procedure was to start with
vertices set to their 1-loop values and propagators set to some naive
initial guess.  Then we perform a conjugate gradient extremization with
respect to $G$, then a conjugate gradient extremization with respect to
$\Delta$ (with opposite sign on the gradient).%
\footnote{%
    In practice we also accelerate this procedure as follows.  We
    evaluate the self-energy diagrams $\Pi(p)$ at a sample of $p$ values holding
    $G$ fixed.  Then we conjugate-gradient
    extremize $G$ in \Eq{eq:intSD} but treating the self-energy $\Pi$ as
    fixed.  We insert the new value of $G$ into the evaluation of the
    self-energy and iterate.  This minimizes the number of evaluations
    of $\Pi(p)$, the most numerically expensive part of the procedure,
    needed to converge to the extremum.  But the extremum obtained by
    this procedure is the one which satisfies \Eq{eq:intSD}, as desired.
    }
Iteratively extremizing
$G$ and $\Delta$ solves the 3-loop 2PI problem.
Then we use gradient descent to
extremize $\Gamma$ with respect to the three-gluon vertex
functions. Finally, the ghost vertex corrections are improved
using the Newton-Raphson method.  Then the procedure is iterated
(propagators and gluon vertices, then ghost vertices) until
convergence is achieved.

\begin{figure}
\centering
\subfigure[~Initially...]
{\includegraphics[scale=0.6]{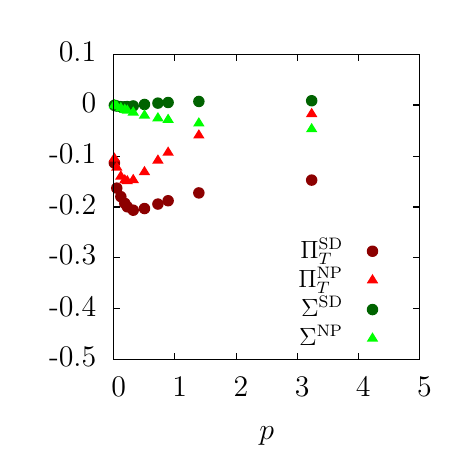}}
\subfigure[~5 Iterations]
{\includegraphics[scale=0.6]{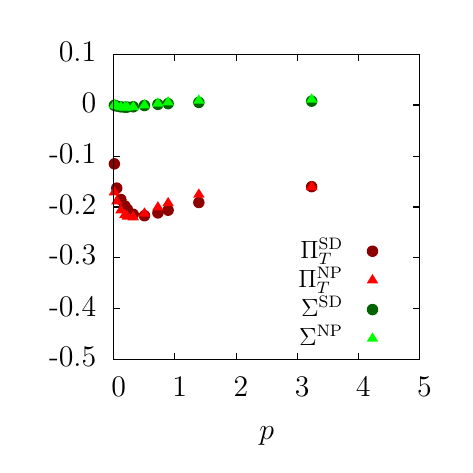}}\\
\subfigure[~100 Iterations]
{\includegraphics[scale=0.6]{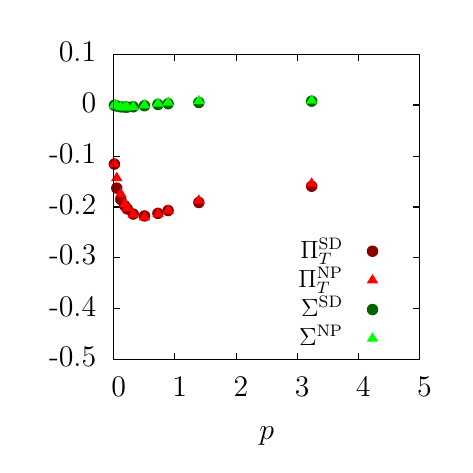}}
\subfigure[~200 Iterations]
{\includegraphics[scale=0.6]{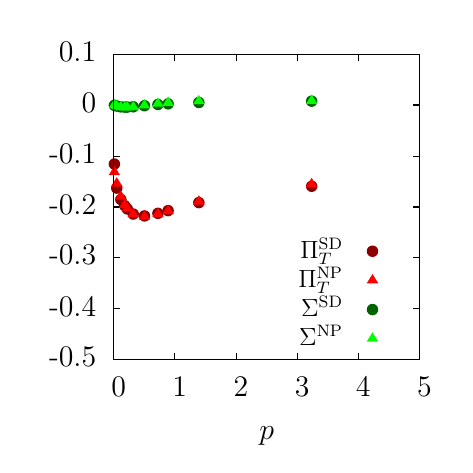}}
\caption{\label{fig:convergence}Evolution of the self-energy under
the gradient descent algorithm.  Here $\Pi^{\text{NP}}$ is the
nonperturbative self-energy according to the \ansatz, while
$\Pi^{\text{SD}}$ is the value as determined by evaluating the
self-energy diagrams.  As the algorithm is iterated, the \ansatz\
approaches a correct reproduction of the self-energy.
}
\end{figure}

The convergence of the algorithm, when applied to the 3-loop 2PI problem in Landau gauge, is depicted in
Fig.~\ref{fig:convergence}.  The figure compares the \ansatz\ value
for the self-energies, $\Pi_T^{\text{NP}}$ and $\Sigma^{\text{NP}}$
as defined in \Eq{eq:GT}, \Eq{eq:DP}, to
the values directly evaluated
by summing the self-energy diagrams, $\Pi_T^{\text{SD}}$ and
$\Sigma^{\text{SD}}$; in each case we have removed the 1-loop
linear-in-$p$ contribution.  That is,
\begin{equation}
\label{eq:PiTSD}
\Pi_T^{\text{SD}} \equiv
\Tproj^{\mu\nu}(\Pi^{(1)}_{\mu\nu}
+\Pi^{(2)}_{\mu\nu})/(\Dim - 1) - \calC^{(1)}_T p ,
\end{equation}
and similarly for $\Sigma^{\text{SD}}$.
The extremization procedure for the effective action with respect to one
of the variational coefficients in $G_T$, \Eq{eq:intSD}, then
corresponds to
\begin{equation}
\label{eq:PropConvergence}
\frac{\delta \Gamma}{\delta c_j}
 = \frac{(\Dim -1)(N^2 - 1)}{2} \int \frac{d^3 p}{(2\pi)^3} \frac{\delta G_T(p)}{\delta c_j}
 \left( -\Pi^{\text{NP}}_T(p)
  + \Pi^{\text{SD}}_T(p) \right) .
\end{equation}

Fig.~\ref{fig:convergence} shows two things.  First, even though
the initial guess for the self-energy falls quite far off
the actual value, after relatively few iterations the
fitted and true values of the self-energy become similar, and the
eventual convergence is excellent.  Second, the value of the self-energy
$\Pi_T^{\text{SD}}$ actually depends quite weakly on
precise form of $\Pi^{\text{NP}}_T(p)$.  That is why our
procedure of varying $\Gamma$ with respect to individual
functions (rather than trying to do everything at once) works so
effectively. The presence of vertices definitely makes matters more
complicated; however, it is also observed that
$\Pi^{\text{SD}}_T(p)$ is fairly insensitive to their
inclusion.

\begin{figure}
\centering
\subfigure
{\includegraphics[scale=0.6]{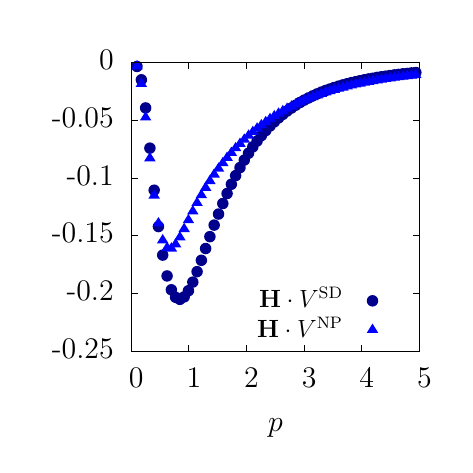}}
\subfigure
{\includegraphics[scale=0.6]{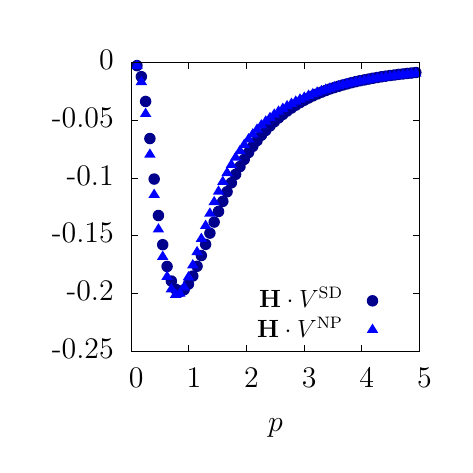}}
\caption{\label{fig:Hconvergence}Vertex analog of
  Fig.~\ref{fig:convergence} in Landau gauge, illustrating in this case,
  convergence of the $H$-function (see \Eq{eq:HConvergence}) and
  the gluon vertex as a whole. The values reside on the curve $k= p / 4$
  and $\cos \theta_{pk} = 1/4$. The figure on the left corresponds to
  $H^{\text{NP}} = 0$, whereas the figure on the right corresponds to
  $H^{\text{NP}}$ which extremizes $\Gamma$. Note that
  the basis, \Eq{BallChiu}, used for the
  vertex is not orthogonal, so $H^{\text{NP}} = 0$ does not imply that
  $\mathbf{H} \cdot V^{\text{NP}} \neq 0$, as illustrated here.}
\end{figure}

\begin{figure}
\centering
\subfigure
{\includegraphics[scale=0.6]{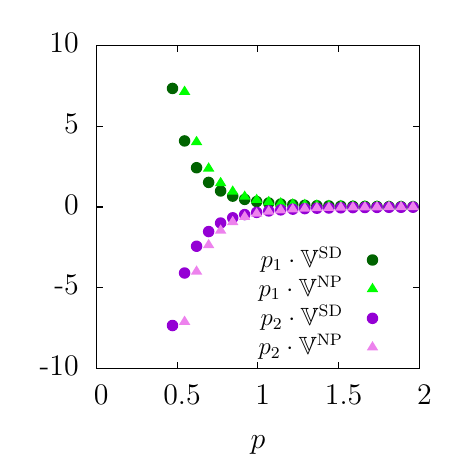}}
\subfigure
{\includegraphics[scale=0.6]{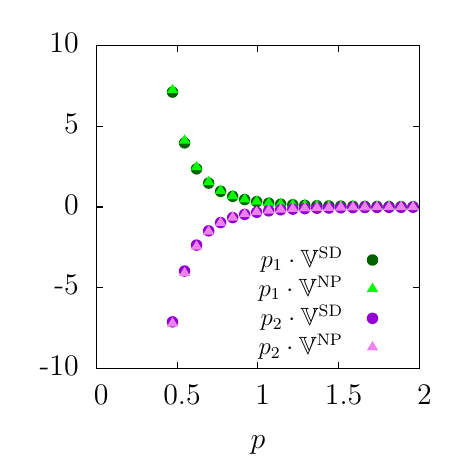}}
\caption{\label{fig:ghostsconvergence} Convergence of the ghost-gluon vertex functions in Landau gauge; similar to Fig.~\ref{fig:Hconvergence}, the figure on the left corresponds to an initial guess of $\mathbb{V}^{\text{NP}} = 0$, whereas that on the right corresponds to the ``solution'' for $\mathbb{V}^{\text{NP}}$ which extremizes $\Gamma$. }
\end{figure}

\Eq{eq:PropConvergence} should be interpreted as an Euler-Lagrange type
of equation for $\Pi$, and a vertex analogue can be defined as
follows. For instance when $c_i$ belongs to the gluon $H$-function, the
variation of $\Gamma$ takes on the following form
\begin{equation}
\label{eq:HConvergence}
\frac{\delta \Gamma}{\delta c_j}
 = \int \frac{d^3p}{(2\pi^3)}\frac{d^3k}{(2\pi)^3} \frac{\delta H^{\text{NP}}(p,k,x)}{\delta c_j}
 \left( - \mathbf{H}\cdot V^{\text{NP}}(p,k,x)
  + \mathbf{H} \cdot V^{\text{SD}}(p,k,x) \right),
\end{equation}
with
\begin{equation}
\label{eq:HdotVNP}
\mathbf{H}\cdot V^{\text{NP}}(p,k,x) \equiv -(N^2-1)
\mathbf{H}_{\mu_1\mu_2\mu_3}
G^{\mu_1\nu_1}(p)G^{\mu_2\nu_2}(k)G^{\mu_3\nu_3}(x)\Big
(\frac{1}{6}V_{\nu_1\nu_2\nu_3} -
\frac{1}{6}V^{(0)}_{\nu_1\nu_2\nu_3}\Big) \,,
\end{equation}
where the overall minus sign comes from the ordering of color
indices. $\mathbf{H}\cdot V^{\text{SD}}(p,k,x)$ is defined in a similar
manner, except that the term in brackets in \Eq{eq:HdotVNP}
contains all of the higher loop terms in the vertex Schwinger-Dyson equation,
\Eq{vert_resum}. The values of $\mathbf{H}\cdot
V^{\text{SD}}(p,k,x)$ and $\mathbf{H}\cdot V^{\text{NP}}(p,k,x)$ along
the curve defined by $k= p / 4$ and $\cos \theta_{pk} = 1/4$ are plotted
in Fig.~\ref{fig:Hconvergence}. Fig.~\ref{fig:ghostsconvergence}
contains plots of a similarly defined set of quantities related to the
ghost vertex.

The \ansatze\ in Figs. \ref{fig:convergence} -
\ref{fig:ghostsconvergence} correspond to $\NPmax = 3$ and $\NVmax = 3$,
from which we observe that the solution is well described by third order
\pade s. However, the size of the \ansatze\ can have a major effect on
the outcome of this technique. With too few coefficients, the numerics
are much simpler, but one is not able to obtain the correct final
answer. However, as the number of coefficients increases, the problem
becomes very numerically difficult (considering the relative ease with
which poles may form in the denominators of the \pade\ approximants),
and furthermore, the data will eventually be over-fitted. We find that
at $\NPmax = 3$ and $\NVmax = 3$ the problem is still not overly
difficult to solve, despite there being a total of 174 coefficients in
Landau gauge (with an additional 33 in other gauges). At the same time
we are not over-fitting. This choice is further motivated by the general
shape of functions we are attempting to converge to; the presence of
additional ``wiggles'' would necessitate larger \ansatze.

\section{Results}

\begin{figure}
\centering
\includegraphics[scale=0.6]{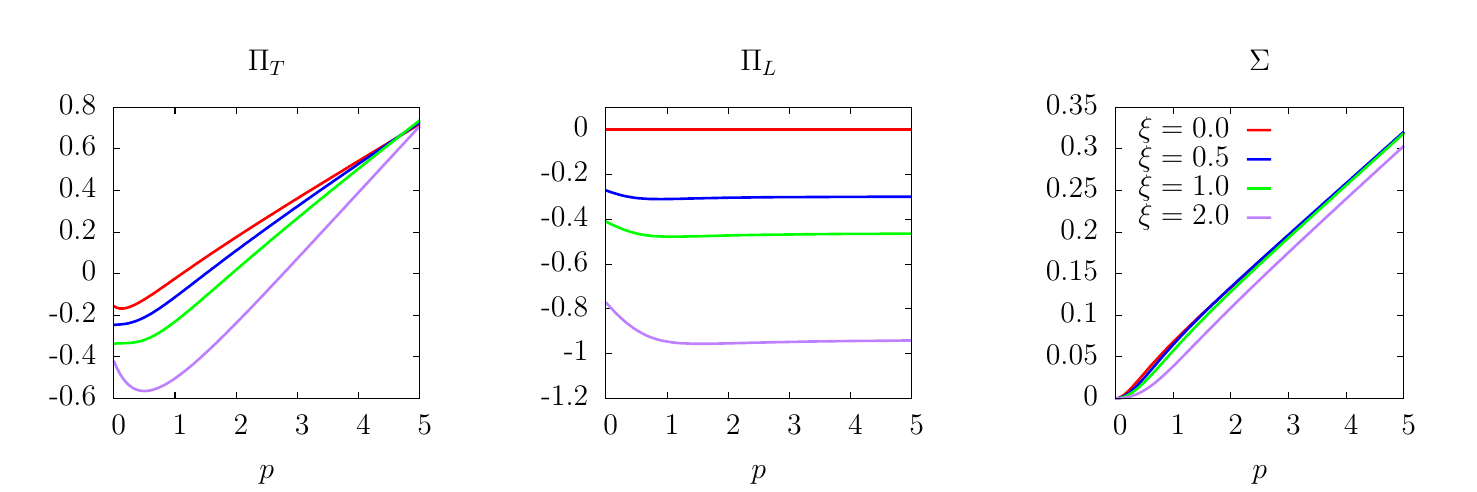}
\caption{\label{fig:SelfEnergies}Self-energies.}
\end{figure}

\begin{figure}[h]
\centering
\includegraphics[scale=0.6]{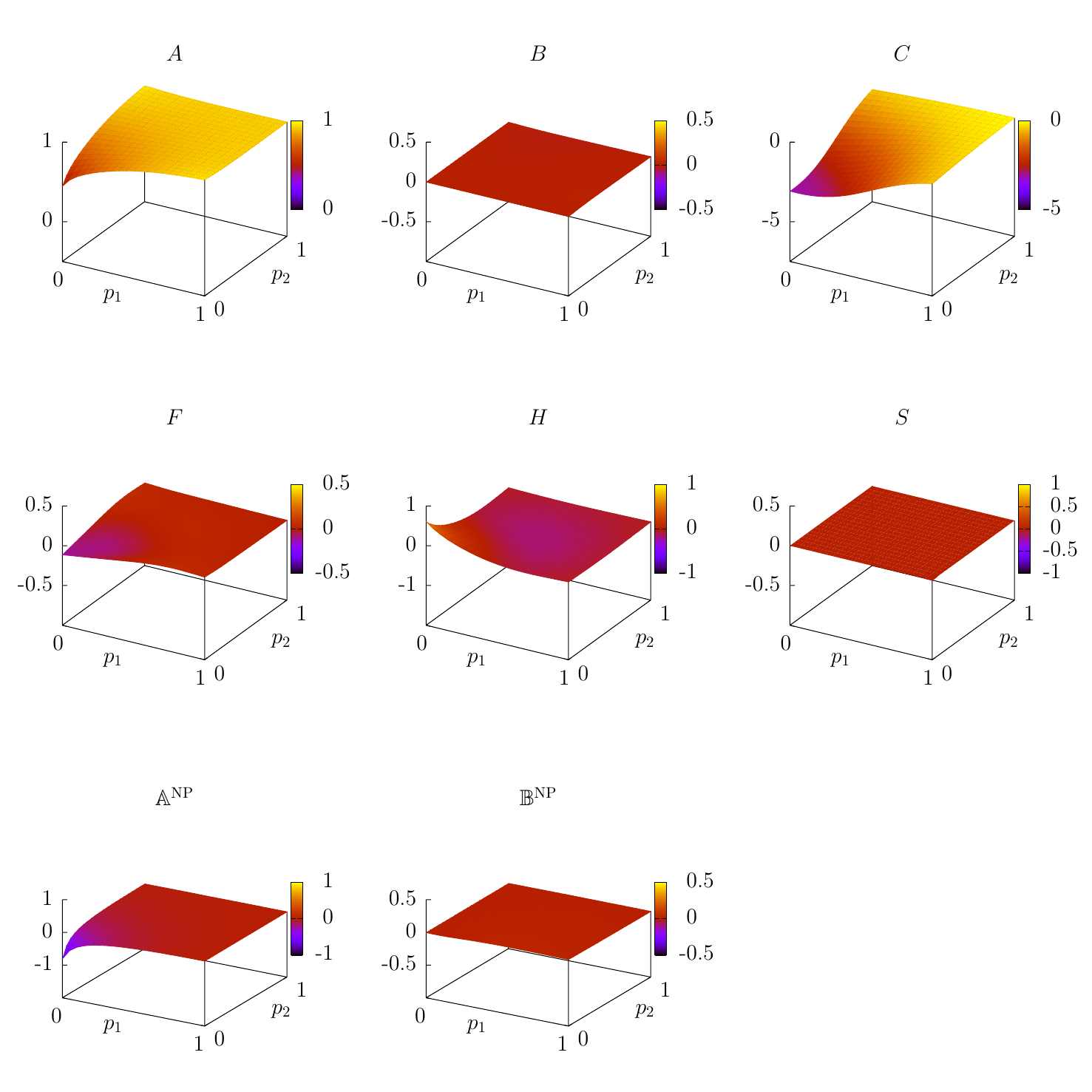}
\caption{\label{fig:xi00Vertices}$\xi = 0.0$, $\NPmax = 3$, $\NVmax = 3$, $\cos\theta_{p_1,p_2} =  1 / 4$.}
\end{figure}

\begin{figure}
\centering
\includegraphics[scale=0.6]{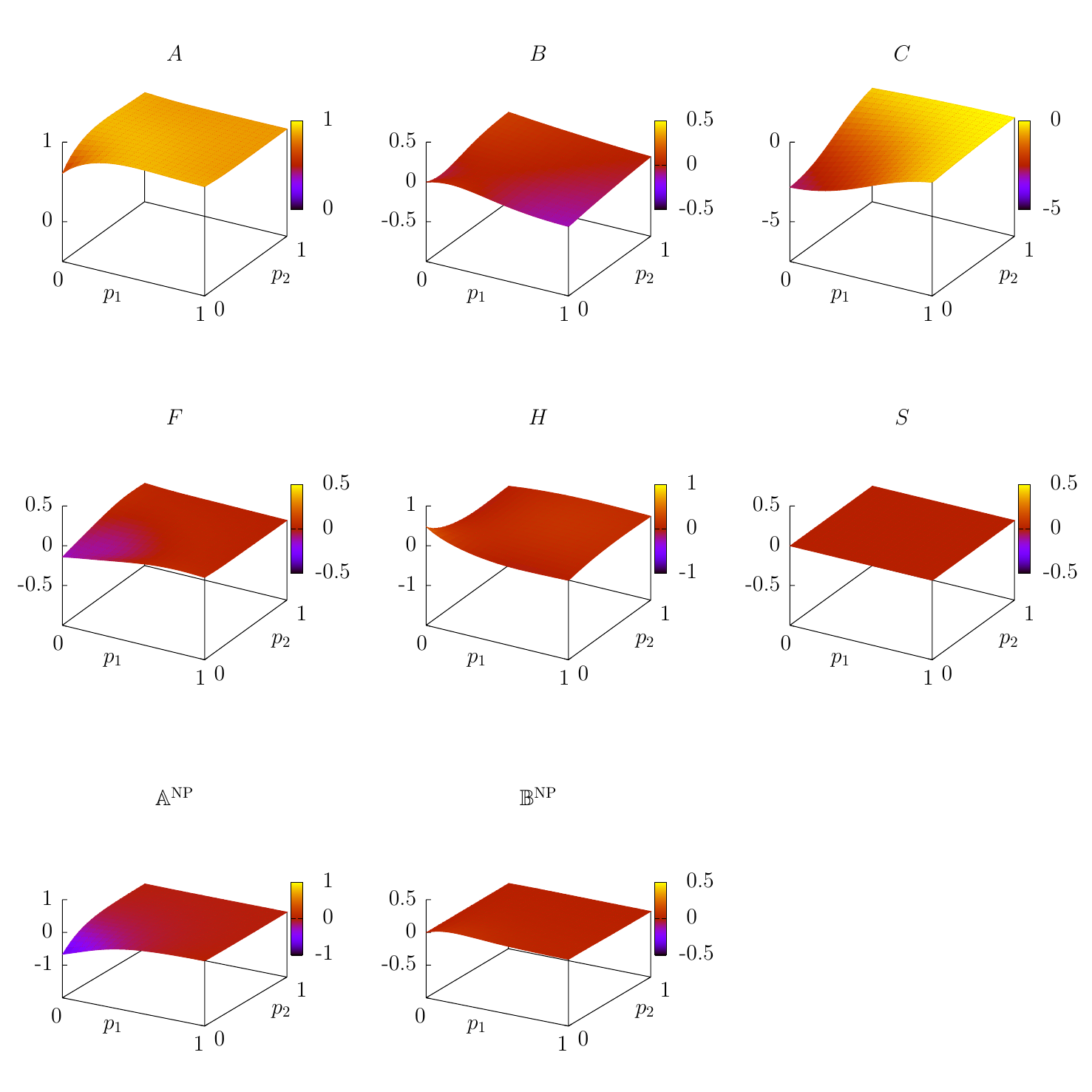}
\caption{\label{fig:xi10Vertices}$\xi = 1.0$, $\NPmax = 3$, $\NVmax = 3$, $\cos\theta_{p_1,p_2} =  1 / 4$.}
\end{figure}

The extremization procedure was carried out for several choices of the
gauge parameter $\xi$, namely 0.0, 0.5, 1.0 and 2.0. The resulting
self-energies are shown in Fig.~\ref{fig:SelfEnergies}, and the gluon
three-vertex functions
$A$ through $S$ and the ghost three-point functions $\Agh^{\text{NP}}$ and
$\Bgh^{\text{NP}}$ are plotted in
Fig.~\ref{fig:xi00Vertices}, which shows the Landau gauge results,
and Fig.~\ref{fig:xi10Vertices}, which shows the results in Feynman gauge.
The Landau gauge variational coefficients
are stated in Table \ref{table:LandauCoeffs}. The results for
$\xi=0.5$ and $\xi=2.0$ are qualitatively similar.

\begin{figure}
\centering
\includegraphics[scale=0.6]{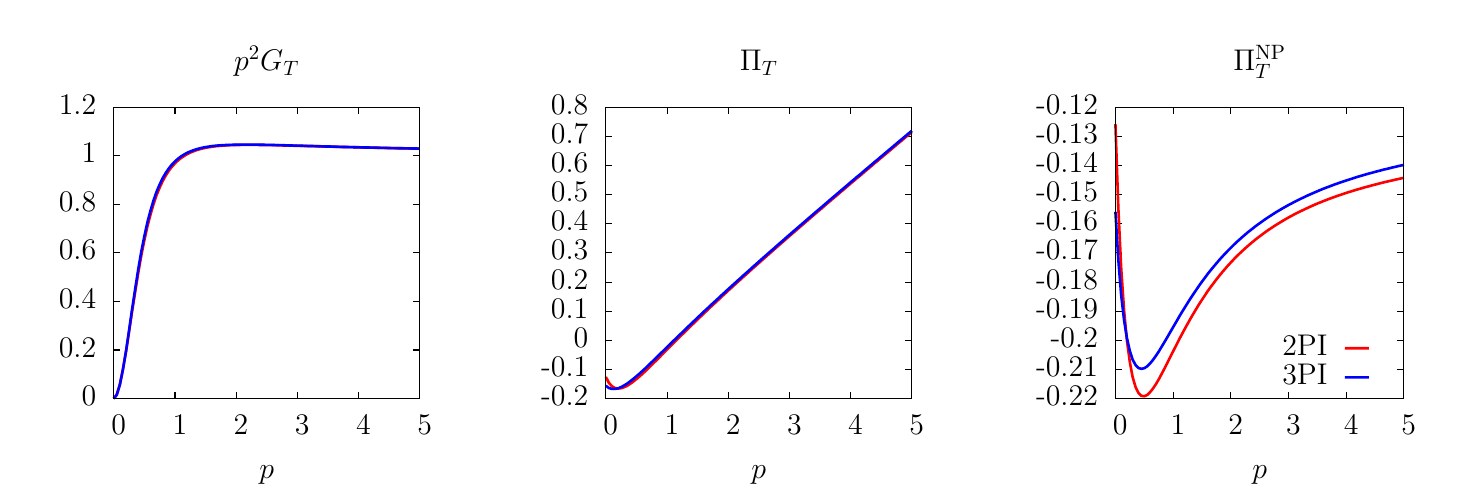}
\caption{\label{fig:2PIvs3PI}Comparison between the 2PI (bare vertices
  only) and 3PI (vertices included) solutions. The correction to $G_T$
  that we obtain when including the vertices is indeed small (in Landau
  gauge).}
\end{figure}

\begin{table}[h]
\centering\scriptsize
\begin{tabular}{l@{\hspace{1mm}}l@{\hspace{1mm}}|@{\hspace{1mm}}l@{\hspace{1mm}}l@{\hspace{1mm}}|@{\hspace{1mm}}l@{\hspace{1mm}}l@{\hspace{1mm}}|@{\hspace{1mm}}l@{\hspace{1mm}}l@{\hspace{1mm}}|@{\hspace{1mm}}l@{\hspace{1mm}}l@{\hspace{1mm}}|@{\hspace{1mm}}l@{\hspace{1mm}}l}
$a^{\{G_T\}}_{0}$ & \tt{-1.56e-01} & $a^{\{G_T\}}_{1}$ & \tt{-1.03e+00} & $a^{\{G_T\}}_{2}$ & \tt{-9.14e-01} & $a^{\{G_T\}}_{3}$ & \tt{-4.47e-01} & $b^{\{G_T\}}_{1}$ & \tt{4.11e+00} & $b^{\{G_T\}}_{2}$ & \tt{4.10e+00} \\
$b^{\{G_T\}}_{3}$ & \tt{3.82e+00} & $a^{\{\Delta\}}_{2}$ & \tt{1.79e-01} & $a^{\{\Delta\}}_{3}$ & \tt{1.13e-01} & $b^{\{\Delta\}}_{1}$ & \tt{1.73e+00} & $b^{\{\Delta\}}_{2}$ & \tt{3.49e+00} & $b^{\{\Delta\}}_{3}$ & \tt{1.75e+00} \\
$a^{\{H\}}_{000}$ & \tt{1.02e-01} & $a^{\{H\}}_{100}$ & \tt{-4.98e-02} & $a^{\{H\}}_{110}$ & \tt{-2.73e-01} & $a^{\{H\}}_{111}$ & \tt{-5.15e-01} & $a^{\{H\}}_{200}$ & \tt{-8.11e-01} & $a^{\{H\}}_{210}$ & \tt{1.40e-01} \\
$a^{\{H\}}_{300}$ & \tt{5.49e-01} & $b^{\{H\}}_{100}$ & \tt{1.19e+00} & $b^{\{H\}}_{110}$ & \tt{1.29e-01} & $b^{\{H\}}_{111}$ & \tt{2.38e-01} & $b^{\{H\}}_{200}$ & \tt{1.04e-01} & $b^{\{H\}}_{210}$ & \tt{-1.77e-01} \\
$b^{\{H\}}_{300}$ & \tt{3.20e-01} & $a^{\{A\}}_{000}$ & \tt{-2.77e-01} & $a^{\{A\}}_{001}$ & \tt{-7.66e-01} & $a^{\{A\}}_{002}$ & \tt{3.70e-01} & $a^{\{A\}}_{003}$ & \tt{8.59e-02} & $a^{\{A\}}_{100}$ & \tt{-9.45e-01} \\
$a^{\{A\}}_{101}$ & \tt{4.90e-01} & $a^{\{A\}}_{102}$ & \tt{-5.26e-02} & $a^{\{A\}}_{110}$ & \tt{7.11e-01} & $a^{\{A\}}_{111}$ & \tt{-8.68e-01} & $a^{\{A\}}_{200}$ & \tt{7.03e-02} & $a^{\{A\}}_{201}$ & \tt{-3.46e-01} \\
$a^{\{A\}}_{210}$ & \tt{-2.06e+00} & $a^{\{A\}}_{300}$ & \tt{-7.44e-02} & $b^{\{A\}}_{001}$ & \tt{1.54e+00} & $b^{\{A\}}_{002}$ & \tt{5.90e-01} & $b^{\{A\}}_{003}$ & \tt{6.73e-01} & $b^{\{A\}}_{100}$ & \tt{6.22e+00} \\
$b^{\{A\}}_{101}$ & \tt{8.78e-01} & $b^{\{A\}}_{102}$ & \tt{7.12e-01} & $b^{\{A\}}_{110}$ & \tt{1.85e+00} & $b^{\{A\}}_{111}$ & \tt{-6.23e-01} & $b^{\{A\}}_{200}$ & \tt{1.93e+00} & $b^{\{A\}}_{201}$ & \tt{-2.77e-01} \\
$b^{\{A\}}_{210}$ & \tt{-1.69e-01} & $b^{\{A\}}_{300}$ & \tt{3.42e-01} & $a^{\{C\}}_{000}$ & \tt{-1.53e+00} & $a^{\{C\}}_{001}$ & \tt{-1.74e+00} & $a^{\{C\}}_{002}$ & \tt{-6.63e-01} & $a^{\{C\}}_{003}$ & \tt{3.15e-02} \\
$a^{\{C\}}_{100}$ & \tt{5.94e-02} & $a^{\{C\}}_{101}$ & \tt{-2.47e-01} & $a^{\{C\}}_{102}$ & \tt{-2.09e-02} & $a^{\{C\}}_{110}$ & \tt{3.44e-01} & $a^{\{C\}}_{111}$ & \tt{-4.98e-01} & $a^{\{C\}}_{200}$ & \tt{-2.97e+00} \\
$a^{\{C\}}_{201}$ & \tt{6.72e-01} & $a^{\{C\}}_{210}$ & \tt{-4.27e-02} & $a^{\{C\}}_{300}$ & \tt{5.05e-01} & $b^{\{C\}}_{001}$ & \tt{4.11e-01} & $b^{\{C\}}_{002}$ & \tt{6.14e-01} & $b^{\{C\}}_{003}$ & \tt{7.04e-01} \\
$b^{\{C\}}_{100}$ & \tt{2.19e-01} & $b^{\{C\}}_{101}$ & \tt{5.53e-01} & $b^{\{C\}}_{102}$ & \tt{6.94e-01} & $b^{\{C\}}_{110}$ & \tt{4.13e-01} & $b^{\{C\}}_{111}$ & \tt{6.53e-01} & $b^{\{C\}}_{200}$ & \tt{3.68e-01} \\
$b^{\{C\}}_{201}$ & \tt{6.20e-01} & $b^{\{C\}}_{210}$ & \tt{5.21e-01} & $b^{\{C\}}_{300}$ & \tt{4.46e-01} & $a^{\{F\}}_{000}$ & \tt{-5.67e-02} & $a^{\{F\}}_{001}$ & \tt{-3.53e-03} & $a^{\{F\}}_{002}$ & \tt{9.77e-03} \\
$a^{\{F\}}_{003}$ & \tt{4.70e-01} & $a^{\{F\}}_{100}$ & \tt{-4.18e-02} & $a^{\{F\}}_{101}$ & \tt{-1.96e-02} & $a^{\{F\}}_{102}$ & \tt{3.20e-01} & $a^{\{F\}}_{110}$ & \tt{-5.84e-02} & $a^{\{F\}}_{111}$ & \tt{2.00e-01} \\
$a^{\{F\}}_{200}$ & \tt{-5.23e-02} & $a^{\{F\}}_{201}$ & \tt{2.57e-01} & $a^{\{F\}}_{210}$ & \tt{1.37e-01} & $a^{\{F\}}_{300}$ & \tt{2.28e-01} & $b^{\{F\}}_{001}$ & \tt{9.96e-01} & $b^{\{F\}}_{002}$ & \tt{9.96e-01} \\
$b^{\{F\}}_{003}$ & \tt{9.97e-01} & $b^{\{F\}}_{100}$ & \tt{9.96e-01} & $b^{\{F\}}_{101}$ & \tt{9.97e-01} & $b^{\{F\}}_{102}$ & \tt{9.98e-01} & $b^{\{F\}}_{110}$ & \tt{9.97e-01} & $b^{\{F\}}_{111}$ & \tt{9.98e-01} \\
$b^{\{F\}}_{200}$ & \tt{9.97e-01} & $b^{\{F\}}_{201}$ & \tt{9.98e-01} & $b^{\{F\}}_{210}$ & \tt{9.99e-01} & $b^{\{F\}}_{300}$ & \tt{9.99e-01} & $a^{\{\Agh\}}_{000}$ & \tt{-8.12e-01} & $a^{\{\Agh\}}_{001}$ & \tt{-1.02e-01} \\
$a^{\{\Agh\}}_{002}$ & \tt{2.02e-01} & $a^{\{\Agh\}}_{003}$ & \tt{2.12e-01} & $a^{\{\Agh\}}_{010}$ & \tt{3.59e-01} & $a^{\{\Agh\}}_{011}$ & \tt{-6.42e-01} & $a^{\{\Agh\}}_{012}$ & \tt{-3.44e-01} & $a^{\{\Agh\}}_{020}$ & \tt{-3.01e-01} \\
$a^{\{\Agh\}}_{021}$ & \tt{-1.23e-01} & $a^{\{\Agh\}}_{030}$ & \tt{2.68e-01} & $a^{\{\Agh\}}_{100}$ & \tt{-1.23e-01} & $a^{\{\Agh\}}_{101}$ & \tt{-9.41e-02} & $a^{\{\Agh\}}_{102}$ & \tt{-1.16e-01} & $a^{\{\Agh\}}_{110}$ & \tt{-4.18e-01} \\
$a^{\{\Agh\}}_{111}$ & \tt{4.75e-01} & $a^{\{\Agh\}}_{120}$ & \tt{2.53e-02} & $a^{\{\Agh\}}_{200}$ & \tt{3.50e-02} & $a^{\{\Agh\}}_{201}$ & \tt{4.01e-01} & $a^{\{\Agh\}}_{210}$ & \tt{-2.83e-01} & $a^{\{\Agh\}}_{300}$ & \tt{-1.65e-01} \\
$b^{\{\Agh\}}_{001}$ & \tt{6.26e+00} & $b^{\{\Agh\}}_{002}$ & \tt{2.89e+00} & $b^{\{\Agh\}}_{003}$ & \tt{1.82e+00} & $b^{\{\Agh\}}_{010}$ & \tt{6.31e+00} & $b^{\{\Agh\}}_{011}$ & \tt{2.81e+00} & $b^{\{\Agh\}}_{012}$ & \tt{1.71e+00} \\
$b^{\{\Agh\}}_{020}$ & \tt{2.96e+00} & $b^{\{\Agh\}}_{021}$ & \tt{1.64e+00} & $b^{\{\Agh\}}_{030}$ & \tt{1.57e+00} & $b^{\{\Agh\}}_{100}$ & \tt{5.13e+00} & $b^{\{\Agh\}}_{101}$ & \tt{2.43e+00} & $b^{\{\Agh\}}_{102}$ & \tt{1.55e+00} \\
$b^{\{\Agh\}}_{110}$ & \tt{2.36e+00} & $b^{\{\Agh\}}_{111}$ & \tt{1.44e+00} & $b^{\{\Agh\}}_{120}$ & \tt{1.36e+00} & $b^{\{\Agh\}}_{200}$ & \tt{2.30e+00} & $b^{\{\Agh\}}_{201}$ & \tt{1.43e+00} & $b^{\{\Agh\}}_{210}$ & \tt{1.31e+00} \\
$b^{\{\Agh\}}_{300}$ & \tt{1.38e+00} & $a^{\{\Bgh\}}_{000}$ & \tt{1.43e-01} & $a^{\{\Bgh\}}_{001}$ & \tt{-3.00e-02} & $a^{\{\Bgh\}}_{002}$ & \tt{1.92e-01} & $a^{\{\Bgh\}}_{010}$ & \tt{1.06e-01} & $a^{\{\Bgh\}}_{011}$ & \tt{-1.90e-01} \\
$a^{\{\Bgh\}}_{020}$ & \tt{2.20e-01} & $a^{\{\Bgh\}}_{100}$ & \tt{2.53e-01} & $a^{\{\Bgh\}}_{101}$ & \tt{3.20e-01} & $a^{\{\Bgh\}}_{110}$ & \tt{-3.44e-02} & $a^{\{\Bgh\}}_{200}$ & \tt{-4.49e-01} & $a^{\{\Bgh\}}_{00}$ & \tt{2.52e-01} \\
$a^{\{\Bgh\}}_{01}$ & \tt{-5.05e-01} & $a^{\{\Bgh\}}_{02}$ & \tt{8.21e-02} & $a^{\{\Bgh\}}_{10}$ & \tt{-4.71e-02} & $a^{\{\Bgh\}}_{11}$ & \tt{2.61e-01} & $a^{\{\Bgh\}}_{20}$ & \tt{2.67e-01} & $b^{\{\Bgh\}}_{001}$ & \tt{1.00e+00} \\
$b^{\{\Bgh\}}_{002}$ & \tt{1.00e+00} & $b^{\{\Bgh\}}_{003}$ & \tt{1.00e+00} & $b^{\{\Bgh\}}_{010}$ & \tt{9.99e-01} & $b^{\{\Bgh\}}_{011}$ & \tt{1.00e+00} & $b^{\{\Bgh\}}_{012}$ & \tt{1.00e+00} & $b^{\{\Bgh\}}_{020}$ & \tt{1.00e+00} \\
$b^{\{\Bgh\}}_{021}$ & \tt{1.00e+00} & $b^{\{\Bgh\}}_{030}$ & \tt{1.00e+00} & $b^{\{\Bgh\}}_{100}$ & \tt{1.00e+00} & $b^{\{\Bgh\}}_{101}$ & \tt{1.00e+00} & $b^{\{\Bgh\}}_{102}$ & \tt{1.00e+00} & $b^{\{\Bgh\}}_{110}$ & \tt{1.00e+00} \\
$b^{\{\Bgh\}}_{111}$ & \tt{1.00e+00} & $b^{\{\Bgh\}}_{120}$ & \tt{1.00e+00} & $b^{\{\Bgh\}}_{200}$ & \tt{1.00e+00} & $b^{\{\Bgh\}}_{201}$ & \tt{1.00e+00} & $b^{\{\Bgh\}}_{210}$ & \tt{1.00e+00} & $b^{\{\Bgh\}}_{300}$ & \tt{1.01e+00} \\
\end{tabular}
\caption{\label{table:LandauCoeffs}$\xi = 0.0$ variational coefficients}
\end{table}

In Landau gauge, $G_L$ is zero, so it is not included in the variation;
hence $G_L$ and $\Pi_L$ are depicted as zero in the
plots. Furthermore, when the tensorial structure
$\mathbb{B}_{\mu_1\mu_2\mu_3}$ is contracted against transverse
propagators on all three legs, the result is zero; therefore
the coefficient $B$ vanishes exactly in Landau gauge, though not in
gauges with nonzero $\xi$.  The function $S$ turns out to vanish in all
gauges.

The effect of including the vertices and allowing them to vary is shown
in Fig.~\ref{fig:2PIvs3PI}, where we see a comparison between the 2PI
and 3PI solutions. The inclusion of the vertices only has a slight
effect the resulting propagators.

The dependence of $G$, $V$, {\it etc.}\ on the choice of the gauge
fixing parameter $\xi$ does not by itself indicate a breakdown or
limitation on the 3PI approach.  The relevant question is, how dependent
are gauge independent quantities on $\xi$, and how closely do such
quantities correspond to the nonperturbative values determined, for
instance, using lattice techniques?  Any $\xi$ dependence in gauge
invariant quantities would be an ambiguity, and any error in their value
in comparison to lattice determinations would be a failure, of the 3PI
technique.  Such comparisons are essential, but they are beyond the
scope of the present manuscript.

\section{Discussion}

\subsection{Comparison with other approaches}

The majority of the literature on this subject is centered around 4D
Yang-Mills theory; however, lattice studies (described shortly) have
shown that Green's functions in 3D and 4D exhibit similar qualitative
behavior. Nevertheless, we will try our best to directly compare our
results with those obtained in 3D, to the extent that they exist.

The gluon propagator is an interesting quantity, despite not being
directly related to any physical observable. $G_T$ as depicted in
Fig.~\ref{fig:2PIvs3PI} violates reflection positivity, that is, it does
not have a K\"all\'en-Lehmann representation in terms of a positive
spectral density. Hence, in (3+1) dimensions (or in our case (2+1)
dimensions) $G_T$ can not describe the correlations of physical
particles. This violation of reflection positivity is allowed despite
its apparent contradiction with the Osterwalder-Schrader axioms; after
all, we are dealing with a confining theory, so there is no one-to-one
correspondence between fields and physical particles. This is further
discussed in greater detail in \cite{Alkofer}, but the main point is
that this behavior signals confinement.

The only propagating degrees of freedom that we can (in principle)
observe are color singlet bound states, \textit{glueballs} for
instance. Hence, gluonic two and three point functions are not
``physical,'' and, in general the results we have presented are $\xi$
dependent, which is not necessarily a bad thing. Indeed our intention is
to use these results to compute
gauge-invariant observables in some later publication.

However, in the mean time, we are stuck with analyzing $\langle
AA\rangle$ and $\langle c\bar c\rangle$ as well as the vertices. Though
arguably our best insight into IR QCD comes from the lattice, there have
been many notable \textit{first-principles} based speculations about the
specific IR form of these functions. In general, the main point of
contention is the exact value of $G_T(0)$. The most popular schools of
thought can be summarized as follows...

\subsubsection{The Gribov-Zwanziger Confinement Hypothesis}

In his study on gauge-fixing and gauge copies in Yang-Mills theory \cite{Gribov}, Gribov proposed that at one loop, the IR behavior of the gluon and ghost propagators is
\begin{equation}
G_T(p) \sim \frac{p^2}{p^4 + m^4},\qquad \Delta(p) |_{p^2\rightarrow 0} \sim \frac{1}{p^4} .
\end{equation}
This form has the generic feature that $G_T$ vanishes at zero momentum, and
moreover, $\Delta$ experiences $1/p^4$ IR enhancement, which can
possibly be interpreted as signaling linear confinement (in $4\Dim$, or
course). This form of $G_T$ and $\Delta$ was later advocated by
Zwanziger, primarily because it vanishes at $p=0$, which is
in accordance with his theorem that $G_T(0)$ (in Landau gauge) must
vanish on any finitely spaced lattice in the infinite volume limit
\cite{Zwanziger}.

This proposal should be regarded as being fairly dated, and it is not in
agreement with any of the more recent lattice data. It also does not
accord with the behavior we determine by solving the 3PI problem.

\subsubsection{Schwinger-Dyson Equations}

These arguments \cite{Smekal} are based on obtaining solutions to a
truncated set of Schwinger-Dyson equations for the gluon and ghost
propagators, and in a sense, are very reminiscent of what we are doing
here. 
If we assume power behavior of the gluon and ghost propagators in
the infrared,
\begin{equation}
p^2G_T(p) \sim (p^2)^{\kappa_G}, \qquad 
p^2\Delta(p) \sim (p^2)^{-\kappa_\Delta} \,,
\end{equation}
then Ref.~\cite{Zwanziger2} claims that $\kappa_G$, $\kappa_\Delta$ must
satisfy
\begin{equation}
\label{eq:kappacondition}
\kappa_G = 2\kappa_\Delta+(4-\Dim)/2 \,,
\end{equation}
and specifically in 3 dimensions $\kappa_G = 0.2952$, implying that the
gauge field propagator goes to zero and the ghost propagator diverges
more strongly than $1/p^2$.  However this result assumes that the loop
integral giving rise to a self-energy at momentum $p$ is dominated by
momenta of order $p$, whereas we find for small $p$ that it is instead
dominated by momenta of order $g^2 N$.  Therefore it is not clear to us
that this result of Ref.~\cite{Zwanziger2} is robust, see also
\cite{Boucaud}.
It is also contradicted my more recent studies
\cite{Aguilar,Maas}, which give results (in 4 dimensions) showing
$G_T(p)$ going over to a constant, and $\Delta(p)\propto p^{-2}$, in the
infrared.  These studies are in at least qualitative agreement with
lattice investigations.
However, since in general
Schwinger-Dyson based approaches are reliant on many
simplifying approximations, they have yet to produce any quantitative
agreement.

\subsubsection{Observations From the Lattice}

There is a wealth of lattice data related to this subject, and
fortunately, different sources are generally in agreement. Simulations
have been performed on very large lattices ($V=96^4$ \cite{Bogolubsky1,Bogolubsky2},
$V=80^4$ \cite{Oliveira}, $V=128^4$ and $V = 320^3$ \cite{Cucchieri1_1,Cucchieri1_2,Cucchieri1_3}),
from which one observes qualitative agreement between the results for 3D
and 4D (hence we will intermittently compare 3D and 4D data, but
\textit{never} 2D). The generic finding is that $p^2 \Delta(p)
|_{p^2\rightarrow 0}$ and $G_T(0)$ are finite and nonzero.

However, $G_T(0)$ is often seen to scale inversely with volume so it
remains an open question as to whether the Zwanziger hypothesis is
observable, and it is not known at what volume one should expect to see
this effect. The results in Landau gauge currently depict a $1/V^\alpha$
scaling for $G_T$ but it is generally not observed that
$G_T(0)\rightarrow 0$ as $V\rightarrow \infty$.

All of the works cited above specifically employ the lattice
implementation of Landau gauge. In fact, it is only fairly recently that
preliminary 3D and 4D results in Feynman gauge have been made available
\cite{Cucchieri3}.

\begin{figure}
\centering
\includegraphics[scale=0.8]{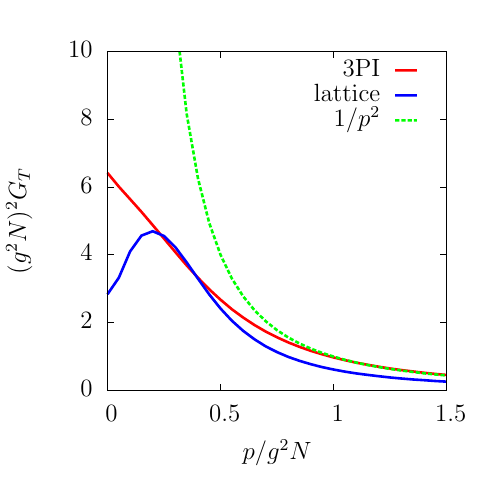}
\caption{\label{fig:LatticePlot}A comparison between the results of our calculation ($G_T(p)$ in Landau gauge) and a reproduction of the plot of $aD(p)$ (appropriately rescaled) in \cite{Cucchieri1_3}. A free $1/p^2$ propagator is shown for reference.}
\end{figure}

In Fig.~\ref{fig:LatticePlot} our data is compared directly to the
results in \cite{Cucchieri1_1,Cucchieri1_2,Cucchieri1_3}. Their calculation was performed for
$\text{SU}(2)$ on an $320^3$ 3D lattice with $\beta = 4/ag^2 = 3.0$.
To facilitate the comparison we have recast their results so momentum is
scaled by $g^2 N$.  Their results are qualitatively
similar to ours but differ quantitatively at the factor-of-2 level in the deep IR.
This might indicate a limitation of the large-$N$ expansion, or it might
simply indicate a failure of the 3PI method.

\subsection{Slavnov-Taylor Identities}

\begin{figure}
\centering
\includegraphics[scale=0.8]{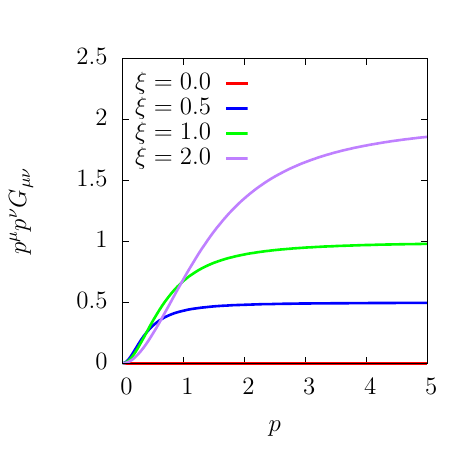}
\caption{\label{fig:PropagatorWardID}Propagator Ward Identity}
\end{figure}

Planar diagrams on their own form a gauge invariant subset of the full loop
expansion \cite{Cvitanovic}. One may hope that in resumming a ``dominant'' or ``important'' set
of planar diagrams (which is what hope to be doing here) gauge invariance is
approximately conserved. This can be measured seeing to what extent
our resulting two and three-point functions violate the
Ward-Slavnov-Taylor (WST) identities. For the gluon propagator, we have
\begin{equation}\label{eq:WSTProp}
p^\mu p^\nu G_{\mu\nu} = \xi ,
\end{equation}
with deviations from this identity shown in Fig.~\ref{fig:PropagatorWardID}.

\begin{figure}
\centering
\includegraphics[scale=0.8]{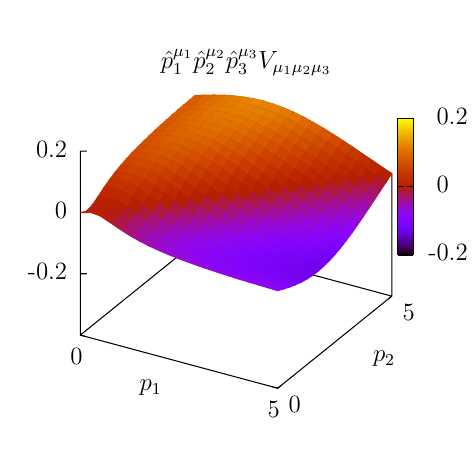}
\caption{\label{fig:VertexWardID}Vertex Ward Identity}
\end{figure}

In its most general form, the WST identity for the gluon three-vertex is
\begin{eqnarray}
p_1^{\mu_1} V_{\mu_1\mu_2\mu_3}(p_1,p_2,p_3) &=& \frac{F(p_1)}{J(p_3)}\left(
p_3^2 g_{\alpha \mu_3} -
p_{3\alpha}p_{3\mu_3}\right)\Vgh^{\alpha}_{\mu_2}(p_3,p_1,p_2)
\nonumber\\
&& -  \frac{F(p_1)}{J(p_2)}\left(p_2^2 g_{\alpha \mu_2} -
p_{2\alpha}p_{2\mu_2}\right)\Vgh^{\alpha}_{\mu_3}(p_2,p_1,p_3),
\end{eqnarray}
where $F$ and $J$ are defined in this context as $J(p) = p^2 G_T(p)$ and
$F(p) = p^2 \Delta(p)$. $\Vgh$ with two Lorentz indices is given by
\begin{eqnarray}
\Vgh^{\alpha}_{\mu_3}(p_1,p_2,p_3) &=& g^{\alpha}_{\mu_3} a(p_3,p_2,p_1) -
p_3^{\alpha}p_{2\mu_3}b(p_3,p_2,p_1) + p_1^{\alpha}p_{3\mu_3} c(p_3,p_2,p_1)
\nonumber\\
&& +~ p_3^{\alpha}p_{1\mu_3}d(p_3,p_2,p_1) +
p_1^{\alpha}p_{1\mu_3}e(p_3,p_2,p_1)
\end{eqnarray}
which is related to the usual ghost-gluon vertex via $\Vgh_{\mu_3}(p_1,p_2,p_3) = p_{1\alpha}\Vgh^\alpha_{\mu_3}(p_1,p_2,p_3)$ (following once again with the decomposition in \cite{Ball}). From this identity one obtains
\begin{equation}\label{eq:WSTVertex}
p_1^{\mu_1}p_2^{\mu_2}p_3^{\mu_3}V_{\mu_1 \mu_2 \mu_3}(p_1,p_2,p_3) = 0.
\end{equation}
With $\hat p^\mu \equiv p^\mu / p$, the deviation from the vertex ward
identity is show in Fig.~\ref{fig:VertexWardID} for Landau gauge. As
previously, the vertex is a function of 3 variables, so to make a 2+1
dimensional plot we have fixed an angular variable to
$\cos \theta_{p_1p_2} = 1/4$.

\section{Conclusions}

We have successfully found the propagators and vertices which extremize
the 3-loop, 3-particle-irreducible action of QCD in 3 dimensions
(the all-threes problem).  We did so by writing a nonlinear variational
\ansatz\ for three propagators (ghosts and the transverse and
longitudinal components of gluons) and for eight vertices (two tensor
structures for ghost-gluon vertices and six tensor structures for
three-gluon vertices).  To avoid divergences it was necessary to add and
subtract terms to 1 and 2-loop self-energies; the added terms are
computed in MS, the subtraction renders the remaining numerical
integrals finite.  It was also necessary to compute the
first loop corrections to 2-point and 3-point functions at large
momentum explicitly and to incorporate these corrections into our
\ansatze\ for those functions.

The most urgent task is to test the resulting resummation against exact
nonperturbative results in 3D QCD by comparing the values of gauge
invariant questions.  For instance, one should be able to
evaluate the $\langle F^2(x) F^2(0) \rangle$ correlator, whose Fourier
transform gives the lowest glueball mass.  It might also be possible to
evaluate the correlator of two field strengths connected by a Wilson
line, which is of interest in evaluating the Debye screening mass in
full QCD \cite{Debyemass_1,Debyemass_2}.  Slightly extending our treatment to include
a fundamental representation scalar, it should be possible to explore
the 3D SU(2)+Higgs phase diagram, which can also be found
nonperturbatively on the lattice \cite{KLRS}.

Unfortunately it is not possible to compute the pressure of 3D
Yang-Mills theory at the 3PI level, because the nontrivial contributions
to the pressure arise at 4 loops.  Evaluating the pressure would require
a solution to the 4-loop 3PI or 4PI problem.
Extending our approach to the 4-loop 4PI treatment would not raise any
new conceptual issues, since all potential UV divergences in the
extremization procedure are already encountered at the level of the
3-loop 3PI problem.  It would be interesting to do so because the
nonperturbative contribution to the pressure
of 3D Yang-Mills theory is needed to compute the $g^6$ term in
the pressure for full QCD \cite{Schroder,g_to_6}.  However the extension to 4
loops and 4PI would be prohibitively difficult because the diagram
generation and loop integration would become even more cumbersome and
the number of possible tensor structures for the 4-point function is
large.

Having studied the 3D theory, it appears to us that the extension to 4
or 3+1 dimensions will be extremely difficult.  The problem is that one
is simultaneously solving nonperturbative infrared physics and
(perturbative) ultraviolet physics.  The effective action is extremely
sensitive to the ultraviolet form of the propagators and vertices; a
procedure along the lines of what we have done here encounters quadratic UV
divergences at every loop order when evaluating self-energies, and
quartic divergences when varying the propagators in a way which changes
their UV behavior -- at every loop order.
It would be much harder to ``cover up'' gauge non-invariance
in 4D because divergently large gauge boson masses would arise at every
loop order, whereas we only encountered them at one loop (at two loops
there were logs but they all cancel).  Thus it is not clear what
additional techniques would have to be developed to successfully extend
our procedure to 4 dimensions.  We will leave this for future
investigation after we know whether the 3-loop treatment is successful
in describing nonperturbative physics.

\section*{Acknowledgments}

We are grateful to J\"urgen Berges and Mikko Laine for useful
conversations.  This work was supported in part by the Natural Sciences
and Engineering Research Council of Canada.

\appendix

\section{Computations involving Bare Diagrams}

\label{sec:appendix}

Since we
are working in three Euclidean dimensions, it is useful to now state the bare
Feynman rules (with $V^{(0)a_1a_2a_3}_{\mu_1\mu_2\mu_3} = F^{a_1a_2a_3}V^{(0)}_{\mu_1\mu_2\mu_3}$ and $\Vgh^{(0)a_1a_2a_3}_{\mu_1\mu_2\mu_3} = F^{a_1a_2a_3}\Vgh^{(0)}_{\mu_1\mu_2\mu_3}$)
\begin{eqnarray}
 gV^{(0)}_{\mu_1\mu_2\mu_3}(p_1,p_2,p_3) &=& g \big(
(p_2-p_3)_{\mu_1}g_{\mu_2\mu_3} + (p_3-p_1)_{\mu_2} g_{\mu_1\mu_3} +
(p_1-p_2)_{\mu_3} g_{\mu_1\mu_2} \big) ~~~~~~\\
g\mathbb{V}^{(0)}_{\mu_3}(p_1,p_2,p_3) &=& g p_{1\mu_3}\\
\nonumber g^2 V^{(0)abcd}_{\mu\nu\rho\tau} & = &
g^2\left(F^{abe}F^{cde}(g_{\mu\rho}g_{\nu\tau} - g_{\mu\tau}g_{\nu\rho})
     \right. \\
\nonumber &&{}+ F^{ace}F^{dbe}(g_{\mu\tau}g_{\nu\rho}
                                -g_{\mu\nu}g_{\rho\tau})\\
&& \left. {}+ F^{ade}F^{bce}(g_{\mu\nu}g_{\rho\tau}
                             -g_{\mu\rho}g_{\nu\tau})\right)
\end{eqnarray}
noting especially the overall sign
of the four vertex. $F^{abc}$ are the generators of the adjoint representation,
defined in terms of the usual SU($N$) structure factors by $F^{abc} = -i
f^{abc}$. $F^{iab}F^{jba} = C_A\delta_{ij}$, and for SU($N$), $C_A = N$. The
bare propagators are simply given by
\begin{eqnarray}
G^{(0)}_{\mu\nu}(p) &=&\frac{1}{p^2}\left(  \Tproj_{\mu\nu} + \xi \Lproj_{\mu\nu}\right)\\
\Delta^{(0)}(p) &=& \frac{1}{p^2}
\end{eqnarray}
with the transverse and longitudinal polarization tensors as defined earlier
(color indices are suppressed). Finally the loop integral in
$\Dim=3+2\epsilon$ dimensions is defined as
\begin{equation}
\int_q \equiv \left(  \frac{4\pi}{\mu^2 e^\gamma} \right) ^\epsilon
\int \frac{d^\Dim q}{(2\pi)^\Dim}.
\end{equation}

\subsection{Gluon Self-Energy}

\label{sec:gluonself}

\begin{fmffile}{gluonselfenergy}The gluon self-energy contains diagrams that
superficially diverge linearly (one loop) and logarithmically (two loops);
however, when computed in
dimensional regularization, one finds that the sum of all one and two-loop
contributions to the self-energy is UV finite.

At one loop, there are three diagrams, only two of which are nonzero in
dimensional regularization.
\begin{equation}
\Pi^{\text{B}(1,\epsilon)}_{\mu\nu} = \frac{1}{2}~
\parbox{20mm}{\begin{fmfgraph}(20,20)
\fmfleft{l1}\fmfright{r1}
\fmfforce{0.25w,0.5h}{v1}\fmfforce{0.75w,0.5h}{v3}
\fmf{photon}{l1,v1}\fmf{photon,left=1}{v1,v3,v1}\fmf{photon}{r1,v3}
\end{fmfgraph}}
~ + \frac{1}{2}~
\parbox{20mm}{\begin{fmfgraph}(20,20)
\fmfleft{l1}\fmfright{r1}\fmftop{t1}\fmfforce{0.5w,0.5h}{v1}
\fmf{photon}{l1,v1}\fmf{phantom,tension=5.0}{t1,v2}
\fmf{photon,left=1,tension=0.4}{v1,v2,v1}
\fmf{photon}{r1,v1}
\end{fmfgraph}}
~ - ~
\parbox{20mm}{\begin{fmfgraph}(20,20)
\fmfleft{l1}\fmfright{r1}
\fmfforce{0.25w,0.5h}{v1}\fmfforce{0.75w,0.5h}{v3}
\fmf{photon}{l1,v1}\fmf{scalar,left=1}{v1,v3,v1}
\fmf{photon}{r1,v3}
\end{fmfgraph}}
\end{equation}
Keeping terms up to $\bigO{(\epsilon)}$ for convenience, the result is found
to be
\begin{eqnarray}
\label{eq:Pi1emunu}\Pi^{\text{B}(1,\epsilon)}_{\mu\nu}&=&\frac{g^2 N}{64}\frac{p^{1+2\epsilon}}{\mu^{2\epsilon}} \big [
(\xi^2+2\xi+11)(1-2\epsilon\log 2)+
\epsilon(12-12\xi-2\xi^2) \big ] \Tproj_{\mu\nu}\\
\label{eq:Pi1e}&=& p \left( \frac{p^{2\epsilon}}{\mu^{2\epsilon}}\right)  \Pi^{\text{B}(1,\epsilon)}
\Tproj_{\mu\nu}
\end{eqnarray}
UV divergences only start to arise at the two-loop level, and on a diagram by
diagram basis, these are all proportional to $g_{\mu\nu}$. The two-loop self-energy also contains diagrams that are potentially IR divergent, and these will
be handled separately. Defining
\begin{eqnarray}
\nonumber \Pi^{\text{B}(2,\epsilon) \text{UV}}_{\mu\nu} &=& \frac{1}{6}~
\parbox{20mm}{\begin{fmfgraph}(20,20)
\fmfleft{l1}\fmfright{r1}\fmfforce{0.25w,0.5h}{v1}\fmfforce{0.75w,0.5h}{v2}
\fmf{photon}{l1,v1}\fmf{photon,left=1}{v1,v2,v1}
\fmf{photon}{v1,v2}
\fmf{photon}{r1,v2}
\end{fmfgraph}}
~ + \frac{1}{2}~
\parbox{20mm}{\begin{fmfgraph}(20,20)
\fmfleft{l1}\fmfright{r1}\fmfforce{0.25w,0.5h}{v1}\fmfforce{0.75w,0.5h}{v2}
\fmfforce{0.5w,0.75h}{vt}\fmfforce{0.5w,0.25h}{vb}
\fmf{photon}{l1,v1}
\fmf{photon,left=0.5}{v1,vt,v2,vb,v1}
\fmf{photon}{r1,v2}\fmf{photon}{vt,vb}
\end{fmfgraph}}
~ - ~
\parbox{20mm}{\begin{fmfgraph}(20,20)
\fmfleft{l1}\fmfright{r1}\fmfforce{0.25w,0.5h}{v1}\fmfforce{0.75w,0.5h}{v2}
\fmfforce{0.5w,0.75h}{vt}\fmfforce{0.5w,0.25h}{vb}
\fmf{photon}{l1,v1}
\fmf{scalar,left=0.5}{v1,vt}
\fmf{dashes,left=0.5}{vt,v2}
\fmf{scalar,left=0.5}{v2,vb}
\fmf{dashes,left=0.5}{vb,v1}
\fmf{photon}{r1,v2}\fmf{photon}{vt,vb}
\end{fmfgraph}}\\
\nonumber & - &
\parbox{20mm}{\begin{fmfgraph}(20,20)
\fmfleft{l1}\fmfright{r1}\fmfforce{0.25w,0.5h}{v1}\fmfforce{0.75w,0.5h}{v2}
\fmfforce{0.5w,0.75h}{vt}\fmfforce{0.5w,0.25h}{vb}
\fmf{photon}{l1,v1}
\fmf{dashes,left=0.5}{v1,vt}
\fmf{dashes,left=0.5}{vb,v1}
\fmf{photon,left=0.5}{vt,v2,vb}
\fmf{photon}{r1,v2}\fmf{scalar}{vt,vb}
\end{fmfgraph}}
~ - ~
\parbox{20mm}{\begin{fmfgraph}(20,20)
\fmfleft{l1}\fmfright{r1}\fmfforce{0.25w,0.5h}{v1}\fmfforce{0.75w,0.5h}{v2}
\fmfforce{0.5w,0.75h}{vt}\fmfforce{0.5w,0.25h}{vb}
\fmf{photon}{l1,v1}
\fmf{dashes,right=0.5}{v2,vt}
\fmf{dashes,right=0.5}{vb,v2}
\fmf{photon,right=0.5}{vt,v1,vb}
\fmf{photon}{r1,v2}\fmf{scalar}{vt,vb}
\end{fmfgraph}}
~ - 2~
\parbox{20mm}{\begin{fmfgraph}(20,20)
\fmfleft{l1}\fmfright{r1}\fmfforce{0.25w,0.5h}{v1}\fmfforce{0.75w,0.5h}{v2}
\fmftop{vt}\fmfforce{0.5w,0.25h}{vb}
\fmf{photon}{l1,v1}
\fmf{photon}{r1,v2}
\fmf{dashes,left=0.5,tension=3}{v1,v3}\fmf{dashes,left=0.5,tension=3}{v4,v2}
\fmf{photon,left=1,tension=0.4}{v3,v4}\fmf{dashes,left=1,tension=0.4}{v4,v3}
\fmf{scalar,left=1}{v2,v1}
\fmf{phantom,tension=3}{v3,vt,v4}
\end{fmfgraph}}\\
& + & \frac{1}{2}~
\parbox{20mm}{\begin{fmfgraph}(20,20)
\fmfleft{l1}\fmfright{r1}\fmfforce{0.25w,0.5h}{v1}\fmfforce{0.75w,0.5h}{v2}
\fmfforce{0.5w,0.75h}{vt}\fmfforce{0.5w,0.25h}{vb}
\fmf{photon}{l1,v1}
\fmf{photon,left=0.5}{v1,vt,v2,vb,v1}
\fmf{photon}{r1,v2}
\fmf{photon,right=0.5}{vt,v2}
\end{fmfgraph}}
~ + \frac{1}{2}~
\parbox{20mm}{\begin{fmfgraph}(20,20)
\fmfleft{l1}\fmfright{r1}\fmfforce{0.25w,0.5h}{v1}\fmfforce{0.75w,0.5h}{v2}
\fmfforce{0.5w,0.75h}{vt}\fmfforce{0.5w,0.25h}{vb}
\fmf{photon}{l1,v1}
\fmf{photon,left=0.5}{v1,vt,v2,vb,v1}
\fmf{photon}{r1,v2}
\fmf{photon,right=0.5}{v1,vt}
\end{fmfgraph}}
~ + \frac{1}{4}~
\parbox{20mm}{\begin{fmfgraph}(20,20)
\fmfleft{l1}\fmfright{r1}\fmfforce{0.25w,0.5h}{v1}\fmfforce{0.75w,0.5h}{v3}
\fmfforce{0.5w,0.5h}{v2}
\fmf{photon}{l1,v1}
\fmf{photon,left=1}{v1,v2,v1}
\fmf{photon}{r1,v3}
\fmf{photon,left=1}{v3,v2,v3}
\end{fmfgraph}}
\end{eqnarray}
where it should be noted that the ``figure eight'' diagram (proportional to
$1/4$) is strictly finite, but is included for completeness. An actual
computation yields
\begin{eqnarray}
\label{eq:2loopbare}
 \nonumber &&\Pi^{\text{B}(2,\epsilon)\text{UV}}_{\mu\nu}=\frac{g^4 N^2}{\pi^2}
\frac{p^{4\epsilon}}{\mu^{4\epsilon}} \Bigg[ \frac{(\xi + 2)(\xi^2 + 2\xi +
1)}{768\epsilon}g_{\mu\nu} \\
\nonumber &&\quad+~ \frac{8(7\xi^3 + 75\xi^2 + 221\xi + 233) -
18\zeta(2)(\xi^2+3)(\xi^2 + 2\xi^2 + 17)}{12288}
\Tproj_{\mu\nu} \\
&&\quad-~\frac{7\xi^3 + 32\xi^2 + 79\xi + 42}{768}\Lproj_{\mu\nu} \Bigg ].
\end{eqnarray}
The IR regulation is achieved via the introduction of a fictitious mass $m^2$ in
the denominators of both of the divergent diagrams. In
$\Dim=3+2\epsilon$, the
IR regulated integrals have the following form
\begin{eqnarray}
\label{eq:Snowcone}
\parbox{20mm}{\begin{fmfgraph*}(20,20)
\fmfleft{l1}\fmfright{r1}\fmftop{t1}\fmfforce{0.5w,0.5h}{v1}
\fmf{photon}{l1,v1}\fmf{phantom,tension=5.0}{t1,v2}
\fmf{photon,left=1,tension=0.4}{v1,v2,v1}
\fmf{photon}{r1,v1}
\fmfIR{v2}
\end{fmfgraph*}}
&=&
g^2\int_{q} V^{(0)}_ {\mu\nu\alpha\beta}
\frac{\Pi^{\text{B}(1,\epsilon)}}{\mu^{2\epsilon}}\frac{g^{\alpha\beta} - \frac{q^\alpha
q^\beta}{q^2+m^2}}{(q^2+m^2)^{\frac{3}{2}
- \epsilon}}\\
 \label{eq:Clam}
\parbox{20mm}{\begin{fmfgraph*}(20,20)
\fmfleft{l1}\fmfright{r1}
\fmfforce{0.25w,0.5h}{v1}\fmfforce{0.75w,0.5h}{v3}
\fmfforce{0.5w,0.25h}{v2}\fmfforce{0.5w,0.75h}{v4}
\fmf{photon}{l1,v1}\fmf{photon,left=0.5,tension=0.4}{v1,v4,v3,v2,v1}
\fmf{photon}{r1,v3}
\fmfIR{v4}
\end{fmfgraph*}}
&=& g^2\int_{q}
V^{(0)}_{\mu\alpha\delta}V^{(0)}_{\nu\beta\kappa}
\frac{\Pi^{\text{B}(1,\epsilon)}}{\mu^{2\epsilon}}\frac{g^{\alpha\beta} -
\frac{q^\alpha q^\beta}{q^2+m^2}}{(q^2+m^2)^{\frac{3}{2} - \epsilon}}\frac{
g^{\delta\kappa} - (1-\xi)\frac{(p+q)^\delta (p+q)^\kappa}{(p+q)^2}}{(p+q)^2}.
\end{eqnarray}
Including the IR regulator $m^2$ in exactly this manner was found to yield
\textit{relatively} simple and compact final expressions. Adding together these two
diagrams,
\begin{equation}
 \Pi^{\text{B}(2,\epsilon) \text{IR Reg}}_{\mu\nu} = ~
\parbox{20mm}{\begin{fmfgraph*}(20,20)
\fmfleft{l1}\fmfright{r1}
\fmfforce{0.25w,0.5h}{v1}\fmfforce{0.75w,0.5h}{v3}
\fmfforce{0.5w,0.25h}{v2}\fmfforce{0.5w,0.75h}{v4}
\fmf{photon}{l1,v1}\fmf{photon,left=0.5,tension=0.4}{v1,v4,v3,v2,v1}
\fmf{photon}{r1,v3}
\fmfIR{v4}
\end{fmfgraph*}}
~+ \frac{1}{2}~
\parbox{20mm}{\begin{fmfgraph*}(20,20)
\fmfleft{l1}\fmfright{r1}\fmftop{t1}\fmfforce{0.5w,0.5h}{v1}
\fmf{photon}{l1,v1}\fmf{phantom,tension=5.0}{t1,v2}
\fmf{photon,left=1,tension=0.4}{v1,v2,v1}
\fmf{photon}{r1,v1}
\fmffreeze
\fmfIR{v2}
\end{fmfgraph*}}
\end{equation}\end{fmffile}%
we obtain
\begin{eqnarray}
\nonumber &&\Pi^{\text{B}(2,\epsilon) \text{IR Reg}}_{\mu\nu} = \frac{g^4 N^2}{\pi^2}
\frac{p^{4\epsilon}}{\mu^{4\epsilon}} \Bigg [
-\frac{(\xi + 2)(\xi^2 + 2\xi + 1)}{768\epsilon}g_{\mu\nu}\\
\nonumber &&\quad+~ \frac{1}{2304} \bigg (
3(\xi+2)(\xi^2+2\xi+11) \log \frac{16p^4}{m^4}+\frac{m^4}{p^4}
(9\xi^3 + 15\xi^2 + 93\xi -33)\\
\nonumber &&\qquad+~ \frac{m^2}{p^2}(-6\xi^3+48\xi^2+54\xi+660) +
(15\xi^3+148\xi^2+299\xi+830)\\
\nonumber &&\qquad-~ 3(\xi^2+2\xi+11) \Big [
\frac{(3\xi-1)m^6 +(11-3\xi)(m^4p^2+m^2p^4) + (2\xi-2)p^6}{p^4(p^2+m^2)}\\
\nonumber &&\quad\qquad+~\frac{\tanh^{-1}\sqrt{\frac{p^2}{p^2+m^2}}}{p^3(p^2+m^2)^{3/2}}
\big ( (4\xi+8)m^6 + (9\xi+43)p^2m^4 + (4\xi+34)p^4m^2 \big )
\Big ] \bigg ) \Tproj_{\mu\nu} \\
\nonumber &&\quad+~ \frac{1}{2304}\bigg ( 3(\xi + 2)(\xi^2+2\xi+11)
\log \frac{16p^4}{m^4} - \frac{m^2}{p^2}(24\xi^3 + 96\xi^2 + 360\xi + 528)\\
\nonumber &&\qquad+~ (21\xi^3 + 100\xi^2 + 245\xi + 170)\\
\nonumber
&&\qquad+~6(\xi^2+2\xi+11)
\frac{\tanh^{-1}\sqrt{\frac{p^2}{p^2+m^2}}}{p^3(p^2+m^2)^{1/2}} \big (
(4\xi + 8)m^4 \\
&&\quad\qquad-~ (\xi-1) m^2 p^2 -(2\xi+4) p^4 \big ) \bigg ) \Lproj_{\mu\nu} \Bigg].
\label{eq:Pi2IRReg}
\end{eqnarray}
As expected all of the UV divergences cancel between diagrams, and hence we can
safely take the $\epsilon \rightarrow 0$ limit and be left with
something finite,
\begin{equation}
\lim_{\epsilon \rightarrow 0} \left( \Pi^{\text{B}(2,\epsilon)\text{UV}}_{\mu\nu} +
\Pi^{\text{B}(2,\epsilon)\text{IR Reg}}_{\mu\nu} \right)  \neq \pm \infty .
\end{equation}
To be able to ensure that $G$ reproduces the correct subleading
$\bigO(p^{-3})$ behavior, we simply need to know the
one-loop gluon self-energy.
$\Pi^{\text{B}(1)}$ as referenced earlier in this paper is given by
\begin{equation}
 g^2\Pi^{\text{B}(1)} = \Pi^{\text{B}(1,0)} .
\end{equation}

\subsection{Ghost Self-Energy}

\label{sec:ghostself}

Knowledge of the one-loop ghost self-energy is necessary
to guarantee the correct $\bigO(p^{-3})$ UV behavior of $\Delta$. There is only a single diagram,
\begin{fmffile}{ghostselfenergy}
\begin{equation}
 \Sigma^{\text{B}(1,\epsilon)} = ~
\parbox{20mm}{\begin{fmfgraph}(20,20)
\fmfleft{l1}\fmfright{r1}
\fmfforce{0.25w,0.5h}{v1}\fmfforce{0.75w,0.5h}{v3}
\fmf{dashes}{l1,v1}
\fmf{photon,left=1}{v1,v3}\fmf{scalar,right=1}{v1,v3}
\fmf{dashes}{r1,v3}
\end{fmfgraph}}
\end{equation}
which in $\Dim=3$ is independent of $\xi$ and equals
\begin{equation}
 \Sigma^{\text{B}(1,0)} (p) = p g^2 \Sigma^{\text{B}(1)} = p g^2 \frac{N}{16} .
\end{equation}

All of the
bare diagrams that contribute at $\bigO(g^4)$ are UV finite, so there is no need to carry out this calculation to the next order.
\end{fmffile}%
\subsection{Three-Gluon and Ghost-Gluon Vertices}\label{sec:vertex}

In this section we will present our results for the one-loop corrections to the three-gluon and ghost-gluon vertices valid for arbitrary covariant gauge in $3\Dim$. The generalization to arbitrary $\Dim$ is available in the literature \cite{Davydychev}.

When working in covariant gauge, one encounters scalar integrals of the form
\begin{equation}
C^0_{\alpha\beta\gamma} = \int \frac{d^\Dim q}{(2\pi)^\Dim}
\frac{1}{\big [ (q-p_2)^2 \big ]^\alpha \big [(q+p_1)^2 \big ]^\beta
\big [ q^2 \big ]^\gamma}
\end{equation}
where $\alpha$, $\beta$ and $\gamma$ can take on values between $-3$ and
$2$ over the course of the calculation. In three dimensions, all of the
triangle integrals with integer $\alpha$, $\beta$ and $\gamma$ are
finite in DR. Moreover, they can all be expressed in terms of
\begin{eqnarray}
 C^0_{111} &=& \frac{1}{8 p_1 p_2 p_3}\\
 C^0_{011} &=& \frac{1}{8p_1}\\
 C^0_{101} &=& \frac{1}{8p_2}\\
 C^0_{110} &=& \frac{1}{8p_3}
\end{eqnarray}
with $p_3^2 = (p_1+p_2)^2$ and $p_i = \sqrt{p_i^2}$. The relations between triangle integrals with different $\alpha$, $\beta$ and $\gamma$ can be obtained from the generic expression for $C^0_{\alpha\beta\gamma}$ which is known in terms of Appel's hypergeometric function \cite{Boos}
\begin{equation}
 F_4(a,b;c,d \vert x,y) = \sum_{i=0}^\infty \sum_{j=0}^\infty
\frac{(a)_{i+j}(b)_{i+j}}{(c)_i(d)_j} \frac{x^i}{i!} \frac{y^j}{j!}
\end{equation}
making use of the Pockhammer symbol $(a)_i = \Gamma(a+i)/ \Gamma(a)$. In any (Euclidean) dimension, omitting the $\MSbar$ scale, the expression reads
\begin{eqnarray}
\nonumber &&C^0_{\alpha\beta\gamma} = \frac{1}{(4\pi)^{\Dim/2}
\Gamma(\gamma)\Gamma(\beta)\Gamma(\alpha)\Gamma(\Dim-\gamma-\beta-\alpha)}\bigg[ \\
\nonumber && \quad (p_3^2)^{\Dim/2 -
\gamma-\beta-\alpha}\Gamma(\gamma)\Gamma\left(\gamma+\beta+\alpha-\Dim/2
\right)\Gamma\left(\Dim/2-\gamma-\beta\right)\Gamma\left(\Dim/2
-\gamma-\alpha\right) \\
\nonumber && \qquad \times~
F_4\left(\gamma,\gamma+\beta+\alpha-\Dim/2;\gamma+\beta-\Dim/2+1,
\gamma+\alpha-\Dim/2+1 \Big \vert \frac{p_1^2}{p_3^2},\frac{p_2^2}{p_3^2}
\right) \\
\nonumber &&\quad+~(p_2^2)^{\Dim/2 -
\gamma-\alpha}(p_3^2)^{-\beta}\Gamma(\beta)\Gamma\left(\Dim/2-\alpha\right)
\Gamma\left(\Dim/2-\gamma-\beta\right)\Gamma\left(\gamma+\alpha-\Dim/2
\right)\\
\nonumber &&\qquad \times~ F_4\left(
\beta,\Dim/2-\alpha;\gamma+\beta-\Dim/2+1,\Dim/2
-\gamma-\alpha+1 \Big \vert\frac{p_1^2}{p_3^2},\frac{p_2^2}{p_3^2} \right)\\
\nonumber &&\quad+~(p_1^2)^{\Dim/2 -
\gamma-\beta}(p_3^2)^{-\alpha} \Gamma(\alpha)\Gamma\left(\Dim/2-\beta\right)
\Gamma\left(\Dim/2-\gamma-\alpha\right)\Gamma\left(\gamma+\beta-\Dim/2
\right)\\
\nonumber &&\qquad\times~ F_4\left(
\alpha,\Dim/2-\beta;\Dim/2-\gamma-\beta+1,\gamma+\alpha-\Dim/2
+1
\Big \vert\frac{p_1^2}{p_3^2},\frac{p_2^2}{p_3^2} \right) \\
\nonumber &&\quad+~\Gamma(\Dim-\gamma-\beta-\alpha)
(p_1^2)^{\Dim/2-\gamma-\beta}
(p_2^2)^{\Dim/2-\gamma-\alpha}
(p_3^2)^{\gamma-\Dim/2}\\
\nonumber &&\qquad\times~ \Gamma\left(\Dim/2-\gamma\right)
\Gamma\left(\gamma+\beta-\Dim/2\right)\Gamma\left(\gamma+\alpha-\Dim/2
\right) \\
&&\qquad\times~
F_4\left(\Dim-\gamma-\beta-\alpha,\Dim/2-\gamma;\Dim/2-\gamma- \beta
+ 1,\Dim/2-\gamma-\alpha+1 \Big \vert
\frac{p_1^2}{p_3^2},\frac{p_2^2}{p_3^2}\right)\bigg].\qquad\quad
\end{eqnarray}
At the one-loop level, the three-gluon and ghost-gluon vertices have the following form
\begin{eqnarray}
\nonumber gV^{(1)a_1 a_2 a_3}_{\mu_1\mu_2\mu_3} (p_1,p_2,p_3) &=& gF^{a_1a_2a_3} \big[ A^{(1)}(p_1,p_2;p_3)
\mathbf{A}_{\mu_1\mu_2\mu_3} + B^{(1)}(p_1,p_2;p_3)
\mathbf{B}_{\mu_1\mu_2\mu_3} \\
\nonumber &&+ ~C^{(1)}(p_1,p_2;p_3) \mathbf{C}_{\mu_1\mu_2\mu_3}+
F^{(1)}(p_1,p_2;p_3)\mathbf{F}_{\mu_1\mu_2\mu_3}\\
\nonumber &&+ ~H^{(1)}(p_1,p_2,p_3)
\mathbf{H}_{\mu_1\mu_2\mu_3} + S^{(1)}(p_1,p_2,p_3)
\mathbf{S}_{\mu_1\mu_2\mu_3}\\
\label{eq:VAppStrucFuncs} &&+ ~ \text{cyclic perms.} ~ \big] \\
 g\Vgh^{(1) a_1 a_2 a_3}_{\mu_3}(p_1,p_2,p_3) &=& g F^{a_1a_2a_3}\big[ \mathbb{A}^{(1)}(p_1,p_2,p_3) p_{1\mu_3} +
 \mathbb{B}^{(1)}(p_1,p_2,p_3)p_{2\mu_3} \big]
\end{eqnarray}
with the tensors $\mathbf{A}$ through $\mathbf{S}$ as defined in Section \ref{sec:ansatz}. The vertex functions are as follows:
\begin{eqnarray}\label{eq:AFunction}
&&A^{(1)}(p_1,p_2;p_3) = - \frac{g^2 N}{1024 p_1^3 p_2^3 p_3 (p_1 {+} p_2 {+} p_3)^2} \Big[ 16 p_1^2 p_2^2 (p_1 {+} p_2 {+} p_3) \big[4 (p_1 {-} p_2)^2 (p_1 {+} p_2) \nonumber\\
&&\quad + ~(5 p_1^2 {+} 6 p_1 p_2 {+} 5 p_2^2) p_3 + 3 (p_1 {+} p_2) p_3^2 + 6 p_3^3\big] -
   4 \big[(p_1^2 {-} p_2^2)^2 (p_1^4 {+} 2 p_1^3 p_2 \nonumber\\
&&\quad +~ 4 p_1^2 p_2^2 {+} 2 p_1 p_2^3 {+} p_2^4) - 2 p_1 p_2 (p_1 {+} p_2) (p_1^4 {-} 4 p_1^3 p_2 {-} 2 p_1^2 p_2^2 {-} 4 p_1 p_2^3 {+} p_2^4) p_3\nonumber\\
&&\quad {}- (p_1^6 {-} 11 p_1^4 p_2^2 {-} 16 p_1^3 p_2^3 {-} 11 p_1^2 p_2^4 {+} p_2^6) p_3^2 + 8 p_1^2 p_2^2 (p_1 {+} p_2) p_3^3\nonumber\\
&&\quad {}- (p_1 {-} p_2)^2 (p_1^2 {+} 4 p_1 p_2 {+} p_2^2) p_3^4 +
     2 p_1 p_2 (p_1 {+} p_2) p_3^5 + (p_1^2 {+} p_2^2) p_3^6 \big] (1{-}\xi) \nonumber\\
&&\quad{}+ (p_1 {+} p_2 {+} p_3)^2 \big [(p_1 {-} p_2)^2 (p_1^4 {+} 2 p_1^3 p_2 {+} 2 p_1 p_2^3 {+} p_2^4) -
     2 (p_1 {+} p_2)^3 (p_1^2 {-} 3 p_1 p_2 {+} p_2^2) p_3\nonumber\\
&&\quad {}+ 2 (p_1^4 {-} p_1^2 p_2^2 {+} p_2^4) p_3^2 - 2 (p_1^3 {+} p_2^3) p_3^3 + (p_1^2 {+} p_2^2) p_3^4\big] (1{-}\xi)^2 \Big]
\end{eqnarray}
\begin{eqnarray}
&&B^{(1)}(p_1,p_2;p_3) = - \frac{g^2 N(p_1 - p_2)}{1024 p_1^3 p_2^3 p_3^3 (p_1 + p_2 + p_3)^2}
 \Big[16 p_1^2 p_2^2 p_3^2 \big[2 (p_1 + p_2)^3\nonumber\\
&&\quad -~ 9 (p_1 + p_2)^2 p_3 - 20 (p_1 + p_2) p_3^2 - 9 p_3^3\big] - 4 \big[2 p_1^2 (p_1 - p_2)^2 p_2^2 (p_1 + p_2)^3\nonumber\\
&&\quad +~ 4 p_1^2 p_2^2 (p_1^2 - p_2^2)^2 p_3 - (p_1 + p_2) (p_1^6 + 2 p_1^5 p_2 - 3 p_1^4 p_2^2 - 4 p_1^3 p_2^3 - 3 p_1^2 p_2^4 +
        2 p_1 p_2^5 + p_2^6) p_3^2\nonumber\\
&&\quad +~ 2 p_1 p_2 (p_1 + p_2)^2 (p_1^2 - 6 p_1 p_2 + p_2^2) p_3^3 + (p_1 + p_2) (p_1^4 - 26 p_1^2 p_2^2 + p_2^4) p_3^4 - 12 p_1^2 p_2^2 p_3^5
\nonumber\\
&&\quad + (p_1 + p_2)^3 p_3^6 - 2 p_1 p_2 p_3^7 - (p_1 + p_2) p_3^8\big] (\xi-1)\nonumber\\
&&\quad + (p_1 + p_2 + p_3)^2 \big[2 p_1^2 (p_1 - p_2)^2 p_2^2 (p_1 + p_2) - (p_1 + p_2) (p_1^4 + p_2^4) p_3^2\nonumber\\
&&\quad +~
      2 (p_1 - p_2)^2 (p_1^2 + 3 p_1 p_2 + p_2^2) p_3^3 - 2 (p_1 + p_2) (p_1^2 + p_2^2) p_3^4\nonumber\\
&&\quad +~ 2 (p_1^2 + p_1 p_2 + p_2^2) p_3^5 - (p_1 + p_2) p_3^6\big] (1-\xi)^2\Big]
\end{eqnarray}
\begin{eqnarray}
&& C^{(1)}(p_1,p_2;p_3) = \frac{g^2N}{512 p_1^3 p_2^3 (p_1 + p_2) p_3 (p_1 + p_2 + p_3)^2} \Big [48 p_1^2 p_2^2 (p_1 + p_2)^2 p_3
\nonumber\\
&&\quad +~ \big[(p_1 - p_2)^2 (p_1 + p_2)^5 - 2 p_1^3 p_2 (p_1 + p_2)^2 p_3 - 2 p_1 p_2^3 (p_1 + p_2)^2 p_3
 - p_1^4 (p_1 + p_2) p_3^2\nonumber\\
&&\quad -~
   (+p_1 + p_2) p_2^4 p_3^2 - (p_1 + p_2)^3 p_3^4 + 2 p_1 p_2 p_3^5 + (p_1 + p_2) p_3^6\big] (\xi-1) (3 + \xi)
\nonumber\\
&&\quad +~ 2 p_1^2 (p_1 + p_2) p_2^2 p_3^2 (81 + 5 [2 + \xi]\xi)
+4 p_1^2 p_2^2 p_3^3 (29 + [6 + \xi]\xi) \Big]
\end{eqnarray}
\begin{eqnarray}
&& F^{(1)}(p_1,p_2;p_3) = -\frac{g^2N}{512 p_1^3 p_2^3 (p_1 + p_2) p_3^3 (p_1 + p_2 + p_3)^3} \bigg[ \big[-(p_1 - p_2)^2 (p_1 + p_2)^6\nonumber\\
&&\quad-~ 3 (p_1 - p_2)^2 (p_1 + p_2)^5 p_3\big] (\xi-1)^2 (3 + \xi)  +
\Big[(p_1 + p_2) p_3^7 (-3 + \xi) (3 + \xi) \nonumber\\
&&\quad+~ (3 + \xi) (-20 p_1^3 p_2 (p_1 + p_2)^2 p_3^2 - 20 p_1 p_2^3 (p_1 + p_2)^2 p_3^2 - p_1^4 (p_1 + p_2) p_3^3 (9 + \xi)\nonumber\\
&&\quad -~ p_2^4 (p_1 + p_2) p_3^3 (9 + \xi) + p_1^4 p_3^4 (11 + \xi) + 4 p_1^3 p_2 p_3^4 (11 + \xi) + 4 p_1 p_2^3 p_3^4 (11 + \xi) \nonumber\\
&&\quad+~ p_2^4 p_3^4 (11 + \xi)
- p_1^4 (p_1 + p_2)^2 p_3^2 (7 + 3 \xi) - p_2^4 (p_1 + p_2)^2 p_3^2 (7 + 3 \xi))\Big](\xi-1)\nonumber\\
&&\quad +~ 2 p_1^2 p_2^2 (p_1 + p_2)^2 p_3^2 \big(-113 + 3 \xi (-1 + [-5 + \xi] \xi)\big)\nonumber\\
&&\quad +~
 2 p_1^2 p_2^2 (p_1 + p_2) p_3^3 \big(-303 + \xi (-25 + [23 + \xi]\xi)\big)\nonumber\\
&&\quad +~ 2 p_1^2 p_2^2 p_3^4 \big(-215 + \xi (33 +[35 + 3 \xi]\xi)\big)
+  \Big[3 (p_1 + p_2)^3 p_3^5(\xi-1)\nonumber\\
&&\quad +~ 12 (p_1 + p_2) (p_1^2 + p_2^2) p_3^5 - p_3^6 \big[5 p_1^2 + 14 p_1 p_2 + 5 p_2^2\nonumber\\
&&\quad-~ 3 (p_1 + p_2)^2 \xi\big]\Big] (1 - \xi) (-3 - \xi) \bigg]
\end{eqnarray}
\begin{eqnarray}\label{eq:HFunction}
&& H^{(1)}(p_1,p_2,p_3) = \frac{g^2N}{1024 p_1^3 p_2^3 p_3^3 (p_1 + p_2 + p_3)^3} \bigg[ \Big[p_1^9 + 6\big[ p_1^7 p_2 p_3 -  p_1^5 p_2^3 p_3 -  p_1^3 p_2^5 p_3 -  p_1^5 p_2 p_3^3\nonumber\\
&&\quad -~  p_1^3 p_2 p_3^5\big] + 3 p_1^8 (p_2 + p_3) + 3 p_1 p_2^4 (p_2 - p_3)^2 (p_2 + p_3)^2 +
   6 p_1 p_2^3 (p_2 - p_3)^2 p_3 (p_2 + p_3)^2\nonumber\\
&&\quad +~ 6 p_1 p_2 (p_2 - p_3)^2 p_3^3 (p_2 + p_3)^2 + 3 p_1 (p_2 - p_3)^2 p_3^4 (p_2 + p_3)^2 + p_2^4 (p_2 - p_3)^2 (p_2 + p_3)^3\nonumber\\
&&\quad +~ 2 p_2^3 (p_2 - p_3)^2 p_3 (p_2 + p_3)^3 + 2 p_2 (p_2 - p_3)^2 p_3^3 (p_2 + p_3)^3\nonumber\\
&&\quad +~ (p_2 - p_3)^2 p_3^4 (p_2 + p_3)^3\Big] (\xi-1)^2 (3 + \xi) + \Big[ \big[2 p_1^2 p_2^5 p_3 (p_2 + p_3)\nonumber\\
&&\quad +~ 2 p_1^2 p_2 p_3^5 (p_2 + p_3)\big] (13 - \xi) + 2\big[ p_1^3 p_2^6 +  p_1^3 p_3^6 +  p_1^6 p_2^2 (p_2 + p_3)
+  p_1^6 p_3^2 (p_2 + p_3)\big] (7 - \xi)\nonumber\\
&&\quad +~ \big[2 p_1^7 p_2^2 - 4 p_1^5 p_2^4 + 2 p_1^7 p_3^2- 4 p_1^5 p_3^4 - 4 p_1^4 p_2^4 (p_2 + p_3) + 2 p_1^2 p_2^6 (p_2 + p_3)\nonumber\\
&&\quad -~ 4 p_1^4 p_3^4 (p_2 + p_3) +
     2 p_1^2 p_3^6 (p_2 + p_3)
+ 6 p_1 p_2^2 (p_2 - p_3)^2 p_3^2 (p_2 + p_3)^2\nonumber\\
&&\quad +~ 2 p_2^2 (p_2 - p_3)^2 p_3^2 (p_2 + p_3)^3\big] (5 + \xi) +
   2 p_1^6 p_2 p_3 (p_2 + p_3) (11 + \xi)\nonumber\\
&&\quad -~ \big [2 p_1^4 p_2^3 p_3 (p_2 + p_3) + 2 p_1^4 p_2 p_3^3 (p_2 + p_3)\big] (23 + \xi)\Big] (\xi-1) (3 + \xi)\nonumber\\
 &&\quad +~
 4 p_1^4 p_2^2 p_3^2 (p_2 + p_3) (78 - 5 \xi + 7 \xi^3) -
 2\big[ p_1^3 p_2^4 p_3^2 + p_1^3 p_2^2 p_3^4\big] \big(-225 + \xi (53 + [25 - 13 \xi] \xi)\big)\nonumber\\
&&\quad -~ 4 p_1^3 p_2^3 p_3^3 \big(-217 + 3 \xi (17 + [7 - 5 \xi] \xi)\big) + 4 p_1^5 p_2^2 p_3^2 \big(-18 + \xi(13 + [20 + \xi]\xi)\big)\nonumber\\
&&\quad +~
 \big[2 p_1^2 p_2^4 p_3^2 (p_2 + p_3) + 2 p_1^2 p_2^2 p_3^4 (p_2 + p_3)\big] \big(3 + \xi (-3 + [29 + 3 \xi]\xi)\big)\nonumber\\
&&\quad +~  4 p_1^2 p_2^3 p_3^3 (p_2 + p_3) \big(111 + \xi (-25 + [-27 + 5 \xi]\xi)\big) \bigg]
\end{eqnarray}
\begin{equation}
S^{(1)}(p_1,p_2,p_3) = 0
\end{equation}
and, for instance, the regulator $\omega^{(1)}$ mentioned in Section \ref{sec:ansatz} is included in the $A$ function by making the transformation
\begin{equation}
\label{eq:OneLoopVertexIrReg}
\frac{1}{p_1^3p_2^3p_3(p_1+p_2+p_3)^2} \rightarrow \frac{1}{(p_1 + \omega^{(1)})^3(p_2 + \omega^{(1)})^3(p_3+\omega^{(1)})(p_1+p_2+p_3+\omega^{(1)})^2} 
\end{equation}
and likewise for $B$ through $H$. The one-loop ghost vertices are
\begin{eqnarray}\label{eq:GhAFunction}
&& \mathbb{A}^{(1)}(p_1,p_2,p_3) = \frac{g^2N}{512 p_1 p_2 p_3^3 (p_1 + p_2 + p_3)}\Big[16 p_3^2 \big[-p_1^3 + p_1^2 (-p_2 + p_3)\nonumber\\
&&\quad -~ (p_2 - p_3)^2 (p_2 + p_3) + p_1 (3 p_2^2 + 2 p_2 p_3 + p_3^2)\big] +
  4 (p_1 - p_2 - p_3) \big[(p_1^2 - p_2^2)^2\nonumber\\
&&\quad +~ 2 p_1 (p_1 - p_2) (p_1 + p_2) p_3 + (5 p_1 - p_2) (p_1 + p_2) p_3^2 + 2 (p_1 + p_2) p_3^3 - 2 p_3^4\big] (1 - \xi)
\nonumber\\
&&\quad -~ (p_1^2 - p_2^2 - p_3^2) \big[(p_1 - p_2)^2 (p_1 + p_2) + (p_1 - p_2)^2 p_3 \nonumber\\
&&\quad +~ (p_1 + p_2) p_3^2 - 3 p_3^3] (1 - \xi)^2\Big]\,,\qquad\quad
\end{eqnarray}
\begin{eqnarray}\label{eq:GhBFunction}
&& \mathbb{B}^{(1)}(p_1,p_2,p_3) = -\frac{g^2N}{512 p_1 p_2 p_3^3 (p_1 + p_2 + p_3)} \bigg[32 p_1 p_3^2 \big[(p_1 - p_2) p_2 + p_3^2\big]\nonumber\\
&&\quad -~ 4 (p_1^2 - p_2^2 + p_3^2) \Big[p_1^3 + (p_2 - p_3) \big[ (p_2 + p_3)^2 - p_1(p_2 + 3 p_3) -p_1^2 \big]\Big] (1 - \xi)\nonumber\\
&&\quad +~ (p_1^2 - p_2^2 + p_3^2) \big[(p_1 - p_2)^2 (p_1 + p_2) + (p_1 - p_2)^2 p_3 \nonumber\\
&&\quad + ~ (p_1 + p_2) p_3^2 - 3 p_3^3\big] (1-\xi)^2\bigg]\,. \qquad\quad
\end{eqnarray}

\section{Phase space integrations}
\label{phasespace}

\begin{fmffile}{PhaseSpace}
In performing numerical integrals over vacuum diagrams encountered in
this paper we need efficient parametrizations of the phase space
integrals.  At two loops the most interesting diagram is the Setting
Sun, at three loops it is the Mercedes diagram.  All other diagrams can
be solved by being reduced to these two (as we will describe), so we will
concentrate on them.

In $\Dim$ dimensions an $n$-loop diagram involves $n\Dim$ real integrations.
However the symmetry group $\text{O}(\Dim)$ helps reduce this because certain
angular integrations are trivial.  Namely, there are $\Dim(\Dim-1)/2$ global
angular integrations.  Selecting $n$ $\Dim$-vectors reduces
$\text{O}(\Dim)$ to $\text{O}(\Dim-n)$ (for $n\leq \Dim-2$) or reduces it completely
(for $n \geq \Dim-1$).  Therefore, for $n \leq \Dim-2$,
$\Dim(\Dim-1)/2 - (\Dim-n)(\Dim-n-1)/2 = n\Dim - n(n+1)/2$ of the integrals
are global angular integrals which can be performed immediately since
none of the invariants depend on them.  This leaves
$n(n-1)/2$ nontrivial integrations, for $n \leq \Dim-2$.
For $n \geq \Dim-1$ there are $n\Dim - \Dim(\Dim-1)/2$ nontrivial
integrations.

In an $n$-loop connected vacuum diagram built entirely with 3-point
vertices there are $3n-3$ propagators.  For $\Dim=3$ and $n \geq 2$
this happens to equal the number of integration variables.  Therefore, in
$\Dim=3$ dimensions, in diagrams composed using 3-point vertices
and where each propagator has a distinct momentum (which is the case for
2PI or 3PI diagrams), it should be possible to arrange for the
integration variables to be precisely the magnitudes of the momenta on
all propagators.  This is a very convenient choice, provided that all
dot products of propagator momenta have simple enough expressions.

\subsection{Two loops: Setting Sun}

\label{sec:setsun}

We apply these ideas first to the Setting Sun diagram, that is, two
vertices connected by three lines:
\begin{figure}[h]
\centering
\begin{fmfgraph*}(30,30)
\fmfforce{1.5mm,0.5h}{v1}\fmfforce{28.5mm,0.5h}{v2}
\fmf{fermion,left=1,label=$k$}{v1,v2}
\fmf{fermion,left=1,label=$q$}{v2,v1}
\fmf{fermion,label=$p$}{v1,v2}
\end{fmfgraph*}
\end{figure}

The ``natural'' integration variables are
\begin{equation}
\int \frac{d^3 p d^3 k}{(2\pi)^6}
= \frac{8\pi^2}{(2\pi)^6} \int_0^\infty p^2 dp
\int_0^\infty k^2 dk \int_{-1}^{1} d\cos \theta_{pk}
\end{equation}
where we have performed the trivial integral over the Eulerian angles,
in the form of the direction of the $\vec p$ integral and the azimuthal
angle of $\vec k$ if $\vec p$ is taken as the $\vec z$ axis.

The dot product $\vec p \cdot \vec k = pk \cos \theta_{pk}$
and
\begin{equation}
q^2 = (\vec p + \vec k)^2 = p^2 + k^2 + 2 p k \cos \theta_{pk}
\quad \Rightarrow \quad
\cos \theta_{pk} = \frac{q^2 - p^2 - k^2}{2 pk} \,.
\end{equation}
If we change variables from $p,k,\cos \theta_{pk}$ to
$p,k,q$ we should differentiate the above holding $p,k$ fixed, giving
\begin{equation}
d\cos \theta_{pk} = \frac{q}{pk} dq \,.
\end{equation}
Therefore we can rewrite the integration as
\begin{equation}
\int \frac{d^3 p d^3 k}{(2\pi)^6}
= \frac{1}{2^3 \pi^4} \int_0^\infty pdp \int_0^\infty kdk
\int_{|p-k|}^{p+k} qdq
= \frac{1}{2^6 \pi^4} \int_0^\infty dp^2 \int_0^\infty dk^2
\int_{(p-k)^2}^{(p+k)^2} dq^2
\end{equation}
which are a convenient set of integration variables.  In particular, all
dot products we will encounter can be written directly in terms of the
integration variables using
\begin{equation}
-\vec p \cdot \vec q = \frac{k^2-p^2-q^2}{2} \,, \quad
-\vec k \cdot \vec q = \frac{p^2-k^2-q^2}{2} \,, \quad
\vec p \cdot \vec k = \frac{q^2-p^2-k^2}{2} \,.
\end{equation}

The remaining two-loop diagram, the Figure-8, can be performed using the
same integration variables; the two lines have momentum $\vec p$ and
$\vec k$, so the $q$ integral can be done directly,
$\int qdq = 2pk$.  This sort of reduction always works, because we can
always consider a 4-point vertex to be two three-point vertices
connected by a propagator, with the propagator collapsed to a point.  So
diagrams containing 4-point vertices can be written with the same
variables as the diagram containing this ``collapsed'' propagator.

\subsection{Three loops: Mercedes}

\label{sec:mercedes}

Now we seek a similar set of integration variables for the Mercedes
diagram,
\begin{figure}[h]
\centering
\begin{fmfgraph*}(30,30)
\fmfforce{15mm,15mm}{v4}\fmfforce{3.309mm,7.75mm}{v1}
\fmfforce{26.691mm,7.75mm}{v2}
\fmfforce{15mm,28.5mm}{v3}
\fmf{fermion,label=$p$,right=0.566}{v1,v2}
\fmf{fermion,label=$k'$,left=0.566}{v1,v3}
\fmf{fermion,label=$q'$,right=0.566}{v2,v3}
\fmf{fermion,label=$k$}{v1,v4}
\fmf{fermion,label=$q$}{v2,v4}
\fmf{fermion,label=$l$}{v4,v3}
\end{fmfgraph*}
\end{figure}

Note that $\vec l = \vec k + \vec q$ and similarly
$\vec k' = -\vec p - \vec k$ and $\vec q' = \vec p - \vec q$.
The phase space is determined by the triple integral
\begin{equation}
\int \frac{d^3 p \: d^3 k \: d^3 q}{(2\pi)^9}
= \int_0^\infty p^2 dp \: k^2 dk \: q^2 dq \; \frac{16\pi^2}{(2\pi)^9}
\int_{-1}^1 d \cos \theta_{pk} \int_{-1}^1 d\cos \theta_{pq}
\int_0^\pi d\phi_{pk;pq}
\end{equation}
where $\phi$ is the azimuthal angle between the $pk$ plane and the $pq$
plane, we have used the symmetry of the $\phi$ integration to reduce it
from $[0,2\pi]$ to $[0,\pi]$, and Eulerian angles have again been
performed.

Using the same trick as before, we can rewrite this integral as
\begin{equation}
\frac{1}{2^5 \pi^7} \int_0^\infty dp \int_0^\infty kdk
\int_0^\infty qdq \int_{|p-k|}^{p+k} k'dk' \int_{|p-q|}^{p+q} q'dq'
\int_0^\pi d\phi_{pq;pk} \,.
\end{equation}
and we would like to rewrite the $\phi$ integral as an $l$ integral.
To do so, write out an expression for $l^2$:
\begin{eqnarray}
l^2 & = & (\vec k + \vec q)^2 = k^2 + q^2 + 2 \vec k \cdot \vec q \,, \\
\vec k \cdot \vec q & = & k q \left( \cos \theta_{pq} \cos \theta_{pk}
+ \sin \theta_{pq} \sin \theta_{pk} \cos \phi \right)
 = \frac{l^2-k^2-q^2}{2}   \\
k q \cos \theta_{pq} \cos \theta_{pk} & = &
\frac{(p^2 + q^2 - {q'}^2)({k'}^2 - p^2 - k^2)}{4p^2} \\
kq \sin \theta_{pq} \sin \theta_{pk} & = &
   \sqrt{ (k^2 - k^2 \cos^2 \theta_{pk})(q^2 - q^2 \cos^2 \theta_{pq} )}
\end{eqnarray}
and hence
\begin{equation}
\cos \phi = \frac{p^4 + 2 p^2 l^2 + k^2 q^2 + {k'}^2 {q'}^2
                  - ( q^2 {k'}^2 {+} {q'}^2 k^2 )
                  - p^2 ( k^2 {+} q^2 {+} {k'}^2 {+} {q'}^2 ) }
    { \sqrt{ (2p^2 q^2 {+} 2p^2 {q'}^2 {+} 2q^2 {q'}^2 {-} p^4
             {-} q^4 {-} {q'}^4)
             (2p^2 k^2 {+} 2p^2 {k'}^2 {+} 2k^2 {k'}^2 {-} p^4
              {-} k^4 {-} {k'}^4)}} \,.
\end{equation}
Since the range of $\cos \phi$ is from $-1$ to $+1$, we find that the
range of $l^2$ at fixed $p,k,q,k',q'$ is between
\begin{eqnarray}
&&\frac{1}{2p^2} \bigg[ \Big(p^2(k^2{+}q^2{+}{k'}^2{+}{q'}^2)
  + q^2{k'}^2 + k^2{q'}^2 - p^4 - k^2q^2 -{k'}^2{q'}^2 \Big)
  \vphantom{\sqrt{(2p^2)}}
\\ && \quad
\pm \sqrt{ (2p^2 q^2 {+} 2p^2 {q'}^2 {+} 2q^2 {q'}^2 {-}
             p^4 {-} q^4 {-} {q'}^4)
             (2p^2 k^2 {+} 2p^2 {k'}^2 {+} 2k^2 {k'}^2
             {-} p^4 {-} k^4 {-} {k'}^4)} \bigg]\,, \nonumber
\end{eqnarray}
where the $+$($-$) sign represents the maximum (minimum) allowed value
of $l^2$.

Differentiating the expression for $\cos \phi$ holding $p,k,q,k',q'$
fixed, we find
\begin{equation}
\sin \phi d\phi = \frac{4p^2 ldl}{\sqrt{(...)(...)}}
\end{equation}
where $\sqrt{(...)(...)}$ is the same long square root in the above
expressions.  Therefore
\begin{equation}
d\phi = \frac{4p^2 ldl}{\sin \phi \sqrt{(...)(...)}} \,.
\end{equation}
Writing $\sin\phi = \sqrt{1-\cos^2\phi}$ and after significant algebra
we find
\begin{eqnarray}
d\phi &  =  & \frac{2pldl}{\sqrt{X}} \,, \\
X & = & p^2 l^2   (k^2{+}{k'}^2{+}q^2{+}{q'}^2{-}p^2{-}l^2)
       +q^2{k'}^2 (k^2{+}{q'}^2{+}p^2{+}l^2{-}q^2{-}{k'}^2) \nonumber \\ &&
       +k^2{q'}^2 (q^2{+}{k'}^2{+}p^2{+}l^2{-}k^2{-}{q'}^2)
 - k^2 {k'}^2 p^2 {-} q^2 {q'}^2 p^2 {-} k^2 q^2 l^2 {-} {k'}^2 {q'}^2 l^2\,.
\label{eq:XJac}\end{eqnarray}
Note that the expression for $X$ has a symmetry, if hard to see.
The momenta are in three pairs; $(p,l)$, $(q,k')$, and $(q',k)$ which
are ``opposite'' momenta in the diagram (momenta which do not touch at a
vertex).  The first terms involve pairs of ``opposite'' momenta, the
last terms involve triples of momenta meeting at a vertex.

The total integration becomes
\begin{equation}
\frac{1}{2^4 \pi^7}
\int \frac{pdp\: kdk\: qdq\: k'dk'\: q'dq'\: ldl}{\sqrt{X}}
\end{equation}
with integration limits listed previously.  We have not written the
integration limits in a symmetric way, but they are symmetric.

The dot product of a pair of momenta which share a vertex
are of form
\begin{equation}
\vec p \cdot \vec k = \frac{k'^2 - p^2 - k^2}{2} \,, \quad
\vec p \cdot \vec q = \frac{p^2 + q^2 - {q'}^2}{2} \,
\end{equation}
where the sign difference is because in the first case the momenta are
both directed out of the vertex while in the latter case one momentum
enters and one exits the common vertex.
For momenta which do not share a vertex, the dot product is
\begin{equation}
\vec p \cdot \vec l = \vec p \cdot (\vec k + \vec q)
 = \frac{q^2 + {k'}^2 - k^2 - {q'}^2}{2}
\end{equation}
and similarly for $\vec k \cdot \vec q'$ and $\vec q \cdot \vec k'$.
(For a mnemonic, note that $q,k'$ are going from the beginning of one
line to the end of the other; $k,q'$ connect the beginnings of each line
or the ends of each line.)  We see that all dot products, including
those for momenta on lines which do not meet at a vertex, have simple
expressions in terms of momenta on lines.

As mentioned before, we can use the same integration variables for
3-loop diagrams with one or more 4-point vertices.
For instance, when the $l$ propagator is collapsed into a 4-point
vertex, one can immediately do the $l$ integral;
\begin{equation}
\int \frac{ldl}{\sqrt{X}} = \frac{\pi}{2p} \,.
\end{equation}
However, if the integrand involves dot products which depend on $l$ then
we cannot do the $l$ integral immediately; we should instead interpret
it as an angular integration which does not change the magnitudes of any
momenta on the remaining lines, but which does affect some of their dot
products.\end{fmffile}









\end{document}